\documentclass[aps,prd,twocolumn,showkeys,amsmath,amssymb]{revtex4}
\usepackage{graphicx}
\usepackage{epstopdf}   
\usepackage{multirow}
\usepackage{subfigure}
\usepackage{extarrows}
\usepackage{feynmf}
\usepackage[colorlinks,citecolor=blue,anchorcolor=red,menucolor=red,linkcolor=red,filecolor=red,runcolor=red,urlcolor=blue,frenchlinks=red]{hyperref}
\usepackage{enumitem}
\usepackage{color}

\newcommand{\feynp}[1]{#1\kern-0.45em/}
\def\dab{\int^{\alpha_{max}}_{\alpha_{min}}{\rm d}\alpha\int^{\beta_{max}}_{\beta_{min}}{\rm d}\beta}
\def\qq{\langle\bar qq\rangle}
\def\GGa{\langle GG\rangle}
\def\GGb{\langle g_s^2GG\rangle}

\def\qGqa{\langle\bar qGq\rangle}
\def\qGqb{\langle g_s\bar q\sigma Gq\rangle}
\def\FF(s){\left[(\alpha+\beta)m_c^2-\alpha\beta s\right]}
\def\HH(s){\left[m_c^2-\alpha(1-\alpha) s\right]}
\def\non{\\ \nonumber}

\begin{document}

\title{Covalent hadronic molecules induced by shared light quarks}

\author{Hua-Xing Chen}
\email{hxchen@seu.edu.cn}
\affiliation{
School of Physics, Southeast University, Nanjing 210094, China
}

\begin{abstract}
After examining Feynman diagrams corresponding to the $\bar D^{(*)} \Sigma_c^{(*)}$, $\bar D^{(*)} \Lambda_c$, $D^{(*)} \bar K^{*}$, and $D^{(*)} \bar D^{(*)}$ hadronic molecular states, we propose a possible binding mechanism induced by shared light quarks. This mechanism is similar to the covalent bond in chemical molecules induced by shared electrons. We use the method of QCD sum rules to calculate its corresponding light-quark-exchange diagrams, and the obtained results indicate a model-independent hypothesis: the light-quark-exchange interaction is attractive when the shared light quarks are totally antisymmetric so that obey the Pauli principle. We build a toy model with four parameters to formulize this picture, and estimate binding energies of some possibly-existing covalent hadronic molecules. A unique feature of this picture is that binding energies of the $(I)J^P = (0)1^+$ $D\bar B^*/D^* \bar B$ hadronic molecules are much larger than those of the $(I)J^P = (0)1^+$ $DD^*/\bar B \bar B^*$ ones, while the $(I)J^P = (1/2)1/2^+$ $\bar D \Sigma_c/\bar D \Sigma_b/B \Sigma_c/B \Sigma_b$ hadronic molecules have similar binding energies.
\end{abstract}

\date{\today}
\maketitle

\section{Introduction}
\label{sec:intro}

Since the discovery of the $X(3872)$ by Belle in 2003~\cite{Choi:2003ue}, lots of charmonium-like $XYZ$ states were discovered in the past two decades~\cite{pdg}. Some of these structures may contain four quarks and are good candidates for hidden-charm tetraquark states. In recent years the LHCb Collaboration continually observed six $P_c/P_{cs}$ states~\cite{Aaij:2015tga,Aaij:2019vzc,Aaij:2020gdg,LHCb:2021chn}, which contain five quarks and are good candidates for hidden-charm pentaquark states. Although there is still a long way to fully understand how the strong interaction binds these quarks and antiquarks together, the above exotic structures have become one of the most intriguing research topics in hadron physics. Their theoretical and experimental studies are significantly improving our understanding of the non-perturbative behaviors of the strong interaction at the low energy region. We refer to the reviews~\cite{Chen:2016qju,Liu:2019zoy,Chen:2022asf,Hosaka:2016pey,Richard:2016eis,Lebed:2016hpi,Esposito:2016noz,Ali:2017jda,Guo:2017jvc,Olsen:2017bmm,Karliner:2017qhf,Albuquerque:2018jkn,Guo:2019twa,Brambilla:2019esw,Yang:2020atz,Dong:2021juy} and references therein for detailed discussions.

\begin{figure*}[]
\begin{center}
\subfigure[~$\Pi_M$]{\includegraphics[width=0.3\textwidth]{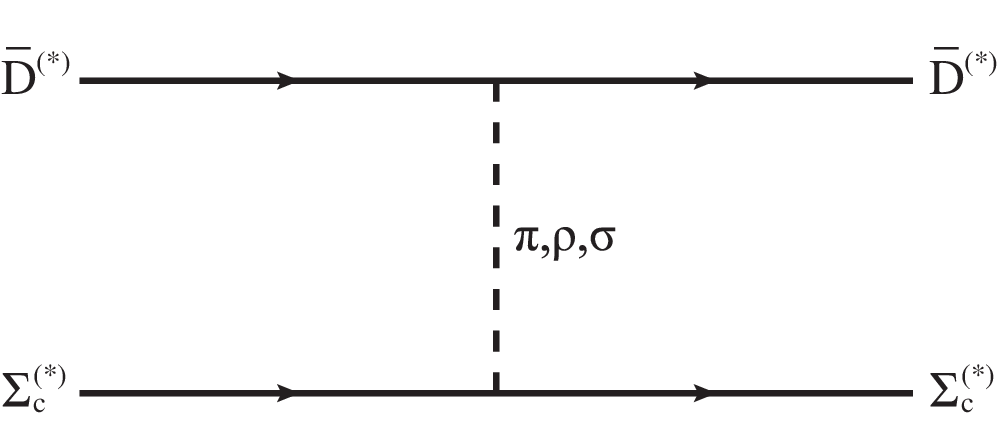}}
~~~~~
\subfigure[~$\Pi_G$]{\includegraphics[width=0.3\textwidth]{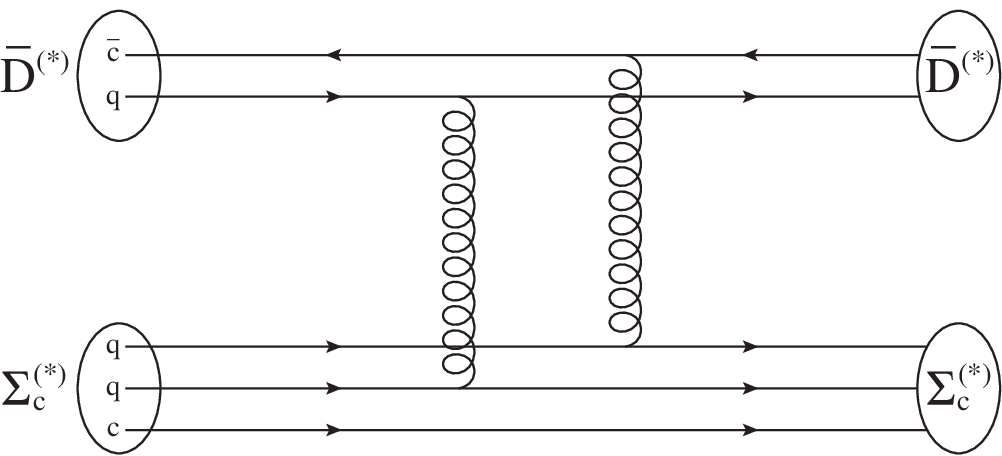}}
~~~~~
\subfigure[~$\Pi_Q$]{\includegraphics[width=0.3\textwidth]{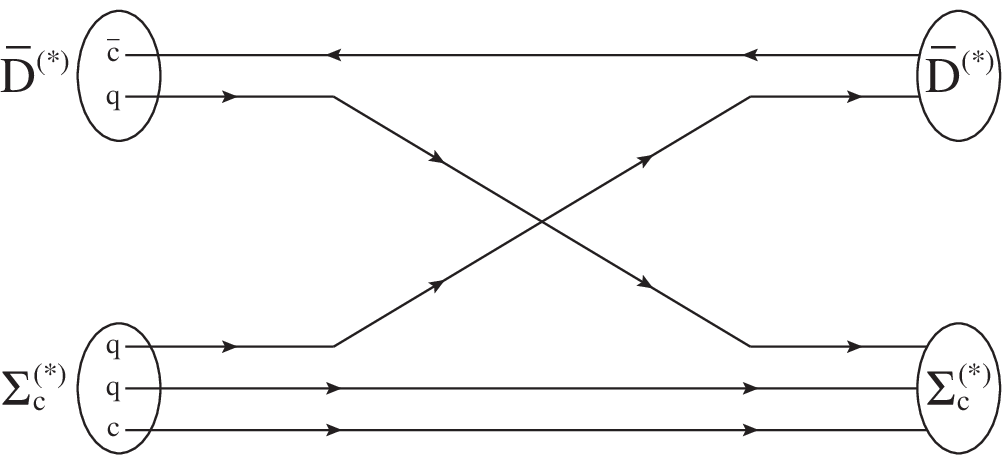}}
\caption{Feynman diagrams between $\bar D^{(*)}$ and $\Sigma_c^{(*)}$ corresponding to : a) the one-meson-exchange interaction $\Pi_{M}$, b) the double-gluon-exchange interaction $\Pi_{G}$, and c) the light-quark-exchange interaction $\Pi_{Q}$. Here $q$ denotes a light up/down quark.}
\label{fig:feynman}
\end{center}
\end{figure*}

Some of the $XYZ$ and $P_c/P_{cs}$ states can be interpreted as hadronic molecular states, which consist of two conventional hadrons~\cite{Weinberg:1965zz,Voloshin:1976ap,DeRujula:1976zlg,Tornqvist:1993ng,Voloshin:2003nt,Close:2003sg,Wong:2003xk,Braaten:2003he,Swanson:2003tb,Tornqvist:2004qy}. For example, the $P_c$ states were proposed to be the $\bar D^{(*)} \Sigma_c^{(*)}$ hadronic molecular states in Refs.~\cite{Wu:2010jy,Wang:2011rga,Yang:2011wz,Karliner:2015ina,Chen:2015loa,Chen:2015moa,Liu:2019tjn} bound by the one-meson-exchange interaction $\Pi_{M}$, as depicted in Fig.~\ref{fig:feynman}(a). Besides, we know from QCD that there can be the double-gluon-exchange interaction $\Pi_{G}$ between $\bar D^{(*)}$ and $\Sigma_c^{(*)}$, as depicted in Fig.~\ref{fig:feynman}(b).

\begin{figure}[]
\begin{center}
\includegraphics[width=0.4\textwidth]{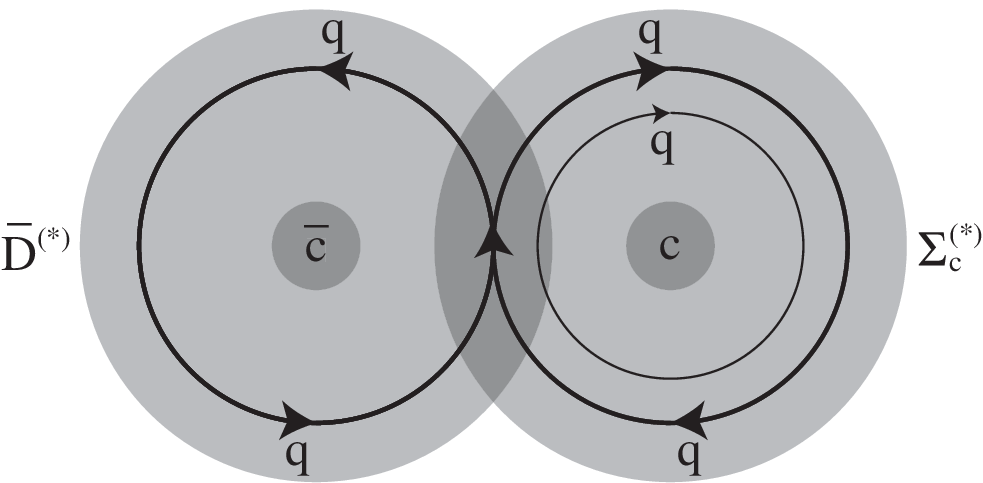}
\caption{Possible binding mechanism induced by shared light quarks, described by the light-quark-exchange term $\Pi_{Q}$. Here $q$ denotes a light up/down quark.}
\label{fig:sharing}
\end{center}
\end{figure}

In this paper we propose another possible interaction between $\bar D^{(*)}$ and $\Sigma_c^{(*)}$ induced by the light-quark-exchange term $\Pi_{Q}$, as depicted in Fig.~\ref{fig:feynman}(c). This term indicates that $\bar D^{(*)}$ and $\Sigma_c^{(*)}$ are exchanging and so sharing two light up/down quarks, as depicted in Fig.~\ref{fig:sharing}. It can induce an interaction between $\bar D^{(*)}$ and $\Sigma_c^{(*)}$, either attractive or repulsive. Note that the two interactions, $\Pi_{M}$ at the hadron level and $\Pi_{Q}$ at the quark-gluon level, can overlap with each other. The quark-exchange effect has been studied in Ref.~\cite{Hoodbhoy:1986fn} by Hoodbhoy and Jaffe to explain the European Muon Collaboration (EMC) effect in three-nucleon systems, and later used in Refs.~\cite{Modarres:1988sb,Modarres:2006ry} to study some other nuclei. We also refer to Ref.~\cite{wang} for some relevant discussions.

%
%

%
%

In this paper we shall systematically examine the Feynman diagrams corresponding to the $\bar D^{(*)} \Sigma_c^{(*)}$, $\bar D^{(*)} \Lambda_c$, $D^{(*)} \bar K^{*}$, and $D^{(*)} \bar D^{(*)}$ hadronic molecular states. We shall apply the method of QCD sum rules to investigate the light-quark-exchange term $\Pi_{Q}$, and study its contributions to these states. Based on the obtained results, we shall study the binding mechanism induced by shared light quarks. This mechanism is somewhat similar to the covalent bond in chemical molecules induced by shared electrons, so we call such hadronic molecules ``covalent hadronic molecules''.

Based on the obtained results, we shall further propose a model-independent hypothesis: {\it the light-quark-exchange interaction is attractive when the shared light quarks are totally antisymmetric so that obey the Pauli principle}. We shall apply this hypothesis to predict some possibly-existing covalent hadronic molecules. We shall also build a toy model to formulize this picture and estimate their binding energies. Our model has four parameters, which are fixed by considering the $P_c/P_{cs}$ and the recently observed $T_{cc}^+$~\cite{LHCb:2021auc,LHCb:2021vvq} as possible covalent hadronic molecules.

Take the $X(3872)$ as another example. We shall find that the light-quark-exchange term $\Pi_{Q}$ does not contribute to the $D^{(*)} \bar D^{(*)}$ molecules, suggesting the $D^{(*)} \bar D^{(*)}$ covalent hadronic molecules not to exist. However, there can still be the $D^{(*)} \bar D^{(*)}$ hadronic molecules induced by some other binding mechanisms, such as the one-meson-exchange interaction~\cite{Weinberg:1965zz,Voloshin:1976ap,DeRujula:1976zlg,Tornqvist:1993ng,Voloshin:2003nt,Close:2003sg,Wong:2003xk,Braaten:2003he,Swanson:2003tb,Tornqvist:2004qy}. Especially, the $X(3872)$ can be interpreted as such a $D \bar D^{*}$ hadronic molecule, while it was suggested to be a compact tetraquark state in Refs.~\cite{Maiani:2004vq,Maiani:2014aja,Hogaasen:2005jv,Ebert:2005nc,Barnea:2006sd}, a conventional $c \bar c$ state in Refs.~\cite{Barnes:2003vb,Eichten:2004uh}, and the mixture of a $c \bar c$ state with the $D \bar D^{*}$ component in Refs.~\cite{Meng:2005er,Meng:2014ota}.

This paper is organized as follows. In Sec.~\ref{sec:correlation} we systematically investigate correlation functions of the $D^- \Sigma_c^{++}$, $\bar D^0 \Sigma_c^+$, $I=1/2$ $\bar D \Sigma_c$, and $I=3/2$ $\bar D \Sigma_c$ hadronic molecules. In Sec.~\ref{sec:sumrule} we apply the method of QCD sum rules to investigate the light-quark-exchange term $\Pi_{Q}$, and study its contributions to these molecules. In Sec.~\ref{sec:more} we follow the same procedures and study the $\bar D^* \Sigma_c/\bar D \Sigma_c^*/\bar D^* \Sigma_c^*$, $\bar D^{(*)} \Lambda_c$, $D^{(*)} \bar K^{*}$, and $D^{(*)} \bar D^{(*)}$ hadronic molecules. Based on the obtained QCD sum rule results, we propose the above hypothesis in Sec.~\ref{sec:covalent}, and predict more possibly-existing covalent hadronic molecules. In Sec.~\ref{sec:model} we build a toy model to formulize the covalent hadronic molecule picture, and the obtained results are summarized and discussed in Sec.~\ref{sec:summary}.

\section{Correlation functions of $D^- \Sigma_c^{++}/\bar D^0 \Sigma_c^+/\bar D \Sigma_c$ molecules}
\label{sec:correlation}

In this section we investigate correlation functions of the $D^- \Sigma_c^{++}$, $\bar D^0 \Sigma_c^+$, $I=1/2$ $\bar D \Sigma_c$, and $I=3/2$ $\bar D \Sigma_c$ hadronic molecular states.

\subsection{$D^- \Sigma_c^{++}$ correlation function}
\label{sec:piDm}

In this subsection we investigate correlation function of the $D^- \Sigma_c^{++}$ molecule. Firstly, we investigate correlation functions of the $D^-$ meson and the $\Sigma_c^{++}$ baryon. Their corresponding interpolating currents are
\begin{eqnarray}
J^{D^-}(x) &=& \bar c_a(x) \gamma_5 d_a(x) \, ,
\\[1mm] J^{\Sigma_c^{++}}(x) &=& {1\over\sqrt2} \epsilon^{abc} u^{\rm T}_a(x) \mathbb{C} \gamma^\mu u_b(x) \gamma_\mu\gamma_{5} c_c(x) \, ,
\label{def:Sigma}
\end{eqnarray}
where $a \cdots c$ are color indices; $\mathbb{C} = {\rm i}\gamma_2 \gamma_0$ is the charge-conjugation operator; the coefficient $1/\sqrt2$ is an isospin factor.

\begin{figure*}[]
\begin{center}
\includegraphics[width=0.8\textwidth]{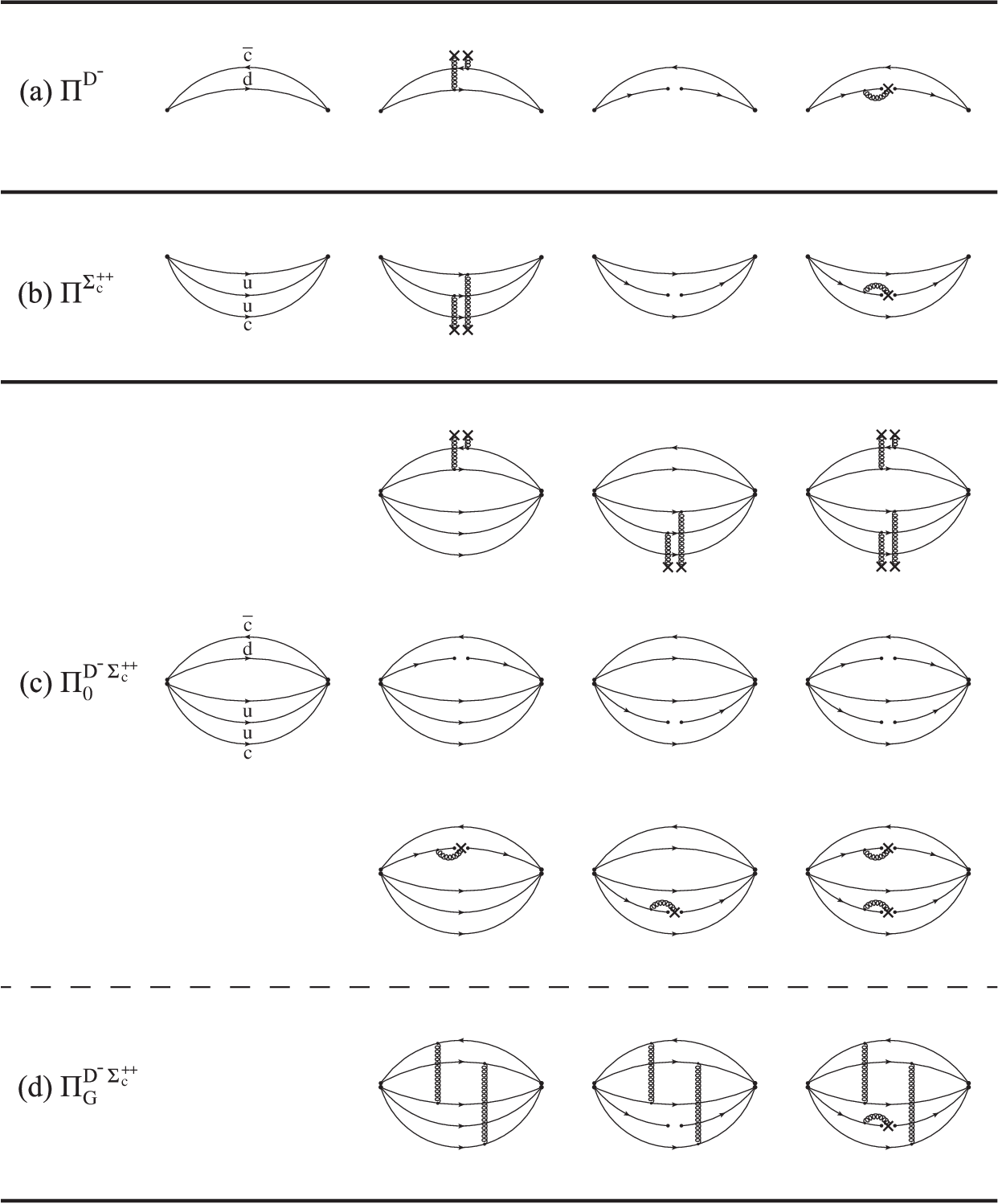}
\caption{Feynman diagrams corresponding to: a) $\Pi^{D^-}(x)$, b) $\Pi^{\Sigma_c^{++}}(x)$, c) $\Pi^{D^- \Sigma_c^{++}}_0(x)$, and d) $\Pi^{D^- \Sigma_c^{++}}_{G}(x)$. $\Pi^{D^-}(x)$ and $\Pi^{\Sigma_c^{++}}(x)$ are correlation functions of the $D^-$ meson and the $\Sigma_c^{++}$ baryon, respectively. The correlation function of the $D^- \Sigma_c^{++}$ molecule satisfies $\Pi^{D^- \Sigma_c^{++}}(x) = \Pi^{D^- \Sigma_c^{++}}_0(x) + \Pi^{D^- \Sigma_c^{++}}_{G}(x)$ with $\Pi^{D^- \Sigma_c^{++}}_0(x) = \Pi^{D^-}(x) \times \Pi^{\Sigma_c^{++}}(x)$.}
\label{fig:feynman1}
\end{center}
\end{figure*}

We write down two-point correlation functions of the $D^-$ meson and the $\Sigma_c^{++}$ baryon in the coordinate space:
\begin{eqnarray}
\Pi^{D^-}(x) &=& \langle 0 | \mathbb{T}\left[J^{D^-}(x) J^{D^-,\dagger}(0)\right] | 0 \rangle
\\ \nonumber &=& - {\bf Tr}\left[{\bf iS}_q^{aa^\prime}(x) \gamma_5 {\bf iS}_c^{a^\prime a}(-x) \gamma_5\right] \, ,
\\[2mm] \Pi^{\Sigma_c^{++}}(x) &=& \langle 0 | \mathbb{T}\left[J^{\Sigma_c^{++}}(x) \bar J^{\Sigma_c^{++}}(0)\right] | 0 \rangle
\\ \nonumber &=& \epsilon^{abc}\epsilon^{a^\prime b^\prime c^\prime }{\bf Tr}\left[{\bf iS}_q^{bb^\prime}(x) \gamma^{\mu^\prime} \mathbb{C}({\bf iS}_q^{aa^\prime}(x))^{\rm T}\mathbb{C} \gamma^\mu\right]
\\ \nonumber && ~~~~~~~~~~~~~~~~~~~~~~~~~~~ \times \gamma_\mu\gamma_{5} {\bf iS}_c^{cc^\prime}(x) \gamma_{\mu^\prime}\gamma_{5} \, .
\end{eqnarray}
Their corresponding Feynman diagrams are depicted in Fig.~\ref{fig:feynman1}(a) and Fig.~\ref{fig:feynman1}(b), respectively. In the present study we do not differentiate propagators of the light up and down quarks, and use ${\bf iS}_q^{ab}(x) = {\bf iS}_{\rm up}^{ab}(x) = {\bf iS}_{\rm down}^{ab}(x)$ to denote both of them in the coordinate space; besides, we use ${\bf iS}_c^{ab}(x)$ to denote propagator of the heavy charm quark in the coordinate space:
\begin{eqnarray}
{\bf iS}_q^{ab}(x) &=& \langle 0 | \mathbb{T}\left[ u^a(x) \bar u^b(0) \right] | 0 \rangle
\\ \nonumber       &=& \langle 0 | \mathbb{T}\left[ d^a(x) \bar d^b(0) \right] | 0 \rangle \, ,
\\[1mm] {\bf iS}_c^{ab}(x) &=& \langle 0 | \mathbb{T}\left[ c^a(x) \bar c^b(0) \right] | 0 \rangle \, .
\end{eqnarray}

%
%

Then we put the $D^-$ meson and the $\Sigma_c^{++}$ baryon at the same location, and construct a composite current corresponding to the $D^- \Sigma_c^{++}$ molecule,
\begin{eqnarray}
J^{D^-\Sigma_c^{++}}(x) &=& J^{D^-}(x) \times J^{\Sigma_c^{++}}(x)
\label{def:currentDm}
\\ \nonumber &=& [\bar c_d(x) \gamma_5 d_d(x)]
\\ \nonumber &\times& {1\over\sqrt2} [\epsilon^{abc} u^{\rm T}_a(x) \mathbb{C} \gamma^\mu u_b(x) \gamma_\mu\gamma_{5} c_c(x)] \, .
\end{eqnarray}
Its correlation function in the coordinate space is
\begin{eqnarray}
\nonumber && \Pi^{D^-\Sigma_c^{++}}(x)
\\ \nonumber &=& \langle 0 | \mathbb{T}\left[J^{D^-\Sigma_c^{++}}(x) \bar J^{D^-\Sigma_c^{++}}(0)\right] | 0 \rangle
\\ \nonumber &=& - {\bf Tr}\left[{\bf iS}_q^{dd^\prime}(x) \gamma_5 {\bf iS}_c^{d^\prime d}(-x) \gamma_5\right]
\\ \nonumber &\times& \epsilon^{abc}\epsilon^{a^\prime b^\prime c^\prime }{\bf Tr}\left[{\bf iS}_q^{bb^\prime}(x) \gamma^{\mu^\prime} \mathbb{C}({\bf iS}_q^{aa^\prime}(x))^{\rm T}\mathbb{C} \gamma^\mu\right]
\\ \nonumber && ~~~~~~~~~~~~~~~~~~~~~~~~~~~ \times \gamma_\mu\gamma_{5} {\bf iS}_c^{cc^\prime}(x) \gamma_{\mu^\prime}\gamma_{5}
\\ &=& \Pi^{D^-\Sigma_c^{++}}_0(x) + \Pi^{D^-\Sigma_c^{++}}_{G}(x) \, ,
\label{pi:Dm}
\end{eqnarray}
where
\begin{equation}
\Pi^{D^-\Sigma_c^{++}}_0(x) = \Pi^{D^-}(x) \times \Pi^{\Sigma_c^{++}}(x) \, ,
\end{equation}
is the leading term contributed by non-correlated $D^-$ and $\Sigma_c^{++}$, and $\Pi^{D^-\Sigma_c^{++}}_{G}(x)$ describes the double-gluon-exchange interaction between them. Their corresponding Feynman diagrams are depicted in Fig.~\ref{fig:feynman1}(c) and Fig.~\ref{fig:feynman1}(d), respectively.

The double-gluon-exchange term $\Pi^{D^-\Sigma_c^{++}}_{G}(x)$ is at the $\mathcal{O}(\alpha_s^2)$ order, and it is expected to be suppressed according to the OZI rule. In the present study we shall not pay much attention to it, while we shall investigate another more important term from the next subsection.

\subsection{$\bar D^0 \Sigma_c^{+}$ correlation function}
\label{sec:piD0}

\begin{figure*}[]
\begin{center}
\includegraphics[width=0.8\textwidth]{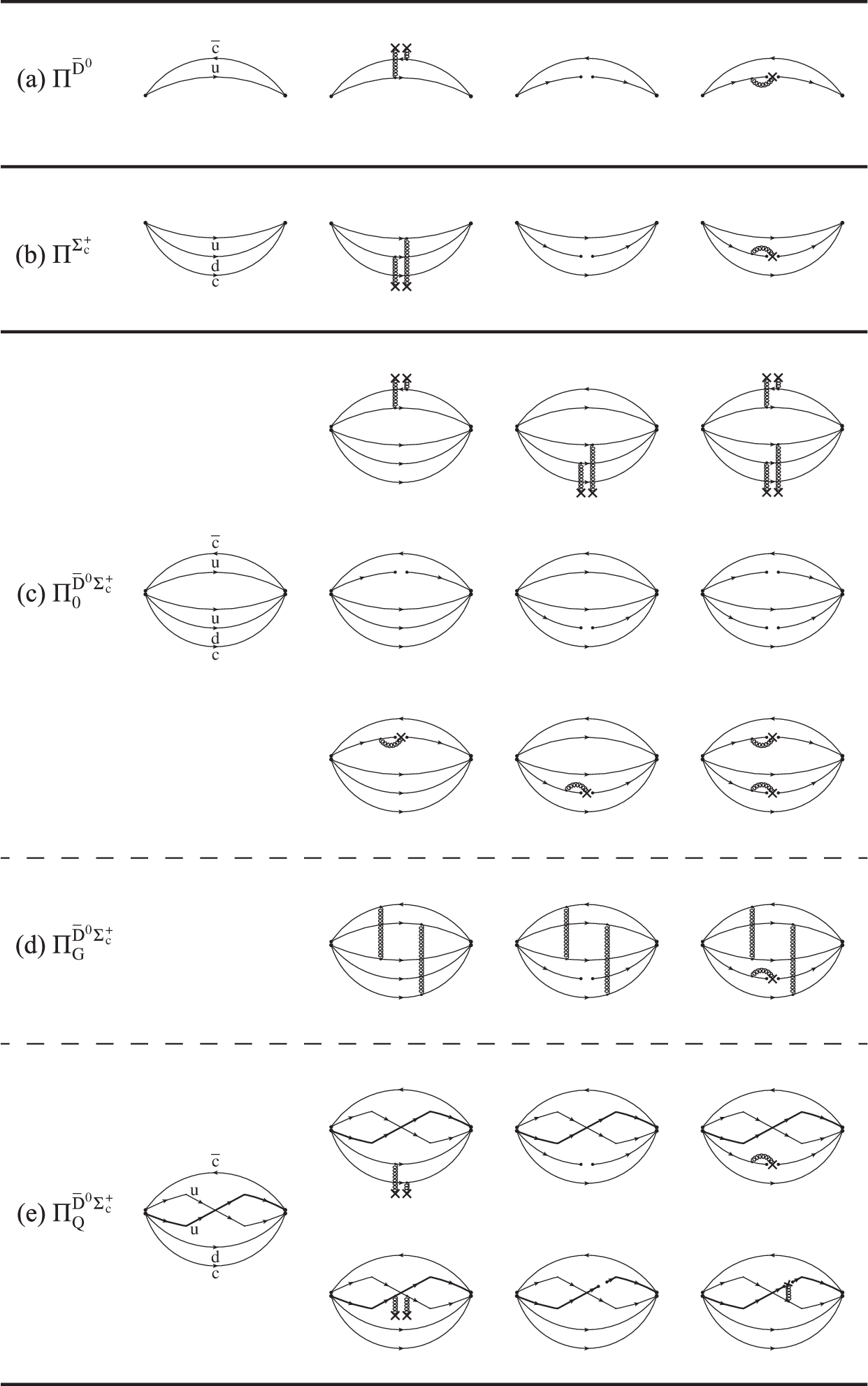}
\caption{Feynman diagrams corresponding to: a) $\Pi^{\bar D^0}(x)$, b) $\Pi^{\Sigma_c^+}(x)$, c) $\Pi^{\bar D^0 \Sigma_c^{+}}_0(x)$, d) $\Pi^{\bar D^0 \Sigma_c^{+}}_{G}(x)$, and e) $\Pi^{\bar D^0 \Sigma_c^{+}}_{Q}(x)$. $\Pi^{\bar D^0}(x)$ and $\Pi^{\Sigma_c^+}(x)$ are correlation functions of the $\bar D^0$ meson and the $\Sigma_c^{+}$ baryon, respectively. The correlation function of the $\bar D^0 \Sigma_c^{+}$ molecule satisfies $\Pi^{\bar D^0 \Sigma_c^{+}}(x) = \Pi^{\bar D^0 \Sigma_c^{+}}_0(x) + \Pi^{\bar D^0 \Sigma_c^{+}}_{G}(x) + \Pi^{\bar D^0 \Sigma_c^{+}}_{Q}(x)$ with $\Pi^{\bar D^0 \Sigma_c^{+}}_0(x) = \Pi^{\bar D^0}(x) \times \Pi^{\Sigma_c^+}(x)$.}
\label{fig:feynman2}
\end{center}
\end{figure*}

In this subsection we investigate correlation function of the $\bar D^0 \Sigma_c^{+}$ molecule. The interpolating currents corresponding to the $\bar D^0$ meson, the $\Sigma_c^{+}$ baryon, and the $\bar D^0 \Sigma_c^{+}$ molecule are
\begin{eqnarray}
J^{\bar D^0}(x) &=& \bar c_a(x) \gamma_5 u_a(x) \, ,
\\[1mm] J^{\Sigma_c^{+}}(x) &=& \epsilon^{abc} u^{\rm T}_a(x) \mathbb{C} \gamma^\mu d_b(x) \gamma_\mu\gamma_{5} c_c(x) \, ,
\\[1mm] J^{\bar D^0 \Sigma_c^{+}}(x) &=& J^{\bar D^0}(x) \times J^{\Sigma_c^{+}}(x)
\label{def:currentD0}
\\ \nonumber &=& [\bar c_d(x) \gamma_5 u_d(x)]
\\ \nonumber &\times& [\epsilon^{abc} u^{\rm T}_a(x) \mathbb{C} \gamma^\mu d_b(x) \gamma_\mu\gamma_{5} c_c(x)] \, .
\end{eqnarray}
\begin{widetext}
Their correlation functions in the coordinate space are:
\begin{eqnarray}
\Pi^{\bar D^0}(x) &=& - {\bf Tr}\left[{\bf iS}_q^{aa^\prime}(x) \gamma_5 {\bf iS}_c^{a^\prime a}(-x) \gamma_5\right] \, ,
\\[2mm] \Pi^{\Sigma_c^+}(x) &=& \epsilon^{abc}\epsilon^{a^\prime b^\prime c^\prime }{\bf Tr}\left[{\bf iS}_q^{bb^\prime}(x) \gamma^{\mu^\prime} \mathbb{C}({\bf iS}_q^{aa^\prime}(x))^{\rm T}\mathbb{C} \gamma^\mu\right]
\times \gamma_\mu\gamma_{5} {\bf iS}_c^{cc^\prime}(x) \gamma_{\mu^\prime}\gamma_{5} \, ,
\\[2mm] \nonumber \Pi^{\bar D^0 \Sigma_c^{+}}(x) &=& - {\bf Tr}\left[{\bf iS}_q^{dd^\prime}(x) \gamma_5 {\bf iS}_c^{d^\prime d}(-x) \gamma_5\right]
\times \epsilon^{abc}\epsilon^{a^\prime b^\prime c^\prime }{\bf Tr}\left[{\bf iS}_q^{bb^\prime}(x) \gamma^{\mu^\prime} \mathbb{C}({\bf iS}_q^{aa^\prime}(x))^{\rm T}\mathbb{C} \gamma^\mu\right]
\gamma_\mu\gamma_{5} {\bf iS}_c^{cc^\prime}(x) \gamma_{\mu^\prime}\gamma_{5}
\\ \nonumber && + {\bf Tr}\left[{\bf iS}_q^{da^\prime}(x) \gamma^{\mu^\prime} \mathbb{C}({\bf iS}_q^{bb^\prime}(x))^{\rm T}\mathbb{C} \gamma^\mu {\bf iS}_q^{ad^\prime}(x) \gamma_5 {\bf iS}_c^{d^\prime d}(-x) \gamma_5 \right]
\gamma_\mu\gamma_{5} {\bf iS}_c^{cc^\prime}(x) \gamma_{\mu^\prime}\gamma_{5}
\\ &=& \Pi^{\bar D^0 \Sigma_c^{+}}_0(x) + \Pi^{\bar D^0 \Sigma_c^{+}}_{G}(x) + \Pi^{\bar D^0 \Sigma_c^{+}}_{Q}(x) \, .
\label{pi:D0}
\end{eqnarray}
\end{widetext}
In the above expressions, $\Pi^{\bar D^0}(x)$ and $\Pi^{\Sigma_c^+}(x)$ are correlation functions of $\bar D^0$ and $\Sigma_c^{+}$, respective; $\Pi^{\bar D^0 \Sigma_c^{+}}_0(x) = \Pi^{\bar D^0}(x) \times \Pi^{\Sigma_c^+}(x)$ is the leading term contributed by non-correlated $\bar D^0$ and $\Sigma_c^{+}$, and $\Pi^{\bar D^0 \Sigma_c^{+}}_{G}(x)$ describes the double-gluon-exchange interaction between them. Their corresponding Feynman diagrams are depicted in Fig.~\ref{fig:feynman2}(a-d).

%
%

Compared to the case of $D^- \Sigma_c^{++}$, we find
\begin{eqnarray}
\Pi^{\bar D^0}(x) &=& \Pi^{D^-}(x) \, ,
\\[1mm] \Pi^{\Sigma_c^+}(x) &=& \Pi^{\Sigma_c^{++}}(x) \, ,
\\[1mm] \Pi^{\bar D^0 \Sigma_c^{+}}_0(x) &=& \Pi^{D^- \Sigma_c^{++}}_0(x) \, ,
\\[1mm] \Pi^{\bar D^0 \Sigma_c^{+}}_{G}(x) &=& \Pi^{D^- \Sigma_c^{++}}_{G}(x) \, ,
\end{eqnarray}
while there exists an extra term,
\begin{eqnarray}
\nonumber \Pi^{\bar D^0 \Sigma_c^{+}}_{Q}(x) &=& {\bf Tr}\Big[{\bf iS}_q^{da^\prime}(x) \gamma^{\mu^\prime} \mathbb{C}({\bf iS}_q^{bb^\prime}(x))^{\rm T}\mathbb{C} \gamma^\mu
\\ \nonumber && ~~~~~~~~~~~~ {\bf iS}_q^{ad^\prime}(x) \gamma_5 {\bf iS}_c^{d^\prime d}(-x) \gamma_5 \Big]
\\ && \times \gamma_\mu\gamma_{5} {\bf iS}_c^{cc^\prime}(x) \gamma_{\mu^\prime}\gamma_{5} \, .
\end{eqnarray}
Its corresponding Feynman diagrams are depicted in Fig.~\ref{fig:feynman2}(e). This term describes the light-quark-exchange effect between $\bar D^0$ and $\Sigma_c^{+}$, {\it i.e.}, $\bar D^0$ and $\Sigma_c^{+}$ are exchanging and so sharing two light up quarks. It can induce an interaction between $\bar D^0$ and $\Sigma_c^{+}$, either attractive or repulsive. This term is color-unconfined, {\it i.e.}, its contribution decreases as the distance between $\bar D^0$ and $\Sigma_c^{+}$ increases; besides, the two exchanged/shared light quarks have the same color. We shall further study it numerically in Sec.~\ref{sec:sumruleD0}.

\subsection{$I=1/2$ $\bar D \Sigma_c$ correlation function}
\label{sec:Dhalf}

In this subsection we investigate correlation function of the $I=1/2$ $\bar D \Sigma_c$ molecule. Its corresponding interpolating current is
\begin{equation}
J^{\bar D \Sigma_c}(x) = \sqrt{1\over3} J^{\bar D^0 \Sigma_c^{+}}(x) - \sqrt{2\over3} J^{D^- \Sigma_c^{++}}(x) \, ,
\label{def:currentDhalf}
\end{equation}
with the correlation function in the coordinate space
\begin{widetext}
\begin{eqnarray}
\nonumber \Pi^{\bar D \Sigma_c}(x) &=& - {\bf Tr}\left[{\bf iS}_q^{dd^\prime}(x) \gamma_5 {\bf iS}_c^{d^\prime d}(-x) \gamma_5\right]
\times \epsilon^{abc}\epsilon^{a^\prime b^\prime c^\prime }{\bf Tr}\left[{\bf iS}_q^{bb^\prime}(x) \gamma^{\mu^\prime} \mathbb{C}({\bf iS}_q^{aa^\prime}(x))^{\rm T}\mathbb{C} \gamma^\mu\right]
\gamma_\mu\gamma_{5} {\bf iS}_c^{cc^\prime}(x) \gamma_{\mu^\prime}\gamma_{5}
\\ \nonumber && - {\bf Tr}\left[{\bf iS}_q^{da^\prime}(x) \gamma^{\mu^\prime} \mathbb{C}({\bf iS}_q^{bb^\prime}(x))^{\rm T}\mathbb{C} \gamma^\mu {\bf iS}_q^{ad^\prime}(x) \gamma_5 {\bf iS}_c^{d^\prime d}(-x) \gamma_5 \right]
\gamma_\mu\gamma_{5} {\bf iS}_c^{cc^\prime}(x) \gamma_{\mu^\prime}\gamma_{5}
\\ &=& \Pi^{\bar D \Sigma_c}_0(x) + \Pi^{\bar D \Sigma_c}_{G}(x) + \Pi^{\bar D \Sigma_c}_{Q}(x) \, .
\label{pi:Dhalf}
\end{eqnarray}
\end{widetext}
Compared to the case of $\bar D^0 \Sigma_c^{+}$ given in Eq.~(\ref{pi:D0}), only the terms $\Pi^{\bar D \Sigma_c}_{Q}(x)$ and $\Pi^{\bar D^0 \Sigma_c^{+}}_{Q}(x)$ are different/oppisite:
\begin{eqnarray}
\Pi^{\bar D \Sigma_c}_0(x) &=& \Pi^{\bar D^0 \Sigma_c^{+}}_0(x) \, ,
\\[1mm] \Pi^{\bar D \Sigma_c}_{G}(x) &=& \Pi^{\bar D^0 \Sigma_c^{+}}_{G}(x) \, ,
\\[1mm] \Pi^{\bar D \Sigma_c}_{Q}(x) &=& - \Pi^{\bar D^0 \Sigma_c^{+}}_{Q}(x) \, .
\end{eqnarray}
Accordingly, if the term $\Pi_{Q}(x)$ induces a repulsive interaction to the $\bar D^0 \Sigma_c^{+}$ molecule, it would induce an attractive interaction to the $I=1/2$ $\bar D \Sigma_c$ molecule; and vice verse. We shall further study it numerically in Sec.~\ref{sec:sumruleDhalf}. For completeness, we explicitly list that the term $\Pi^{\bar D \Sigma_c}_{Q}(x)$ is contributed separately by:
\begin{align}
\nonumber & \langle 0 | \mathbb{T}\left[\sqrt{1\over3}J^{\bar D^0 \Sigma_c^{+}}(x) \sqrt{1\over3}\bar J^{\bar D^0 \Sigma_c^{+}}(0)\right] | 0 \rangle \rightarrow - {1\over3}\Pi^{\bar D \Sigma_c}_{Q}(x) \, ,
\\ \nonumber & \langle 0 | \mathbb{T}\left[\sqrt{2\over3}J^{D^-\Sigma_c^{++}}(x) \sqrt{2\over3}\bar J^{D^-\Sigma_c^{++}}(0)\right] | 0 \rangle \rightarrow 0 \, ,
\\ \nonumber & - \langle 0 | \mathbb{T}\left[{J^{\bar D^0 \Sigma_c^{+}}(x)\over\sqrt3} \sqrt{2\over3}\bar J^{D^-\Sigma_c^{++}}(0)\right] | 0 \rangle \rightarrow {2\over3}\Pi^{\bar D \Sigma_c}_{Q}(x) \, ,
\\ \nonumber & - \langle 0 | \mathbb{T}\left[\sqrt{2\over3}J^{D^-\Sigma_c^{++}}(x) {\bar J^{\bar D^0 \Sigma_c^{+}}(0)\over\sqrt3}\right] | 0 \rangle \rightarrow {2\over3}\Pi^{\bar D \Sigma_c}_{Q}(x) \, .
\\
\end{align}

\subsection{$I=3/2$ $\bar D \Sigma_c$ correlation function}
\label{sec:D3half}

For completeness, we investigate correlation function of the $I=3/2$ $\bar D \Sigma_c$ molecule in this subsection. Its corresponding interpolating current is
\begin{equation}
J_{I=3/2}^{\bar D \Sigma_c}(x) = \sqrt{2\over3} J^{\bar D^0 \Sigma_c^{+}}(x) + \sqrt{1\over3} J^{D^- \Sigma_c^{++}}(x) \, ,
\label{def:currentD3half}
\end{equation}
with the correlation function to be
\begin{equation}
\Pi_{I=3/2}^{\bar D \Sigma_c}(x) = \Pi^{\bar D \Sigma_c}_0(x) + \Pi^{\bar D \Sigma_c}_{G}(x) - 2\Pi^{\bar D \Sigma_c}_{Q}(x) \, .
\label{pi:D3half}
\end{equation}
In the above expression $\Pi^{\bar D \Sigma_c}_0(x)$, $\Pi^{\bar D \Sigma_c}_{G}(x)$, and $\Pi^{\bar D \Sigma_c}_{Q}(x)$ are taken from the $I=1/2$ $\bar D \Sigma_c$ molecule.

Compared to Eq.~(\ref{pi:Dhalf}), if the term $\Pi_{Q}(x)$ induces an attractive interaction to the $I=1/2$ $\bar D \Sigma_c$ molecule, it would induce a repulsive interaction to the $I=3/2$ $\bar D \Sigma_c$ molecule; and vice verse.

\section{QCD sum rule studies of $D^- \Sigma_c^{++}/\bar D^0 \Sigma_c^+/\bar D \Sigma_c$ molecules}
\label{sec:sumrule}

In this section we apply the method of QCD sum rules~\cite{Shifman:1978bx,Reinders:1984sr} to investigate the light-quark-exchange term $\Pi_{Q}(x)$, and study its contributions to the $D^- \Sigma_c^{++}$, $\bar D^0 \Sigma_c^+$, $I=1/2$ $\bar D \Sigma_c$, and $I=3/2$ $\bar D \Sigma_c$ hadronic molecular states.

In QCD sum rule analyses we consider the two-point correlation function in the momentum space:
%
\begin{eqnarray}
\nonumber \Pi(q^2) &=& {\rm i} \int {\rm d}^4x {\rm e}^{{\rm i}qx} \langle 0 | \mathbb{T}\left[ J(x) \bar J(0)\right] | 0 \rangle
\\ &=& {\rm i} \int {\rm d}^4x {\rm e}^{{\rm i}qx}~\Pi(x) \, ,
\label{def:pi}
\end{eqnarray}
%
where $J(x)$ is an interpolating current. We generally assume it to be a composite current, coupling to the molecular state $X \equiv |YZ\rangle$ through
\begin{equation}
\langle 0 | J | X \rangle = f_X u_X \, .
\end{equation}

We write $\Pi(q^2)$ in the form of dispersion relation as
%
\begin{equation}
\Pi(q^2)=\int^\infty_{s_<}\frac{\rho(s)}{s-q^2-{\rm i}\varepsilon}{\rm d}s \, ,
\label{eq:disper}
\end{equation}
where $s_< = 4 m_c^2$ is the physical threshold and $\rho(s) \equiv {\rm Im}\Pi(s)/\pi$ is the spectral density.

At the quark-gluon level we calculate $\Pi(q^2)$ using the method of operator product expansion (OPE) up to certain order. According to Eqs.~(\ref{pi:Dm}), (\ref{pi:D0}), (\ref{pi:Dhalf}), and (\ref{pi:D3half}), we further separate it into
\begin{equation}
\Pi(q^2) \approx \Pi_0(q^2) + \Pi_{Q}(q^2) \, ,
\label{eq:pimomentum}
\end{equation}
and define their imaginary parts to be $\rho_0(s)$ and $\rho_{Q}(s)$. Here we have omitted the other term $\Pi_{G}(q^2)$, since the light-quark-exchange term $\Pi_{Q}(q^2)$ is much larger.

At the hadron level we evaluate the spectral density by inserting intermediate hadron states $\sum_n|n\rangle\langle n|$:
%
\begin{eqnarray}
\nonumber \rho(s) &=& \sum_n\delta(s-M^2_n)\langle 0 |J| n \rangle \langle n|\bar J|0\rangle
\\ &=& f^2_X\delta(s-M^2_X)+ \rm{continuum} \, ,
\label{eq:rho}
\end{eqnarray}
%
where we have adopted a parametrization of one pole dominance for the ground state $X$ together with a continuum contribution.

Given $X \equiv |YZ\rangle$ to be a molecular state, its mass $M_X$ can be expanded as
\begin{eqnarray}
M_X &=& M_Y + M_Z + \Delta M
\\ \nonumber &\equiv& M_0 + \Delta M \, .
\end{eqnarray}
Then we insert Eq.~(\ref{eq:rho}) into Eq.~(\ref{eq:disper}), and expand it as
\begin{eqnarray}
\Pi(q^2) &=& {f_X^2 \over M_X^2 - q^2} + \cdots
\\ \nonumber &\approx& {f_X^2 \over M_0^2 - q^2} - {2 M_0 f_X^2 \over \left(M_0^2 - q^2\right)^2} \Delta M + \cdots  \, .
\end{eqnarray}
The former term is contributed by the non-correlated $Y$ and $Z$, and the latter term is contributed by their interactions. Compared to Eq.~(\ref{eq:pimomentum}), we further obtain:
\begin{eqnarray}
\Pi_0(q^2) &=& {f_X^2 \over M_0^2 - q^2} + \cdots \, ,
\\[1mm] \Pi_{Q}(q^2) &=& - {2 M_0 f_X^2 \over \left(M_0^2 - q^2\right)^2} \Delta M + \cdots \, .
\end{eqnarray}

We perform the Borel transformation to the above correlation functions at both hadron and quark-gluon levels. After assuming contributions from the continuum to be approximated by the OPE spectral densities $\rho_0(s)$ and $\rho_{Q}(s)$ above a threshold value $s_0$, we arrive at two sum rule equations:
%
\begin{widetext}
\begin{eqnarray}
f^2_X {\rm e}^{-M_0^2/M_B^2} &=& \Pi_0(M_B^2, s_0) = \int^{s_0}_{s_<} {\rm e}^{-s/M_B^2}\rho_0(s){\rm d}s \, ,
\label{eq:sumrule1}
\\[1mm] - {2 M_0 f_X^2 \over M_B^2} \Delta M {\rm e}^{-M_0^2/M_B^2} &=& \Pi_Q(M_B^2, s_0) = \int^{s_0}_{s_<} {\rm e}^{-s/M_B^2}\rho_Q(s){\rm d}s \, .
\label{eq:sumrule2}
\end{eqnarray}
\end{widetext}
%
There are two free parameters in Eqs.~(\ref{eq:sumrule1}) and (\ref{eq:sumrule2}): the Borel mass $M_B$ and the threshold value $s_0$.
Differentiating Eq.~(\ref{eq:sumrule1}) with respect to $1/M_B^2$, we obtain
%
\begin{equation}
M^2_0 = \frac{\int^{s_0}_{s_<} {\rm e}^{-s/M_B^2}s\rho(s){\rm d}s}{\int^{s_0}_{s_<} {\rm e}^{-s/M_B^2}\rho(s){\rm d}s} \, .
\label{eq:mass}
\end{equation}
%
Given $M_0 = M_Y + M_Z$, this equation can be used to relate $M_B$ and $s_0$, so that there is only one free parameter left.

Dividing Eq.~(\ref{eq:sumrule2}) by Eq.~(\ref{eq:sumrule1}), we obtain
\begin{equation}
- {2 M_0 \over M_B^2} \Delta M = {\Pi_Q \over \Pi_0} = \frac{\int^{s_0}_{s_<} {\rm e}^{-s/M_B^2}\rho_Q(s){\rm d}s}{\int^{s_0}_{s_<} {\rm e}^{-s/M_B^2}\rho_0(s){\rm d}s} \, .
\label{eq:binding}
\end{equation}
This equation can be used to calculate $\Delta M$.

We shall use Eq.~(\ref{eq:binding}) to study contributions of the term $\Pi_{Q}(x)$ to the $D^- \Sigma_c^{++}$, $\bar D^0 \Sigma_c^+$, $I=1/2$ $\bar D \Sigma_c$, and $I=3/2$ $\bar D \Sigma_c$ hadronic molecular states, separately in the following subsections. Before doing this, we note that $\Delta M$ is actually not the binding energy, but relates to some potential $V(r)$ between $Y$ and $Z$ induced by exchanged/shared light quarks. We use the $I=1/2$ $\bar D \Sigma_c$ hadronic molecular state as an example to qualitatively discuss this. After transforming the local current $J^{\bar D \Sigma_c}(x)$ defined in Eq.~(\ref{def:currentDhalf}) into its non-local form:
\begin{eqnarray}
&& J^{\bar D \Sigma_c}(x,y)
\\ \nonumber &=& \sqrt{1\over3} J^{\bar D^0 \Sigma_c^{+}}(x,y) - \sqrt{2\over3} J^{D^- \Sigma_c^{++}}(x,y)
\\ \nonumber &=& \sqrt{1\over3} J^{\bar D^0}(x) J^{\Sigma_c^{+}}(y) - \sqrt{2\over3} J^{D^-}(x)J^{\Sigma_c^{++}}(y) \, ,
\end{eqnarray}
we calculate its non-local correlation function in the coordinate space to be:
\begin{widetext}
\begin{eqnarray}
\nonumber && \Pi^{\bar D \Sigma_c}(x,y;x^\prime,y^\prime)
\equiv \langle 0 | \mathbb{T}\left[J^{\bar D \Sigma_c}(x,y) \bar J^{\bar D \Sigma_c}(x^\prime,y^\prime)\right] | 0 \rangle
\\ \nonumber &=& - {\bf Tr}\left[{\bf iS}_q^{dd^\prime}(\Delta x) \gamma_5 {\bf iS}_c^{d^\prime d}(-\Delta x) \gamma_5\right]
\times \epsilon^{abc}\epsilon^{a^\prime b^\prime c^\prime }{\bf Tr}\left[{\bf iS}_q^{bb^\prime}(\Delta x) \gamma^{\mu^\prime} \mathbb{C}({\bf iS}_q^{aa^\prime}(\Delta x))^{\rm T}\mathbb{C} \gamma^\mu\right]
\gamma_\mu\gamma_{5} {\bf iS}_c^{cc^\prime}(\Delta x) \gamma_{\mu^\prime}\gamma_{5}
\\ \nonumber && - {\bf Tr}\left[{\bf iS}_q^{da^\prime}(\Delta x - r) \gamma^{\mu^\prime} \mathbb{C}({\bf iS}_q^{bb^\prime}(\Delta x))^{\rm T}\mathbb{C} \gamma^\mu {\bf iS}_q^{ad^\prime}(\Delta x + r) \gamma_5 {\bf iS}_c^{d^\prime d}(-\Delta x) \gamma_5 \right]
\gamma_\mu\gamma_{5} {\bf iS}_c^{cc^\prime}(\Delta x) \gamma_{\mu^\prime}\gamma_{5}
\\ &=& \Pi^{\bar D \Sigma_c}_0(\Delta x) + \Pi^{\bar D \Sigma_c}_{G}(\Delta x, r) + \Pi^{\bar D \Sigma_c}_{Q}(\Delta x, r) \, .
\end{eqnarray}
\end{widetext}
In the above expression we have assumed that $y - y^\prime = x - x^\prime = \Delta x$ and $y-x = y^\prime - x^\prime = r$. The leading term $\Pi^{\bar D \Sigma_c}_0(\Delta x)$ does not change with the parameter $r$, while the light-quark-exchange term $\Pi^{\bar D \Sigma_c}_{Q}(\Delta x, r)$ decreases as $|r| \rightarrow \infty$. Accordingly, we arrive at:
\begin{itemize}

\item Because we are using local currents in QCD sum rule analyses,
\begin{equation}
V(|r|=0) = \Delta M \, .
\end{equation}

\item Because the term $\Pi_{Q}(x)$ is color-unconfined, its contribution decreases as $r$ increases:
\begin{equation}
V(|r| \rightarrow \infty) \rightarrow 0 \, .
\end{equation}

\end{itemize}
We may build a model and use the light-quark-exchange potential $V(r)$ to derive the binding energy of $X$, but this will not be done in the present study. Other than this, we shall calculate $\Delta M$ and qualitatively study several hadronic molecules possibly bound by this potential, through which we shall propose a model-independent hypothesis for such molecules.

The binding mechanism induced by the light-quark-exchange potential $V(r)$ is somewhat similar to the covalent bond in chemical molecules induced by shared electrons, so we call such hadronic molecules ``covalent hadronic molecules''.

\subsection{$I=1/2$ $\bar D \Sigma_c$ sum rules}
\label{sec:sumruleDhalf}

In this subsection we apply QCD sum rules to study the $I=1/2$ $\bar D \Sigma_c$ molecule. We have calculated its correlation function at the leading order of $\alpha_s$ and up to the $D({\rm imension}) = 10$ terms, including the perturbative term, the quark condensate $\langle \bar q q \rangle$, the gluon condensate $\langle g_s^2 GG \rangle$, the quark-gluon mixed condensate $\langle g_s \bar q \sigma G q \rangle$, and their combinations $\langle \bar q q \rangle^2$, $\langle \bar q q \rangle\langle g_s \bar q \sigma G q \rangle$, $\langle \bar q q \rangle^3$, and $\langle g_s \bar q \sigma G q \rangle^2$. The extracted spectral density
\begin{equation}
\rho^{\bar D \Sigma_c}(s) = \rho_{0}^{\bar D \Sigma_c}(s) + \rho_{Q}^{\bar D \Sigma_c}(s) \, ,
\end{equation}
is given in Appendix~\ref{app:ope}.

To perform numerical analyses, we use the following values for various QCD sum rule parameters in the present study~\cite{pdg,Yang:1993bp,Ellis:1996xc,Eidemuller:2000rc,Narison:2002pw,Gimenez:2005nt,Jamin:2002ev,Ioffe:2002be,Ovchinnikov:1988gk,colangelo}:
\begin{eqnarray}
\nonumber m_s &=& 96 ^{+8}_{-4} \mbox{ MeV} \, ,
\\ \nonumber m_c &=& 1.275 ^{+0.025}_{-0.035} \mbox{ GeV} \, ,
\\ \nonumber  \langle\bar qq \rangle &=& -(0.240 \pm 0.010)^3 \mbox{ GeV}^3 \, ,
\\ \langle\bar ss \rangle &=& (0.8\pm 0.1)\times \langle\bar qq \rangle \, ,
\label{eq:condensates}
\\ \nonumber  \langle g_s^2GG\rangle &=& 0.48\pm 0.14 \mbox{ GeV}^4 \, ,
\\
\nonumber \langle g_s\bar q\sigma G q\rangle &=& - M_0^2\times\langle\bar qq\rangle \, ,
\\
\nonumber \langle g_s\bar s\sigma G s\rangle &=& - M_0^2\times\langle\bar ss\rangle \, ,
\\
\nonumber M_0^2 &=& 0.8 \pm 0.2 \mbox{ GeV}^2 \, ,
\end{eqnarray}
where the running mass in the $\overline{MS}$ scheme is used for the charm quark.

\begin{figure}[hbtp]
\begin{center}
\includegraphics[width=0.45\textwidth]{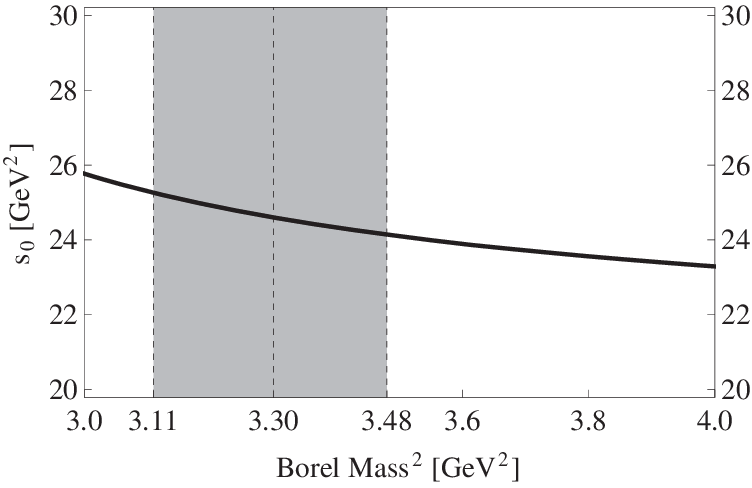}
\caption{Relation between the Borel mass $M_B$ and the threshold value $s_0$, constrained by Eq.~(\ref{eq:mass}).}
\label{fig:relation}
\end{center}
\end{figure}

There are two free parameters in Eqs.~(\ref{eq:sumrule1}) and (\ref{eq:sumrule2}): the Borel mass $M_B$ and the threshold value $s_0$. We use Eq.~(\ref{eq:mass}) to constrain them by setting~\cite{pdg}:
\begin{eqnarray}
\nonumber M_0^{\bar D \Sigma_c} &=& {1\over3}\times\left(M_{\bar D^0} + M_{\Sigma_c^{+}}\right) + {2\over3}\times\left(M_{D^-} + M_{\Sigma_c^{++}}\right)
\\ &=& 4321.66~{\rm MeV} \, .
\end{eqnarray}
The derived relation between $M_B$ and $s_0$ is depicted in Fig.~\ref{fig:relation}, which will be used in the following calculations.

\begin{figure}[hbtp]
\begin{center}
\includegraphics[width=0.45\textwidth]{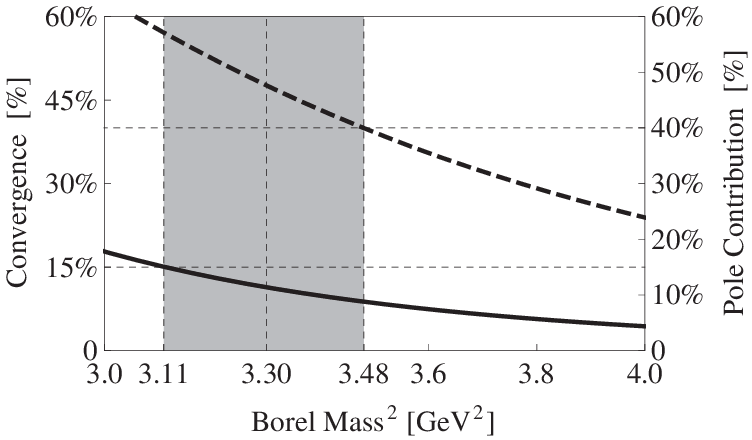}
\caption{Convergence (solid) and Pole-Contribution (dashed) as functions of the Borel mass $M_B$.}
\label{fig:cvgpole}
\end{center}
\end{figure}

There are two criteria to constrain the Borel mass $M_B$. The first criterion is to insure the convergence of OPE series, by requiring the $D=10$ terms $m_c \langle \bar q q \rangle^3$ and $\langle g_s \bar q \sigma G q \rangle^2$ to be less than 15\%:
%
\begin{equation}
\mbox{Convergence} \equiv \left|\frac{ \Pi^{D=10}(\infty, M_B) }{ \Pi(\infty, M_B) }\right| \leq 15\% \, .
\label{eq:cvg}
\end{equation}
%
This criterion determines the lower limit of $M_B$. As shown in Fig.~\ref{fig:cvgpole} using the solid curve, we find it to be $\left(M_B^{min}\right)^2 = 3.11$~GeV$^2$.

The second criterion is to insure the validity of one-pole parametrization, by requiring the pole contribution to be larger than 40\%:
%
\begin{equation}
\mbox{Pole-Contribution} \equiv \frac{ \Pi(s_0(M_B), M_B) }{ \Pi(\infty, M_B) } \geq 40\% \, .
\label{eq:pole}
\end{equation}
%
This determines the upper limit of $M_B$. As shown in Fig.~\ref{fig:cvgpole} using the dashed curve, we find it to be $\left(M_B^{max}\right)^2 = 3.48$~GeV$^2$.

\begin{figure}[hbtp]
\begin{center}
\includegraphics[width=0.45\textwidth]{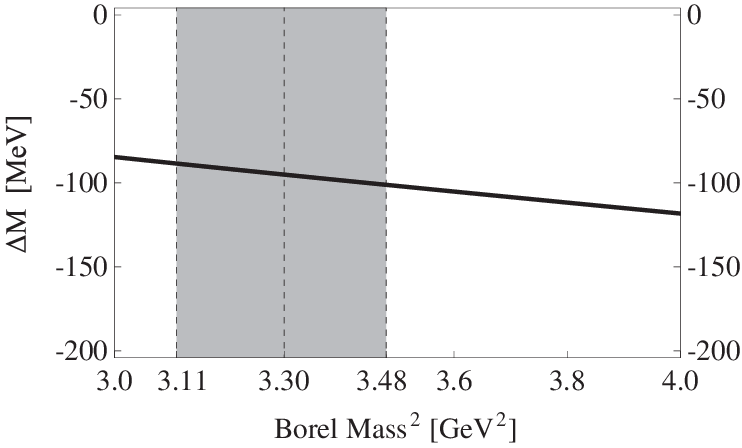}
\caption{The mass correction $\Delta M^{\bar D \Sigma_c}_{I=1/2,J=1/2}$ as a function of the Borel mass $M_B$.}
\label{fig:binding}
\end{center}
\end{figure}

Altogether we extract the working region of $M_B$ to be $3.11$~GeV$^2< M_B^2 < 3.48$~GeV$^2$, where we use Eq.~(\ref{eq:binding}) to calculate the mass correction to be
\begin{equation}
\Delta M^{\bar D \Sigma_c}_{I=1/2,J=1/2} = - 95~{\rm MeV} \, .
\label{result:Dhalf}
\end{equation}
We show its variation in Fig.~\ref{fig:binding} with respect to the Borel mass $M_B$. It is shown in a broader region $3.0$~GeV$^2\leq M_B^2 \leq 4.0$~GeV$^2$, and we find it quite stable inside the above Borel window.

The mass correction $\Delta M^{\bar D \Sigma_c}_{I=1/2,J=1/2}$ given in Eq.~(\ref{result:Dhalf}) is negative, suggesting the light-quark-exchange potential $V^{\bar D \Sigma_c}_{I=1/2,J=1/2}(r)$ to be attractive, so there can be the $\bar D \Sigma_c$ covalent molecule of $I=1/2$ and $J=1/2$.

\subsection{$D^- \Sigma_c^{++}$ sum rules}
\label{sec:sumruleDm}

As shown in Eq.~(\ref{pi:Dm}), the light-quark-exchange term $\Pi_{Q}(x)$ does not contribute to the correlation function of the $D^- \Sigma_c^{++}$ molecule, so the $D^- \Sigma_c^{++}$ covalent molecule does not exist.

\subsection{$\bar D^0 \Sigma_c^{+}$ sum rules}
\label{sec:sumruleD0}

We follow Sec.~\ref{sec:sumruleDhalf} to study the $\bar D^0 \Sigma_c^{+}$ molecule, and calculate its mass correction to be
\begin{equation}
\Delta M^{\bar D^0 \Sigma_c^{+}}_{J=1/2} \approx - \Delta M^{\bar D \Sigma_c}_{I=1/2,J=1/2} = 95~{\rm MeV} \, .
\end{equation}
This result suggests $V^{\bar D^0 \Sigma_c^{+}}_{J=1/2}(r)$ to be repulsive, so the $\bar D^0 \Sigma_c^{+}$ covalent molecule does not exist.

\subsection{$I=3/2$ $\bar D \Sigma_c$ sum rules}
\label{sec:sumruleD3half}

We follow Sec.~\ref{sec:sumruleDhalf} to study the $I=3/2$ $\bar D \Sigma_c$ molecule, and calculate its mass correction to be
\begin{equation}
\Delta M^{\bar D \Sigma_c}_{I=3/2,J=1/2} \approx - 2\Delta M^{\bar D \Sigma_c}_{I=1/2,J=1/2} = 190~{\rm MeV} \, .
\label{eq:isospin3half}
\end{equation}
This result suggests $V^{\bar D \Sigma_c}_{I=3/2,J=1/2}(r)$ to be repulsive, so the $\bar D \Sigma_c$ covalent molecule of $I=3/2$ and $J=1/2$ does not exist.

\section{More hadronic molecules}
\label{sec:more}

In this section we follow the procedures used in Sec.~\ref{sec:sumruleDhalf} and study more possibly-existing covalent hadronic molecules. We shall investigate the $\bar D^* \Sigma_c/\bar D \Sigma_c^*/\bar D^* \Sigma_c^*$ molecules, the $\bar D^{(*)} \Lambda_c$ molecules, the $D^{(*)} \bar K^{*}$ molecules, and the $D^{(*)} \bar D^{(*)}$ molecules, separately in the following subsections.

\subsection{$\bar D^* \Sigma_c/\bar D \Sigma_c^*/\bar D^* \Sigma_c^*$ molecules}
\label{sec:sumruleDSigma}

In this subsection we investigate the light-quark-exchange term $\Pi_{Q}(x)$ and study its contributions to the $\bar D^* \Sigma_c/\bar D \Sigma_c^*/\bar D^* \Sigma_c^*$ molecules.

The current corresponding to the $\bar D^* \Sigma_c$ molecule is
\begin{eqnarray}
&& J_\alpha^{\bar D^* \Sigma_c}(x) = J_\alpha^{\bar D^*}(x) \times J^{\Sigma_c}(x)
\\ \nonumber &=& [\bar c_d(x) \gamma_\alpha q_d(x)] \times [\epsilon^{abc} q^{\rm T}_a(x) \mathbb{C} \gamma^\mu q_b(x) \gamma_\mu\gamma_{5} c_c(x)] \, .
\end{eqnarray}
Its correlation function is
\begin{eqnarray}
&& \Pi_{\alpha\beta}^{\bar D^*\Sigma_c}(q^2)
\\ \nonumber &=& {\rm i} \int {\rm d}^4x {\rm e}^{{\rm i}qx} \langle 0 | \mathbb{T}\left[J_\alpha^{\bar D^*\Sigma_c}(x) \bar J_\beta^{\bar D^*\Sigma_c}(0)\right] | 0 \rangle
\\ \nonumber &=& \mathcal{G}^{3/2}_{\alpha\beta}(q^2)~\Pi^{\bar D^*\Sigma_c}_{J=3/2}\left(q^2\right) + \mathcal{G}^{1/2}_{\alpha\beta}(q^2)~\Pi^{\bar D^*\Sigma_c}_{J=1/2}\left(q^2\right) \, ,
\end{eqnarray}
where $\Pi^{\bar D^*\Sigma_c}_{J=3/2}\left(q^2\right)$ and $\Pi^{\bar D^*\Sigma_c}_{J=1/2}\left(q^2\right)$ are contributed by the spin-3/2 and spin-1/2 components, respectively. $\mathcal{G}^{3/2}_{\alpha\beta}\left(q^2\right)$ and $\mathcal{G}^{1/2}_{\alpha\beta}\left(q^2\right)$ are coefficients of the spin-3/2 and spin-1/2 propagators, respectively:
\begin{eqnarray}
\nonumber \mathcal{G}^{3/2}_{\alpha\beta}(q^2) &=& \big( g_{\alpha\beta} - {\gamma_\alpha \gamma_\beta \over3}  - {q_\alpha \gamma_\beta - q_\beta \gamma_\alpha \over 3M} - {2q_\alpha q_\beta \over 3M^2} \big)
\\ && \times\,(q\!\!\!\slash~ + M) \, ,
\\[1mm] \mathcal{G}^{1/2}_{\alpha\beta}(q) &=& q_\alpha q_\beta \times\,(q\!\!\!\slash~ + M) \, .
\end{eqnarray}
In the present study we have only calculated the terms proportional to the Lorentz coefficient $g_{\alpha\beta}$, so we can only extract the mass correction to the spin-3/2 $\bar D^* \Sigma_c$ molecule. We find it negative for the $I=1/2$ one:
\begin{equation}
\Delta M^{\bar D^* \Sigma_c}_{I=1/2,J=3/2} = - 89~{\rm MeV} \, .
\end{equation}
This result suggests $V_{I=1/2,J=3/2}^{\bar D^* \Sigma_c}(r)$ to be attractive, so there can be the $\bar D^* \Sigma_c$ covalent molecule of $I=1/2$ and $J=3/2$.

Similarly, we use the current
\begin{eqnarray}
&& J_\alpha^{\bar D \Sigma_c^*}(x) = J^{\bar D}(x) \times J_\alpha^{\Sigma_c^*}(x)
\\ \nonumber &=& [\bar c_d(x) \gamma_5 q_d(x)] \times [\epsilon^{abc}q^{\rm T}_a(x) \mathbb{C} \gamma^\mu q_b(x) P^{3/2}_{\alpha\mu}c_c(x)] \, ,
\end{eqnarray}
to perform QCD sum rule analyses and study the $\bar D \Sigma_c^*$ molecule. Here $P^{3/2}_{\alpha\mu}$ is the spin-3/2 projection operator
\begin{equation}
P^{3/2}_{\alpha\mu} = g_{\alpha\mu} - {1 \over 4} \gamma_\alpha\gamma_\mu \, .
\end{equation}
We calculate its mass correction to be
\begin{equation}
\Delta M^{\bar D \Sigma_c^*}_{I=1/2,J=3/2} = - 86~{\rm MeV} \, .
\end{equation}
This result suggests $V_{I=1/2,J=3/2}^{\bar D \Sigma_c^*}(r)$ to be attractive, so there can be the $\bar D \Sigma_c^*$ covalent molecule of $I=1/2$ and $J=3/2$.

The current corresponding to the $\bar D^* \Sigma_c^*$ molecule is
\begin{eqnarray}
&& J_{\alpha_1\alpha_2}^{\bar D^* \Sigma_c^*}(x) = J_{\alpha_1}^{\bar D^*}(x) \times J_{\alpha_2}^{\Sigma_c^*}(x)
\\ \nonumber &=& [\bar c_d(x) \gamma_{\alpha_1} q_d(x)] \times [\epsilon^{abc}q^{\rm T}_a(x) \mathbb{C} \gamma^\mu q_b(x) P^{3/2}_{\alpha_2\mu}c_c(x)] \, .
\end{eqnarray}
Its correlation function is
\begin{eqnarray}
&& \Pi_{\alpha_1\alpha_2,\beta_1\beta_2}^{\bar D^*\Sigma_c^*}(q^2)
\\ \nonumber &=& {\rm i} \int {\rm d}^4x {\rm e}^{{\rm i}qx} \langle 0 | \mathbb{T}\left[J_{\alpha_1\alpha_2}^{\bar D^*\Sigma_c^*}(x) \bar J_{\beta_1\beta_2}^{\bar D^*\Sigma_c^*}(0)\right] | 0 \rangle
\\ \nonumber &=& \mathcal{G}^{5/2}_{\alpha_1\alpha_2,\beta_1\beta_2}(q^2)~\Pi^{\bar D^*\Sigma_c^*}_{J=5/2}\left(q^2\right)
\\ \nonumber &+& \mathcal{G}^{3/2}_{\alpha_1\alpha_2,\beta_1\beta_2}(q^2)~\Pi^{\bar D^*\Sigma_c^*}_{J=3/2}\left(q^2\right)
\\ \nonumber &+& \mathcal{G}^{1/2}_{\alpha_1\alpha_2,\beta_1\beta_2}(q^2)~\Pi^{\bar D^*\Sigma_c^*}_{J=1/2}\left(q^2\right) + \cdots \, ,
\end{eqnarray}
where $\Pi^{\bar D^*\Sigma_c^*}_{J=5/2}\left(q^2\right)$, $\Pi^{\bar D^*\Sigma_c^*}_{J=3/2}\left(q^2\right)$, and $\Pi^{\bar D^*\Sigma_c^*}_{J=1/2}\left(q^2\right)$ are contributed by the spin-5/2, spin-3/2, and spin-1/2 components, respectively. $\mathcal{G}^{5/2}_{\alpha_1\alpha_2,\beta_1\beta_2}(q^2)$, $\mathcal{G}^{3/2}_{\alpha_1\alpha_2,\beta_1\beta_2}(q^2)$, and $\mathcal{G}^{1/2}_{\alpha_1\alpha_2,\beta_1\beta_2}(q^2)$ are coefficients of the spin-5/2, spin-3/2, and spin-1/2 propagators, respectively:
\begin{eqnarray}
&& \mathcal{G}^{5/2}_{\alpha_1\alpha_2,\beta_1\beta_2}(q^2) = (q\!\!\!\slash~ + M)
\\ \nonumber && ~~~~~~~~~ \times \big( {g_{\alpha_1\beta_1} g_{\alpha_2\beta_2} \over 2} + {g_{\alpha_1\beta_2} g_{\alpha_2\beta_1} \over 2} - {g_{\alpha_1\alpha_2} g_{\beta_1\beta_2} \over 4} \big) \, ,
\\[1mm] && \mathcal{G}^{3/2}_{\alpha_1\alpha_2,\beta_1\beta_2}(q^2) = (q\!\!\!\slash~ + M)
\\ \nonumber && ~~~~~~~~~ \times \big( {g_{\alpha_1\beta_1} g_{\alpha_2\beta_2} \over 2} - {g_{\alpha_1\beta_2} g_{\alpha_2\beta_1} \over 2}  \big) \, ,
\\[1mm] && \mathcal{G}^{1/2}_{\alpha_1\alpha_2,\beta_1\beta_2}(q^2) = (q\!\!\!\slash~ + M) \times g_{\alpha_1\alpha_2} g_{\beta_1\beta_2} \, .
\end{eqnarray}
In the present study we have calculated the terms proportional to the three Lorentz coefficients $g_{\alpha_1\beta_1} g_{\alpha_2\beta_2}$, $g_{\alpha_1\beta_2} g_{\alpha_2\beta_1}$, and $g_{\alpha_1\alpha_2} g_{\beta_1\beta_2}$, so we can  extract mass corrections to all the three $I=1/2$ $\bar D^* \Sigma_c^*$ molecules of $J=5/2$, $J=3/2$, and $J=1/2$:
\begin{eqnarray}
\Delta M^{\bar D^* \Sigma_c^*}_{I=1/2,J=5/2} &=& -107~{\rm MeV} \, ,
\\[1mm] \Delta M^{\bar D^* \Sigma_c^*}_{I=1/2,J=3/2} &=& -47~{\rm MeV} \, ,
\\[1mm] \Delta M^{\bar D^* \Sigma_c^*}_{I=1/2,J=1/2} &=& 3.5~{\rm MeV} \, .
\end{eqnarray}
These results suggest $V_{I=1/2,J=5/2}^{\bar D^* \Sigma_c^*}(r)$ to be attractive, so there can be the $\bar D^* \Sigma_c^*$ covalent molecule of $I=1/2$ and $J=5/2$.

\begin{figure}[hbtp]
\begin{center}
\includegraphics[width=0.45\textwidth]{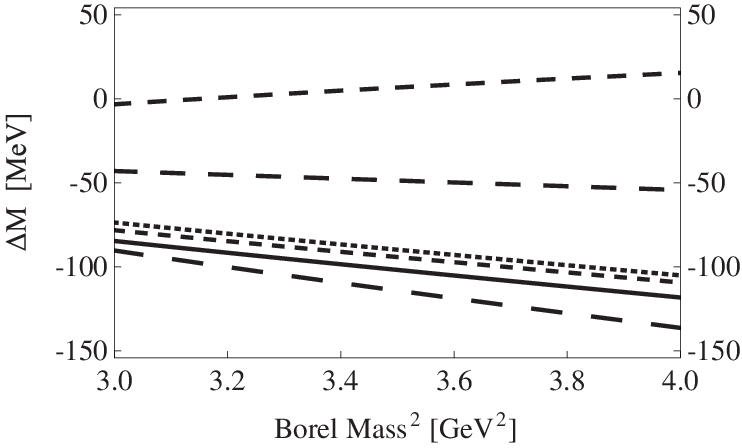}
\caption{Mass corrections $\Delta M^{\bar D^{(*)} \Sigma_c^{(*)}}$ as functions of the Borel mass $M_B$. The curves from top to bottom correspond to $\Delta M^{\bar D^* \Sigma_c^*}_{I=1/2,J=1/2}$, $\Delta M^{\bar D^* \Sigma_c^*}_{I=1/2,J=3/2}$, $\Delta M^{\bar D^* \Sigma_c}_{I=1/2,J=3/2}$, $\Delta M^{\bar D \Sigma_c^*}_{I=1/2,J=3/2}$, $\Delta M^{\bar D \Sigma_c}_{I=1/2,J=1/2}$, and $\Delta M^{\bar D^* \Sigma_c^*}_{I=1/2,J=5/2}$, respectively.}
\label{fig:binding1}
\end{center}
\end{figure}

For completeness, we show variations of $\Delta M^{\bar D^{(*)} \Sigma_c^{(*)}}$ in Fig.~\ref{fig:binding1} with respect to the Borel mass $M_B$. Besides, we study the $I=3/2$ $\bar D^* \Sigma_c/\bar D \Sigma_c^*/\bar D^* \Sigma_c^*$ molecules and obtain
\begin{equation}
\Delta M^{\bar D^* \Sigma_c/\bar D \Sigma_c^*/\bar D^* \Sigma_c^*}_{I=3/2} \approx - 2 \Delta M^{\bar D^* \Sigma_c/\bar D \Sigma_c^*/\bar D^* \Sigma_c^*}_{I=1/2} \, .
\end{equation}

\subsection{$\bar D^{(*)} \Lambda_c$ molecules}
\label{sec:sumruleDLambda}

In this subsection we investigate the light-quark-exchange term $\Pi_{Q}(x)$ and study its contributions to the $\bar D^{(*)} \Lambda_c$ molecules.

The currents corresponding to the $\bar D \Lambda_c$ and $\bar D^{*} \Lambda_c$ molecules are
\begin{eqnarray}
&& J^{\bar D \Lambda_c}(x) = J^{\bar D}(x) \times J^{\Lambda_c}(x)
\\ \nonumber &=& [\bar c_d(x) \gamma_5 q_d(x)] \times [\epsilon^{abc} q^{\rm T}_a(x) \mathbb{C} \gamma_5 q_b(x) c_c(x)] \, ,
\\[1mm] && J_\alpha^{\bar D^* \Lambda_c}(x) = J_\alpha^{\bar D^*}(x) \times J^{\Lambda_c}(x)
\\ \nonumber &=& [\bar c_d(x) \gamma_\alpha q_d(x)] \times [\epsilon^{abc} q^{\rm T}_a(x) \mathbb{C} \gamma_5 q_b(x) c_c(x)] \, .
\end{eqnarray}
We use them to perform QCD sum rule analyses, and calculate mass corrections to the $\bar D \Lambda_c$ molecule of $J=1/2$ and the $\bar D^* \Lambda_c$ molecule of $J=3/2$. We find both of them to be positive. However, we do not obtain the mass correction to the $\bar D^* \Lambda_c$ molecule of $J=1/2$, just like the $\bar D^* \Sigma_c$ molecule of $J=1/2$ . These results suggest $V_{I=1/2,J=1/2}^{\bar D \Lambda_c}(r)$ and $V_{I=1/2,J=3/2}^{\bar D^* \Lambda_c}(r)$ to be both repulsive, so the $J=1/2$ $\bar D \Lambda_c$ and $J=3/2$ $\bar D^* \Lambda_c$ covalent molecules do not exist.

\subsection{$D^{(*)} \bar K^{*}$ molecules}
\label{sec:sumruleDK}

In this subsection we investigate the light-quark-exchange term $\Pi_{Q}(x)$ and study its contributions to the $D^{(*)} \bar K^{*}$ molecules. We do not investigate the $D^{(*)} \bar K$ molecules in the present study, because their bare masses $M_{D^{(*)}} + M_{\bar K}$ can not be easily reached within the present QCD sum rule approach, due to the nature of $K$ mesons as Nambu-Goldstone bosons.

The current corresponding to the $D \bar K^{*}$ molecule is
\begin{eqnarray}
J_{\alpha}^{D \bar K^{*}}(x) &=& J^{\bar D}(x) \times J_{\alpha}^{\bar K^{*}}(x)
\\ \nonumber &=& [\bar q_a(x) \gamma_5 c_a(x)] \times [\bar q_b(x) \gamma_{\alpha} s_b(x)] \, .
\end{eqnarray}
Its correlation function is
\begin{eqnarray}
&& \Pi_{\alpha\beta}^{D \bar K^{*}}(q^2)
\\ \nonumber &=& {\rm i} \int {\rm d}^4x {\rm e}^{{\rm i}qx} \langle 0 | \mathbb{T}\left[J_{\alpha}^{D \bar K^{*}}(x) J_{\beta}^{D \bar K^{*},\dagger}(0)\right] | 0 \rangle
\\ \nonumber &=& \left( g_{\alpha\beta} - {q_\alpha q_\beta \over q^2} \right)~\Pi^{D \bar K^{*}}\left(q^2\right) + \cdots \, .
\end{eqnarray}
We use $J_{\alpha}^{D \bar K^{*}}(x)$ to perform QCD sum rule analyses, and calculate its mass correction to be
\begin{equation}
\Delta M^{D \bar K^{*}}_{I=0,J=1} = -180~{\rm MeV} \, .
\end{equation}
This suggests $V^{D \bar K^{*}}_{I=0,J=1}(r)$ to be attractive, so there can be the $D \bar K^{*}$ covalent molecule of $I=0$ and $J=1$.

The current corresponding to the $D^* \bar K^{*}$ molecule is
\begin{eqnarray}
J_{\alpha_1\alpha_2}^{D^* \bar K^{*}}(x) &=& J_{\alpha_1}^{\bar D^*}(x) \times J_{\alpha_2}^{\bar K^{*}}(x)
\\ \nonumber &=& [\bar q_a(x) \gamma_{\alpha_1} c_a(x)] \times [\bar q_b(x) \gamma_{\alpha_2} s_b(x)] \, .
\end{eqnarray}
Its correlation function is
\begin{eqnarray}
&& \Pi_{\alpha_1\alpha_2,\beta_1\beta_2}^{D^* \bar K^{*}}(q^2)
\\ \nonumber &=& {\rm i} \int {\rm d}^4x {\rm e}^{{\rm i}qx} \langle 0 | \mathbb{T}\left[J_{\alpha_1\alpha_2}^{D^* \bar K^{*}}(x) J_{\beta_1\beta_2}^{D^* \bar K^{*},\dagger}(0)\right] | 0 \rangle
\\ \nonumber &=& \mathcal{G}^{2}_{\alpha_1\alpha_2,\beta_1\beta_2}(q^2)~\Pi^{D^* \bar K^{*}}_{J=2}\left(q^2\right)
\\ \nonumber &+& \mathcal{G}^{1}_{\alpha_1\alpha_2,\beta_1\beta_2}(q^2)~\Pi^{D^* \bar K^{*}}_{J=1}\left(q^2\right)
\\ \nonumber &+& \mathcal{G}^{0}_{\alpha_1\alpha_2,\beta_1\beta_2}(q^2)~\Pi^{D^* \bar K^{*}}_{J=0}\left(q^2\right) + \cdots \, ,
\end{eqnarray}
where $\Pi^{D^* \bar K^{*}}_{J=2}\left(q^2\right)$, $\Pi^{D^* \bar K^{*}}_{J=1}\left(q^2\right)$, and $\Pi^{D^* \bar K^{*}}_{J=0}\left(q^2\right)$ are contributed by the spin-2, spin-1, and spin-0 components, respectively. $\mathcal{G}^{2}_{\alpha_1\alpha_2,\beta_1\beta_2}(q^2)$, $\mathcal{G}^{1}_{\alpha_1\alpha_2,\beta_1\beta_2}(q^2)$, and $\mathcal{G}^{0}_{\alpha_1\alpha_2,\beta_1\beta_2}(q^2)$ are coefficients of the spin-2, spin-1, and spin-0 propagators, respectively:
\begin{eqnarray}
&& \mathcal{G}^{2}_{\alpha_1\alpha_2,\beta_1\beta_2}(q^2) =
\\ \nonumber && ~~~~~~~~~ {g_{\alpha_1\beta_1} g_{\alpha_2\beta_2} \over 2} + {g_{\alpha_1\beta_2} g_{\alpha_2\beta_1} \over 2} - {g_{\alpha_1\alpha_2} g_{\beta_1\beta_2} \over 4} \, ,
\\[1mm] && \mathcal{G}^{1}_{\alpha_1\alpha_2,\beta_1\beta_2}(q^2) = {g_{\alpha_1\beta_1} g_{\alpha_2\beta_2} \over 2} - {g_{\alpha_1\beta_2} g_{\alpha_2\beta_1} \over 2} \, ,
\\[1mm] && \mathcal{G}^{0}_{\alpha_1\alpha_2,\beta_1\beta_2}(q^2) = g_{\alpha_1\alpha_2} g_{\beta_1\beta_2} \, .
\end{eqnarray}
We use $J_{\alpha_1\alpha_2}^{D^* \bar K^{*}}(x)$ to perform QCD sum rule analyses, and extract mass corrections to the three isoscalar $D^* \bar K^{*}$ molecules of $J=2$, $J=1$, and $J=0$:
\begin{eqnarray}
\Delta M^{D^* \bar K^{*}}_{I=0,J=2} &=& -119~{\rm MeV} \, ,
\\[1mm] \Delta M^{D^* \bar K^{*}}_{I=0,J=1} &=& -46~{\rm MeV} \, ,
\\[1mm] \Delta M^{D^* \bar K^{*}}_{I=0,J=0} &=& -4.5~{\rm MeV} \, .
\label{eq:DK00}
\end{eqnarray}
These results suggest $V_{I=0,J=2}^{D^* \bar K^{*}}(r)$ to be attractive, so there can be the $D^* \bar K^{*}$ covalent molecule of $I=0$ and $J=2$.

\begin{figure}[hbtp]
\begin{center}
\includegraphics[width=0.45\textwidth]{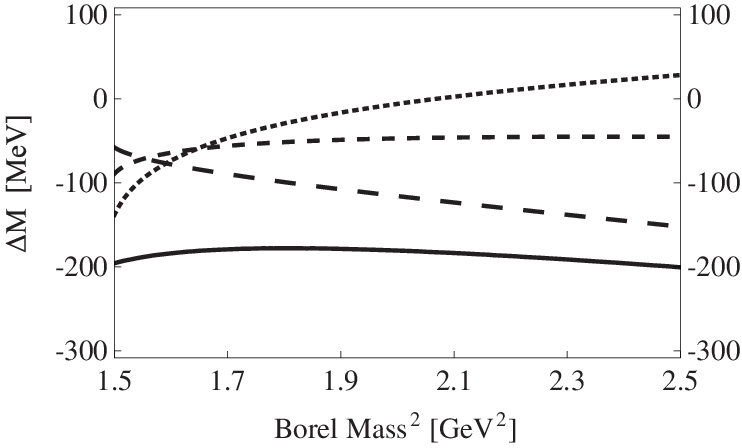}
\caption{Mass corrections $\Delta M^{D^{(*)} \bar K^{*}}$ as functions of the Borel mass $M_B$. The curves from top to bottom correspond to $\Delta M^{D^* \bar K^*}_{I=0,J=0}$ (short-dashed), $\Delta M^{D^* \bar K^*}_{I=0,J=1}$ (middle-dashed), $\Delta M^{D^* \bar K^*}_{I=0,J=2}$ (long-dashed), and $\Delta M^{D \bar K^*}_{I=0,J=1}$ (solid), respectively.}
\label{fig:binding2}
\end{center}
\end{figure}

We show variations of $\Delta M^{D^{(*)} \bar K^{*}}$ in Fig.~\ref{fig:binding2} with respect to the Borel mass $M_B$. Especially, $V_{I=0,J=0}^{D^* \bar K^{*}}(r)$ depends significantly on the Borel mass $M_B$ and so also on the threshold value $s_0$, for which we refer to Sec.~\ref{sec:covalent} for relevant model-independent discussions.

For completeness, we study the isovector $D^{(*)} \bar K^{*}$ molecules and obtained
\begin{equation}
\Delta M^{D \bar K^{*}/D^{*} \bar K^{*}}_{I=1} \approx - \Delta M^{D \bar K^{*}/D^{*} \bar K^{*}}_{I=0} \, .
\end{equation}

\subsection{$D^{(*)} \bar D^{(*)}$ molecules}

The light-quark-exchange term $\Pi_{Q}(x)$ does not contribute to correlation functions of the $D^{(*)} \bar D^{(*)}$ molecules, suggesting the $D^{(*)} \bar D^{(*)}$ covalent molecules not to exist. However, there can still be the $D^{(*)} \bar D^{(*)}$ hadronic molecular states induced by some other binding mechanisms, such as the one-meson-exchange interaction~\cite{Weinberg:1965zz,Voloshin:1976ap,DeRujula:1976zlg,Tornqvist:1993ng,Voloshin:2003nt,Close:2003sg,Wong:2003xk,Braaten:2003he,Swanson:2003tb,Tornqvist:2004qy}.

\section{Covalent hadronic molecule}
\label{sec:covalent}

In the previous section we have applied QCD sum rules to study the binding mechanism induced by shared light quarks. This mechanism is somewhat similar to the covalent bond in chemical molecules induced by shared electrons, so we call such hadronic molecules ``covalent hadronic molecules''. Recalling that the two shared electrons must spin in opposite directions (and so totally antisymmetric obeying the Pauli principle) in order to form a chemical covalent bond, our QCD sum rule results indicate a similar behavior: {\it the light-quark-exchange interaction is attractive when the shared light quarks are totally antisymmetric so that obey the Pauli principle} (these quarks may spin in the same direction given their flavor structure capable of being antisymmetric).

In this section we qualitatively study the above hypothesis. Its logical chain is quite straightforward. We assume the two light quarks $q_A$ inside $Y$ and $q_B$ inside $Z$ are totally antisymmetric. Hence, $q_A$ and $q_B$ obey the Pauli principle, so that they can be exchanged and shared. By doing this, wave-functions of $Y$ and $Z$ overlap with each other, so that they are attracted and there can be the covalent hadronic molecule $X = | YZ \rangle$. This picture has been depicted in Fig.~\ref{fig:sharing}. We believe it better and more important than our QCD sum rule results, given it to be model independent and more easily applicable.

We apply it to study several examples as follows. The two exchanged light quarks have the same color and so the symmetric color structure; besides, we assume their orbital structure to be $S$-wave and so also symmetric; consequently, we only need to investigate their spin and flavor structures.

\paragraph{$\bar D^{(*)0}[\bar c_1 u_2]$--$\Lambda_c^+[u_3 d_4 c_5]$ covalent molecules.} Let us exchange $u_2$ inside $\bar D^{(*)0}$ and $u_3$ inside $\Lambda_c^+$. $u_3$ and $d_4$ inside $\Lambda_c^+$ spin in opposite directions, $u_2$ and $d_4$ also need to spin in opposite directions in order to form another $\Lambda_c^+$, so $u_2$ and $u_3$ spin in the same direction with the symmetric spin structure. The flavor structure of $u_2$ and $u_3$ is also symmetric, so they are totally symmetric (${\bf S}$=symmetric and ${\bf A}$=antisymmetric):
\begin{equation}\nonumber
\begin{array}{cccccc}
\hline
                           & {\rm color} & {\rm flavor} & {\rm spin} & {\rm orbital} & {\rm total}
\\ \hline
u_2 \leftrightarrow u_3    & {\bf S}     & {\bf S}      & {\bf S}    & {\bf S}       & {\bf S}
\\ \hline
\end{array}
\end{equation}
Accordingly, the $\bar D^{(*)} \Lambda_c$ covalent molecules do not exist. This is consistent with our QCD sum rule result obtained in Sec.~\ref{sec:sumruleDLambda}.

\paragraph{$\bar D^{0}[\bar c_1 u_2]$--$\Sigma_c^+[u_3 d_4 c_5]$ covalent molecule.} Let us exchange $u_2$ inside $\bar D^{0}$ and $u_3$ inside $\Sigma_c^+$. $\bar c_1$ and $u_2$ inside $\bar D^{0}$ spin in opposite directions, $\bar c_1$ and $u_3$ also need to spin in opposite directions in order to form another $\bar D^{0}$, so $u_2$ and $u_3$ spin in the same direction with the symmetric spin structure. The flavor structure of $u_2$ and $u_3$ is also symmetric, so they are totally symmetric:
\begin{equation}\nonumber
\begin{array}{cccccc}
\hline
                           & {\rm color} & {\rm flavor} & {\rm spin} & {\rm orbital} & {\rm total}
\\ \hline
u_2 \leftrightarrow u_3    & {\bf S}     & {\bf S}      & {\bf S}    & {\bf S}       & {\bf S}
\\ \hline
\end{array}
\end{equation}
Accordingly, the $\bar D^{0} \Sigma_c^+$ covalent molecule does not exist, consistent with Sec.~\ref{sec:sumruleD0}.

\paragraph{$\bar D[\bar c_1 q_2]$--$\Sigma_c[q_3 q_4 c_5]$ covalent molecule ($q=u/d$).} After including the isospin symmetry, the exchange can take place between up and down quarks. Let us exchange $q_2$ inside $\bar D$ and $q_3$ inside $\Sigma_c$. As discussed above, they have the symmetric spin structure, so they can be totally antisymmetric as long as their flavor structure is antisymmetric:
\begin{equation}\nonumber
\begin{array}{cccccc}
\hline
                           & {\rm color} & {\rm flavor} & {\rm spin} & {\rm orbital} & {\rm total}
\\ \hline
q_2 \leftrightarrow q_3    & {\bf S}     & {\bf A}      & {\bf S}    & {\bf S}       & {\bf A}
\\ \hline
\end{array}
\end{equation}
Accordingly, there can be the $I=1/2$ $\bar D\Sigma_c$ covalent molecule, but not the $I=3/2$ one. This is consistent with our QCD sum rule result obtained in Sec.~\ref{sec:sumruleDhalf} and Sec.~\ref{sec:sumruleD3half}. Similarly, we derive that there can be the $I=1/2$ $\bar D\Sigma_c^*$ and $I=0$ $D \bar K^{*}$ covalent molecules, consistent with  Sec.~\ref{sec:sumruleDSigma} and Sec.~\ref{sec:sumruleDK}.

\paragraph{$J=5/2$ $\bar D^*[\bar c_1 q_2]$--$\Sigma_c^*[q_3 q_4 c_5]$ covalent molecule.} Let us exchange $q_2$ inside $\bar D^*$ and $q_3$ inside $\Sigma_c^*$. In the $J_z = + 5/2$ component, all the quarks/antiquark spin in the same direction, so $q_2$ and $q_3$ have the symmetric spin structure. They can be totally antisymmetric as long as their flavor structure is antisymmetric:
\begin{equation}\nonumber
\begin{array}{cccccc}
\hline
                           & {\rm color} & {\rm flavor} & {\rm spin} & {\rm orbital} & {\rm total}
\\ \hline
q_2 \leftrightarrow q_3    & {\bf S}     & {\bf A}      & {\bf S}    & {\bf S}       & {\bf A}
\\ \hline
\end{array}
\end{equation}
Accordingly, there can be the $I=1/2$ $\bar D^* \Sigma_c^*$ covalent molecule of $J=5/2$, but not the $I=3/2$ one. This is consistent with our QCD sum rule result obtained in Sec.~\ref{sec:sumruleDSigma}. Similarly, we derive that there can be the $D^* \bar K^{*}$ covalent molecule of $I=0$ and $J=2$, consistent with  Sec.~\ref{sec:sumruleDK}.

\paragraph{$J=0$ $D^*[c_1 \bar q_2]$--$\bar K^{*}[s_3 \bar q_4]$ covalent molecule.} Let us exchange $\bar q_2$ inside $D^*$ and $\bar q_4$ inside $\bar K^{*}$. We perform the spin decomposition (see Appendix~\ref{app:spin}):
\begin{eqnarray}
&& | {1}_{c \bar q} \otimes {1}_{s \bar q}; J = 0 \rangle
\\ \nonumber &=& {\sqrt3\over2}~| {0}_{c s} \otimes {0}_{\bar q \bar q}; J = 0 \rangle - {1\over2}~| {1}_{c s} \otimes {1}_{\bar q \bar q}; J = 0 \rangle \, .
\end{eqnarray}
Hence, there exists both ${s}_{\bar q_2 \bar q_4}={0}$ (75\%) and ${s}_{\bar q_2 \bar q_4}={1}$ (25\%) components. The former becomes attractive when $I=1$, and the latter becomes attractive when $I=0$:
\begin{equation}\nonumber
\begin{array}{cccccc}
\hline\hline
                                   & {\rm color} & {\rm flavor} & {\rm spin} & {\rm orbital} & {\rm total}
\\ \hline \hline
\bar q_2 \leftrightarrow \bar q_4  & {\bf S}     & {\bf S}      & {\bf A}    & {\bf S}       & {\bf A}
\\ \hline
\bar q_2 \leftrightarrow \bar q_4  & {\bf S}     & {\bf A}      & {\bf S}    & {\bf S}       & {\bf A}
\\ \hline \hline
\end{array}
\end{equation}
It is well known that there are both the para-hydrogen and ortho-hydrogen, where the two protons spin in opposite directions and in the same direction, respectively. Similarly, there might be two $J=0$ $D^*\bar K^{*}$ covalent molecules, {\it i.e.}, the two components $| {0}_{c s} \otimes {0}_{\bar q \bar q}; J = 0 \rangle$ of $I=1$ and $| {1}_{c s} \otimes {1}_{\bar q \bar q}; J = 0 \rangle$ of $I=0$. However, our QCD sum rule studies performed in Sec.~\ref{sec:sumruleDK} can not differentiate these two hyperfine structures, because there we have summed over the $D^*$ and $\bar K^{*}$ polarizations.

It is useful to generally discuss how many light quarks at most are there in the lowest orbit ($q=u/d$):
\begin{enumerate}

\item In the $\bar D^{(*)}[\bar c_1 q_2] K^{(*)}[\bar s_3 q_4]$ covalent molecule, the two exchanged light quarks $q_2$ and $q_4$ have the same color and so the symmetric color structure; besides, we assume their orbital structure to be $S$-wave and so also symmetric; consequently, there are two possible configurations satisfying the Pauli principle (${\bf S}$=symmetric and ${\bf A}$=antisymmetric):
    \begin{equation}\nonumber
    \begin{array}{cccccc}
    \hline\hline
                            & {\rm color} & {\rm flavor} & {\rm spin} & {\rm orbital} & {\rm total}
    \\ \hline \hline
    q_2 \leftrightarrow q_4 & {\bf S}     & {\bf A}      & {\bf S}    & {\bf S}       & {\bf A}
    \\ \hline
    q_2 \leftrightarrow q_4 & {\bf S}     & {\bf S}      & {\bf A}    & {\bf S}       & {\bf A}
    \\ \hline\hline
    \end{array}
    \end{equation}
    Hence, two antiquarks can share (at most) two quarks, with the quantum numbers either $(I)J^P = (0)1^+$ or $(1)0^+$.

\item In the $\bar D^{(*)}[\bar c_1 q_2]$--$\Sigma_c^{(*)}[q_3 q_4 c_5]$ covalent molecule, there are three light up/down quarks. We assume the two exchanged quarks to be $q_2$ and $q_3$ with the same color. There are also two possible configurations:
    \begin{equation}\nonumber
    \begin{array}{cccccc}
    \hline\hline
                            & {\rm color} & {\rm flavor} & {\rm spin} & {\rm orbital} & {\rm total}
    \\ \hline \hline
    q_2 \leftrightarrow q_3 & {\bf S}     & {\bf A}      & {\bf S}    & {\bf S}       & {\bf A}
    \\
    q_2 \leftrightarrow q_4 & {\bf A}     & {\bf S}      & {\bf S}    & {\bf S}       & {\bf A}
    \\
    q_3 \leftrightarrow q_4 & {\bf A}     & {\bf S}      & {\bf S}    & {\bf S}       & {\bf A}
    \\ \hline
    q_2 \leftrightarrow q_3 & {\bf S}     & {\bf S}      & {\bf A}    & {\bf S}       & {\bf A}
    \\
    q_2 \leftrightarrow q_4 & {\bf A}     & {\bf S}      & {\bf S}    & {\bf S}       & {\bf A}
    \\
    q_3 \leftrightarrow q_4 & {\bf A}     & {\bf S}      & {\bf S}    & {\bf S}       & {\bf A}
    \\ \hline\hline
    \end{array}
    \end{equation}
    They satisfy the condition that any two of the three light quarks are totally antisymmetric so that obey the Pauli principle. Hence, one quark and one antiquark can share (at most) three quarks, with either $(I)J^P = ({1\over2}){3\over2}^+$ or $({3\over2}){1\over2}^+$. However, the $\bar D^{(*)}\Lambda_c$ system does not satisfy this condition.

\item In the $\Sigma_c^{(*)}[q_1 q_2 c_3]$--$\Sigma_b^{(*)}[q_4 q_5 c_6]$ covalent molecule, there are four light up/down quarks. We assume that $q_1$ and $q_4$ can be exchanged with the same color, and $q_2$ and $q_5$ can also be exchanged with the same color. There can be four possible configurations:
    \begin{equation}\nonumber
    \begin{array}{cccccc}
    \hline\hline
                            & {\rm color} & {\rm flavor} & {\rm spin} & {\rm orbital} & {\rm total}
    \\ \hline \hline
    q_1 \leftrightarrow q_4 & {\bf S}     & {\bf A}      & {\bf S}    & {\bf S}       & {\bf A}
    \\
    q_2 \leftrightarrow q_5 & {\bf S}     & {\bf A}      & {\bf S}    & {\bf S}       & {\bf A}
    \\
    q_1 \leftrightarrow q_2 & {\bf A}     & {\bf S}      & {\bf S}    & {\bf S}       & {\bf A}
    \\
    q_1 \leftrightarrow q_5 & {\bf A}     & {\bf S}      & {\bf S}    & {\bf S}       & {\bf A}
    \\
    q_2 \leftrightarrow q_4 & {\bf A}     & {\bf S}      & {\bf S}    & {\bf S}       & {\bf A}
    \\
    q_4 \leftrightarrow q_5 & {\bf A}     & {\bf S}      & {\bf S}    & {\bf S}       & {\bf A}
    \\ \hline
    q_1 \leftrightarrow q_4 & {\bf S}     & {\bf S}      & {\bf A}    & {\bf S}       & {\bf A}
    \\
    q_2 \leftrightarrow q_5 & {\bf S}     & {\bf A}      & {\bf S}    & {\bf S}       & {\bf A}
    \\ \cdots
    \\ \hline
    q_1 \leftrightarrow q_4 & {\bf S}     & {\bf A}      & {\bf S}    & {\bf S}       & {\bf A}
    \\
    q_2 \leftrightarrow q_5 & {\bf S}     & {\bf S}      & {\bf A}    & {\bf S}       & {\bf A}
    \\ \cdots
    \\ \hline
    q_1 \leftrightarrow q_4 & {\bf S}     & {\bf S}      & {\bf A}    & {\bf S}       & {\bf A}
    \\
    q_2 \leftrightarrow q_5 & {\bf S}     & {\bf S}      & {\bf A}    & {\bf S}       & {\bf A}
    \\ \cdots
    \\ \hline\hline
    \end{array}
    \end{equation}
    They satisfy that any two of the four light quarks are totally antisymmetric so that obey the Pauli principle. Hence, two quarks can share (at most) four quarks, with $(I)J^P = (0)2^+/(1)1^+/(2)0^+$. However, neither $\Sigma_c^{(*)}\Lambda_b$ nor $\Lambda_b\Lambda_b$ satisfies this condition.

\end{enumerate}

We apply the above model-independent hypothesis to qualitatively predict more covalent hadronic molecules:
\begin{itemize}

\item Induced by shared light up/down quarks, there can be the $I=0$ $\bar D^{(*)}[\bar c q]$--$B^{(*)}[\bar b q]$, $I=0$ $\bar D^{(*)}[\bar c q]$--$\Xi_c^{(\prime*)}[csq]$, $I=0$ $\Sigma_c^{(*)}[cqq]$--$\Sigma_b^{(*)}[bqq]$, and $I=1/2$ $\Sigma_c^{(*)}[cqq]$--$\Xi_b^{(\prime*)}[bsq]$ covalent molecules, etc. We roughly estimate their binding energies to be at the 10~MeV level, considering the $P_c/P_{cs}$ and $X_0(2900)$~\cite{Aaij:2020hon,Aaij:2020ypa} as possible covalent hadronic molecules. We list them in Table~\ref{tab:result}, which are still waiting to be carefully analysed.

\item If the strange quark is also exchangeable, there can be the $\bar D_s^{(*)}[\bar c s]$--$B^{(*)}[\bar b q]$ and $\bar D_s^{(*)}[\bar c s]$--$\Xi_c^{(\prime*)}[csq]$ covalent molecules, etc. These states are induced by shared light up/down/strange quarks, with the $SU(3)$ light flavor structure to be antisymmetric.


\item If the heavy-quark-exchange interaction is negligible, there might be the $I=0$ $\bar D^{(*)}[\bar c q]$--$\bar D^{(*)}[\bar c q]$, $I=0$ $\Sigma_c^{(*)}[cqq]$--$\Sigma_c^{(*)}[cqq]$, and $I=1/2$ $\Sigma_c^{(*)}[cqq]$--$\Xi_c^{(\prime*)}[csq]$ covalent molecules, etc. These states are still induced by shared light up/down quarks.

\item Especially, we propose to search for the $D^{*-}[\bar c d]$--$K^{*0}[\bar s d]$, $D^{*-}[\bar c d]$--$D^{*-}[\bar c d]$, and $D^{*-}[\bar c d]$--$B^{*0}[\bar b d]$ covalent molecules of $I=1$ and $J=0$. These states might exist, when the two light quarks spin in opposite directions and the two antiquarks also spin in opposite directions, just like the para-hydrogen.

\end{itemize}

\begin{table*}[hptb]
\begin{center}
\renewcommand{\arraystretch}{2}
\caption{Possibly-existing covalent hadronic molecules $|(I)J^P\rangle$ induced by shared light up/down quarks, derived from the hypothesis that {\it the light-quark-exchange interaction is attractive when the shared light quarks are totally antisymmetric so that obey the Pauli principle}. Here, $q$ and $s$ denote the light up/down and strange quarks, respectively; $Q$ and $Q^\prime$ denote two different heavy quarks; the symbols $[AB] = AB - BA$ and $\{AB\} = AB + BA$ denote the antisymmetric and symmetric $SU(3)$ light flavor structures, respectively. The states with $\checkmark$ have been confirmed in QCD sum rule calculations of the present study, but the states with $?$ and $??$ have not. This is because the latter contain relatively-polarized components, while one needs to sum over polarizations of these components within QCD sum rule method and so can not differentiate these hyperfine structures. Moreover, the states with $??$ contain light up/down quarks with the symmetric flavor/isospin structure, whose masses are (probably) considerably larger than their partners with the antisymmetric flavor/isospin structure. The states without any identification are still waiting to be carefully analysed in our future QCD sum rule studies.}
\begin{tabular}{c | c | c | c | c | c | c | c | c }
\hline\hline
                                                     &  $|\bar Q q,{1\over2}0^-\rangle$                    & $|\bar Q q,{1\over2}1^-\rangle$
& $|Q [q q],0{1\over2}^+\rangle$                     & $|Q \{q q\},1{1\over2}^+\rangle$                    & $|Q \{q q\},1{3\over2}^+\rangle$
&  $|Q [s q],{1\over2}{1\over2}^+\rangle$            & $|Q \{s q\},{1\over2}{1\over2}^+\rangle$            & $|Q \{s q\},{1\over2}{3\over2}^+\rangle$
\\ \hline
$|\bar Q^\prime q,{1\over2}0^-\rangle$               & $| (0)0^+ \rangle$                                  & $| (0)1^+ \rangle$~($\checkmark$)
& --                                                 & $| ({1\over2}){1\over2}^- \rangle$~($\checkmark$)   & $| ({1\over2}){3\over2}^- \rangle$~($\checkmark$)
& $| (0){1\over2}^- \rangle$                         & $| (0){1\over2}^- \rangle$                          & $| (0){3\over2}^- \rangle$
\\ \hline
$|\bar Q^\prime q,{1\over2}1^-\rangle$ & &
$\begin{array}{c}
| (0)0^+ \rangle~(?)
\\
| (1)0^+ \rangle~(??)
\\
| (0)1^+ \rangle~(?)
\\
| (1)1^+ \rangle~(??)
\\
| (0)2^+ \rangle~(\checkmark)
\end{array}$
& --
&
$\begin{array}{c}
| ({1\over2}){1\over2}^- \rangle~(?)
\\
| ({3\over2}){1\over2}^- \rangle~(??)
\\
| ({1\over2}){3\over2}^- \rangle~(\checkmark)
\\
| ({3\over2}){3\over2}^- \rangle~(??)
\end{array}$
&
$\begin{array}{c}
| ({1\over2}){1\over2}^- \rangle~(?)
\\
| ({3\over2}){1\over2}^- \rangle~(??)
\\
| ({1\over2}){3\over2}^- \rangle~(?)
\\
| ({3\over2}){3\over2}^- \rangle~(??)
\\
| ({1\over2}){5\over2}^- \rangle~(\checkmark)
\end{array}$
&
$\begin{array}{c}
| (0){1\over2}^- \rangle
\\
| (0){3\over2}^- \rangle
\end{array}$
&
$\begin{array}{c}
| (0){1\over2}^- \rangle
\\
| (1){1\over2}^- \rangle
\\
| (0){3\over2}^- \rangle
\\
| (1){3\over2}^- \rangle
\end{array}$
&
$\begin{array}{c}
| (0){1\over2}^- \rangle
\\
| (1){1\over2}^- \rangle
\\
| (0){3\over2}^- \rangle
\\
| (1){3\over2}^- \rangle
\\
| (0){5\over2}^- \rangle
\end{array}$
\\ \hline
$|Q [q q],0{1\over2}^+\rangle$ & & & -- & -- & -- & -- & -- & --
\\ \hline
$|Q \{q q\},1{1\over2}^+\rangle$ & & & &
$\begin{array}{c}
| (0)1^+ \rangle
\\
| (1)0/1^+ \rangle
\\
| (2)0/1^+ \rangle
\end{array}$
&
$\begin{array}{c}
| (0)1/2^+ \rangle
\\
| (1)1/2^+ \rangle
\\
| (2)1^+ \rangle
\end{array}$
& $| ({1\over2})0/1^+ \rangle$ &
$\begin{array}{c}
| ({1\over2})0/1^+ \rangle
\\
| ({3\over2})0/1^+ \rangle
\end{array}$
&
$\begin{array}{c}
| ({1\over2})1/2^+ \rangle
\\
| ({3\over2})1/2^+ \rangle
\end{array}$
\\ \hline
$|Q \{q q\},1{3\over2}^+\rangle$ & & & & &
$\begin{array}{c}
| (0)1/2/3^+ \rangle
\\
| (1)0/1/2^+ \rangle
\\
| (2)0/1^+ \rangle
\end{array}$
& $| ({1\over2})1/2^+ \rangle$ &
$\begin{array}{c}
| ({1\over2})1/2^+ \rangle
\\
| ({3\over2})1/2^+ \rangle
\end{array}$
&
$\begin{array}{c}
| ({1\over2})0/1/2/3^+ \rangle
\\
| ({3\over2})0/1/2^+ \rangle
\end{array}$
\\ \hline\hline
\end{tabular}
\label{tab:result}
\end{center}
\end{table*}

\section{A toy model to formulize covalent hadronic molecules}
\label{sec:model}

In the previous section we have qualitatively discussed the hypothesis: {\it the light-quark-exchange interaction is attractive when the shared light quarks are totally antisymmetric so that obey the Pauli principle}. In this section we further build a toy model to quantitatively formulize it, and estimate binding energies of some possibly-existing covalent hadronic molecules.

We shall use the following formula to estimate binding energies of the $\bar D^{(*)} \Sigma_c^{(*)}$, $D^{(*)} \bar B^{(*)}$, and $\Sigma_c^{(*)} \Sigma_b^{(*)}$ hadronic molecules:
\begin{equation}
B = N_A A - N_R R - N \epsilon - \kappa \langle \lambda_{c} \cdot \lambda_{\bar c/b} ~ s_{c} \cdot s_{\bar c/b} \rangle \, .
\label{eq:formula}
\end{equation}
This formula will be explained in Sec.~\ref{sec:parameter} in details, and similar formulae are used for some other covalent hadronic molecules. There are altogether four parameters:
\begin{eqnarray}
A &\sim& 30~{\rm MeV} \, ,
\\ R &\sim& 17~{\rm MeV} \, ,
\\ \epsilon &\sim& 6~{\rm MeV} \, ,
\\ \kappa &\sim& 13~{\rm MeV} \, ,
\end{eqnarray}
which are estimated by considering the $P_c/P_{cs}$ and the recently observed $T_{cc}^+$ as possible covalent hadronic molecules.

\subsection{Parameters}
\label{sec:parameter}

%
\begin{figure}[hbtp]
\begin{center}
\includegraphics[width=0.35\textwidth]{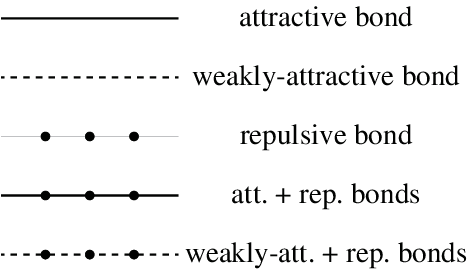}
\caption{(Weakly-)Attractive and repulsive covalent hadronic bonds as well as their combinations. In the present study we take into account the attractive bond and its combination with the repulsive bond, whose bond energies are estimated to be $A \sim 30$~MeV and $A - R \sim 13$~MeV, respectively.}
\label{fig:bond}
\end{center}
\end{figure}
%

As discussed in the previous section, there exists the attractive interaction when exchanging up and down quarks with the configuration:
\begin{equation}\nonumber
\begin{array}{cccccc}
\hline
                           & {\rm color} & {\rm flavor} & {\rm spin} & {\rm orbital} & {\rm total}
\\ \hline
q \leftrightarrow q^\prime & {\bf S}     & {\bf A}      & {\bf S}    & {\bf S}       & {\bf A}
\\ \hline
\end{array}
\end{equation}
These two light up/down quarks have the same color and their relative orbital structure is $S$-wave. Besides, they have the antisymmetric flavor structure and the symmetric spin structure, so with the quantum numbers $(I)J^P = (0)1^+$. We use the attractive bond energy $A$ to describe this attraction, which is estimated to be $A \sim 30$~MeV for each bond, with $N_A$ the number of such bonds. It is illustrated in Fig.~\ref{fig:bond} using the solid curve.

As discussed in the previous section, the two exchanged light up/down quarks can form another configuration of $(I)J^P = (1)0^+$:
\begin{equation}\nonumber
\begin{array}{cccccc}
\hline
                           & {\rm color} & {\rm flavor} & {\rm spin} & {\rm orbital} & {\rm total}
\\ \hline
q \leftrightarrow q^\prime & {\bf S}     & {\bf S}      & {\bf A}    & {\bf S}       & {\bf A}
\\ \hline
\end{array}
\end{equation}
However, its induced interaction is weaker, so we do not take this configuration into account in the present study. It is illustrated in Fig.~\ref{fig:bond} using the dashed curve.

There exists an $(I)J^P = (0)0^+$ up-down quark pair inside the proton/neutron with the configuration:
\begin{equation}\nonumber
\begin{array}{cccccc}
\hline
                        & {\rm color} & {\rm flavor} & {\rm spin} & {\rm orbital} & {\rm total}
\\ \hline
q_1 \leftrightarrow q_2 & {\bf A}     & {\bf A}      & {\bf A}    & {\bf S}       & {\bf A}
\\ \hline
\end{array}
\end{equation}
One can not exchange $q_1$ (nor $q_2$) with another light up/down quark $q_3$, at the same time keeping: a) any two of the three light quarks are totally antisymmetric, and b) the proton/neutron remains unchanged. As an example, we exchange $q_1 \leftrightarrow q_3$ and keep (b), but then (a) is not satisfied:
\begin{equation}\nonumber
\begin{array}{cccccc}
\hline
                        & {\rm color} & {\rm flavor} & {\rm spin} & {\rm orbital} & {\rm total}
\\ \hline
q_1 \leftrightarrow q_2 & {\bf A}     & {\bf A}      & {\bf A}    & {\bf S}       & {\bf A}
\\
q_2 \leftrightarrow q_3 & {\bf A}     & {\bf A}      & {\bf A}    & {\bf S}       & {\bf A}
\\
q_1 \leftrightarrow q_3 & {\bf S}     & {\bf S}      & {\bf S}    & {\bf S}       & {\bf S}
\\ \hline
\end{array}
\end{equation}
Therefore, the above up-down quark pair is in some sense ``saturated''. This suggests that the $(I)J^P = (0)0^+$ up-down quark pairs inside protons/neutrons can not be exchanged, so they are capable of forming repulsive cores in the nucleus.

We use the repulsive bond energy $R$ to describe the repulsion between two repulsive cores, which is estimated to be $R \sim 17$~MeV for each bond, with $N_R$ the number of such bonds. It is illustrated in Fig.~\ref{fig:bond} using the dotted curve. Besides, we shall find in Sec.~\ref{sec:DD} and Sec.~\ref{sec:SS} that the two charm quarks can also form such repulsive cores in $D^{(*)}D^{(*)}$ and $\Sigma_c^{(*)} \Sigma_c^{(*)}$ hadronic molecules, etc.

The third parameter is the residual energy $\epsilon$, which is estimated to be $\epsilon \sim 6$~MeV for each component hadron, with $N$ the number of components. We use it to describe the part of kinetic energy that can not be absorbed into $A$ and $R$.

The fourth parameter $\kappa$ relates to the spin splitting. We use the following term to describe the interaction between the $c$ and $\bar c$ quarks when investigating $\bar D^{(*)} \Sigma_c^{(*)}$ hadronic molecules:
\begin{equation}
\mathcal{H}_{spin} = - \kappa~\langle \lambda_c \cdot \lambda_{\bar c} ~ s_c \cdot s_{\bar c} \rangle \, ,
\label{eq:spin1}
\end{equation}
where $\kappa$ is estimated to be $\kappa \sim 13$~MeV; $s_c$ and $s_{\bar c}$ are spins of the $c$ and $\bar c$ quarks; $\lambda_c$ and $\lambda_{\bar c}$ are their color charges. We use a similar term to describe the interaction between the $c$ and $b$ quarks when investigating $D^{(*)} \bar B^{(*)}$ and $\Sigma_c^{(*)} \Sigma_b^{(*)}$ hadronic molecules:
\begin{equation}
\mathcal{H}^\prime_{spin} = - \kappa~\langle \lambda_c \cdot \lambda_{b} ~ s_c \cdot s_{b} \rangle \, .
\label{eq:spin2}
\end{equation}
However, we do not include such terms when investigating $D^{(*)}D^{(*)}$ and $\Sigma_c^{(*)} \Sigma_c^{(*)}$ hadronic molecules, since the interaction between two charm quarks has been (partly) taken into account in the repulsive bond energy $R$. Note that the spin splitting effect in hadronic molecules still needs to be updated with future experiments, since we do not well understand it at this moment. See Sec.~\ref{sec:spin} for more discussions.

\subsection{Nucleus}
\label{sec:nucleus}

\begin{figure}[hbtp]
\begin{center}
\subfigure[~$^2$H]{\includegraphics[width=0.2\textwidth]{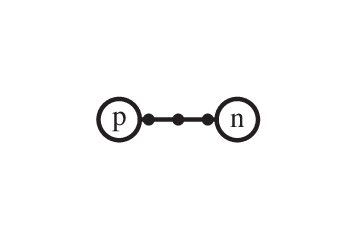}}
\subfigure[~$^3$H]{\includegraphics[width=0.2\textwidth]{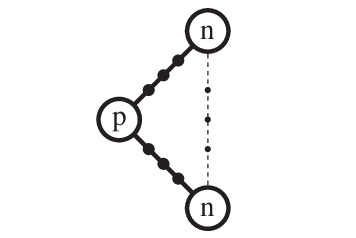}}
\\[3mm]
\subfigure[~$^3$He]{\includegraphics[width=0.2\textwidth]{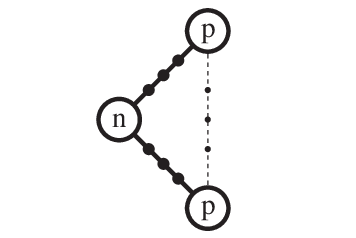}}
\subfigure[~$^4$He]{\includegraphics[width=0.2\textwidth]{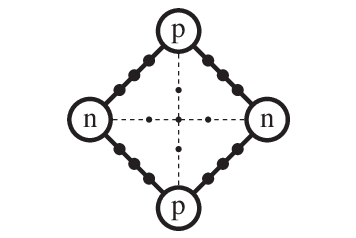}}
\caption{Illustration of the $^2$H, $^3$H, $^3$He, and $^4$He in our model. The shape of $^4$He is a tetrahedron other than a square.}
\label{fig:nucleus}
\end{center}
\end{figure}

Taking the proton/neutron as the combination of an $(I)J^P = (0)0^+$ up-down quark pair together with another up/down quark, we can estimate binding energies of the $^2$H, $^3$H, $^3$He, and $^4$He, as illustrated in Fig.~\ref{fig:nucleus}. In the present study we do not investigate other nuclei consisting of more nucleons, because there are at most two up and two down quarks in the lowest orbit. See Sec.~\ref{sec:strange} for relevant studies on the hypernucleus.

%
%

The $^2$H contains two shared light up/down quarks with the configuration of $(I)J^P = (0)1^+$:
\begin{equation}\nonumber
\begin{array}{cccccc}
\hline
                           & {\rm color} & {\rm flavor} & {\rm spin} & {\rm orbital} & {\rm total}
\\ \hline
q_1 \leftrightarrow q_2    & {\bf S}     & {\bf A}      & {\bf S}    & {\bf S}       & {\bf A}
\\ \hline
\end{array}
\end{equation}
We estimate its binding energy to be
\begin{equation}
B^{^2{\rm H}}_{I=0,J=1} = A - R - 2 \epsilon \sim 1~{\rm MeV} \, .
\end{equation}

The $^3$H and $^3$He both contain three shared light up/down quarks with the configuration of $(I)J^P = (1/2)1/2^+$:
\begin{equation}\nonumber
\begin{array}{cccccc}
\hline
                           & {\rm color} & {\rm flavor} & {\rm spin} & {\rm orbital} & {\rm total}
\\ \hline
q_1 \leftrightarrow q_2    & {\bf S}     & {\bf A}      & {\bf S}    & {\bf S}       & {\bf A}
\\
q_1 \leftrightarrow q_3    & {\bf S}     & {\bf A}      & {\bf S}    & {\bf S}       & {\bf A}
\\
q_2 \leftrightarrow q_3    & {\bf S}     & {\bf S}      & {\bf A}    & {\bf S}       & {\bf A}
\\ \hline
\end{array}
\end{equation}
We estimate their binding energies to be
\begin{equation}
B^{^3{\rm H}/^3{\rm He}}_{I=1/2,J=1/2} = 2A - 2R - 3 \epsilon \sim 8~{\rm MeV} \, .
\end{equation}

The $^4$He contains four shared light up/down quarks with the configuration of $(I)J^P = (0)0^+$:
\begin{equation}\nonumber
\begin{array}{cccccc}
\hline
                           & {\rm color} & {\rm flavor} & {\rm spin} & {\rm orbital} & {\rm total}
\\ \hline
q_1 \leftrightarrow q_2    & {\bf S}     & {\bf S}      & {\bf A}    & {\bf S}       & {\bf A}
\\
q_1 \leftrightarrow q_3    & {\bf S}     & {\bf A}      & {\bf S}    & {\bf S}       & {\bf A}
\\
q_1 \leftrightarrow q_4    & {\bf S}     & {\bf A}      & {\bf S}    & {\bf S}       & {\bf A}
\\
q_2 \leftrightarrow q_3    & {\bf S}     & {\bf A}      & {\bf S}    & {\bf S}       & {\bf A}
\\
q_2 \leftrightarrow q_4    & {\bf S}     & {\bf A}      & {\bf S}    & {\bf S}       & {\bf A}
\\
q_3 \leftrightarrow q_4    & {\bf S}     & {\bf S}      & {\bf A}    & {\bf S}       & {\bf A}
\\ \hline
\end{array}
\end{equation}
We estimate its binding energy to be
\begin{equation}
B^{^4{\rm He}}_{I=0,J=0} = 4A - 4R - 4 \epsilon \sim 28~{\rm MeV} \, .
\end{equation}

\subsection{$D^{(*)}D^{(*)}/D^{(*)} \bar B^{(*)}/\bar B^{(*)} \bar B^{(*)}$ molecules}
\label{sec:DD}

The $(I)J^P=(0)0^+$ $DD$ hadronic molecule does not exist due to the Bose-Einstein statistics. So do the $D^*D^*$ molecules of $(I)J^P=(0)0^+$ and $(I)J^P=(0)2^+$. Actually, we can construct their corresponding currents, and explicitly prove them to be zero.

%
\begin{figure*}[]
\begin{center}
\subfigure[~$\Pi^{D^{(*)}D^{(*)}}_0$]{\includegraphics[width=0.35\textwidth]{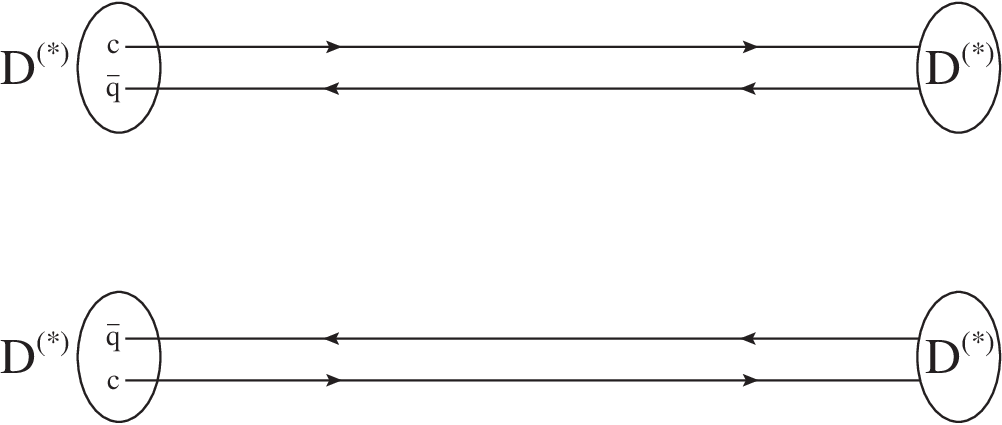}}
~~~~~~~~~~~~~~~~~~~~
\subfigure[~$\Pi^{D^{(*)}D^{(*)}}_q$]{\includegraphics[width=0.35\textwidth]{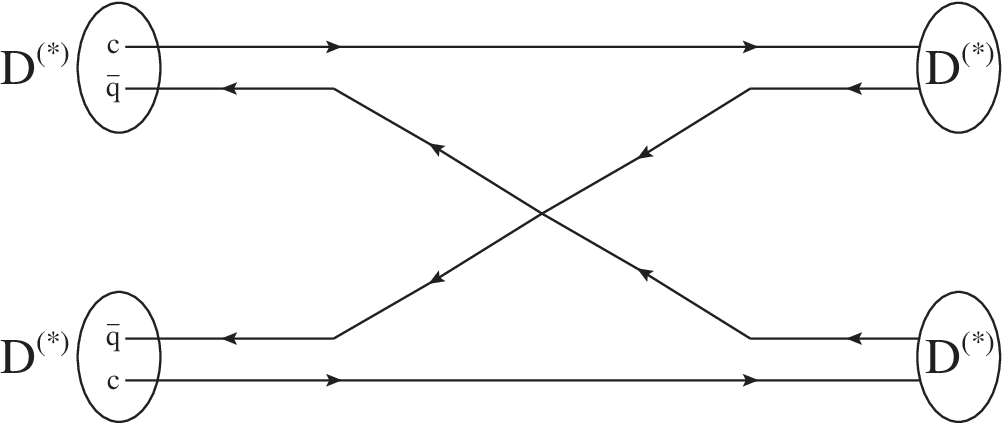}}
\\[10mm]
\subfigure[~$\Pi^{D^{(*)}D^{(*)}}_c$]{\includegraphics[width=0.35\textwidth]{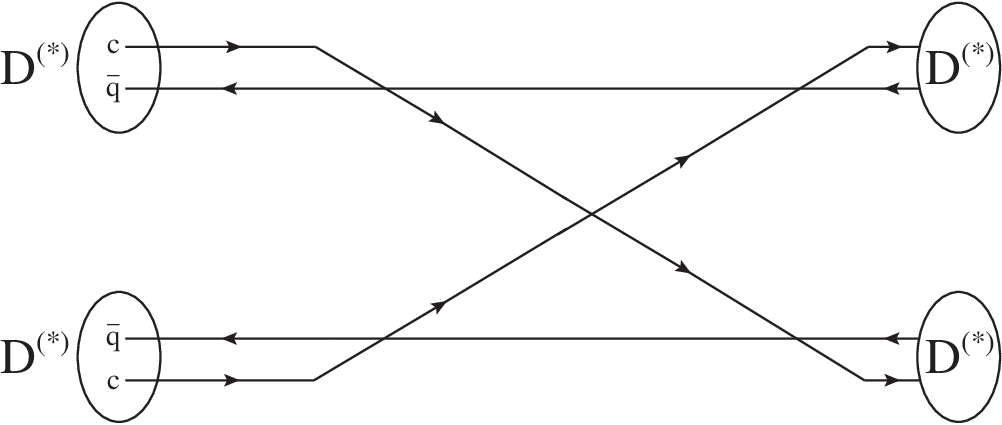}}
~~~~~~~~~~~~~~~~~~~~
\subfigure[~$\Pi^{D^{(*)}D^{(*)}}_{qc}$]{\includegraphics[width=0.35\textwidth]{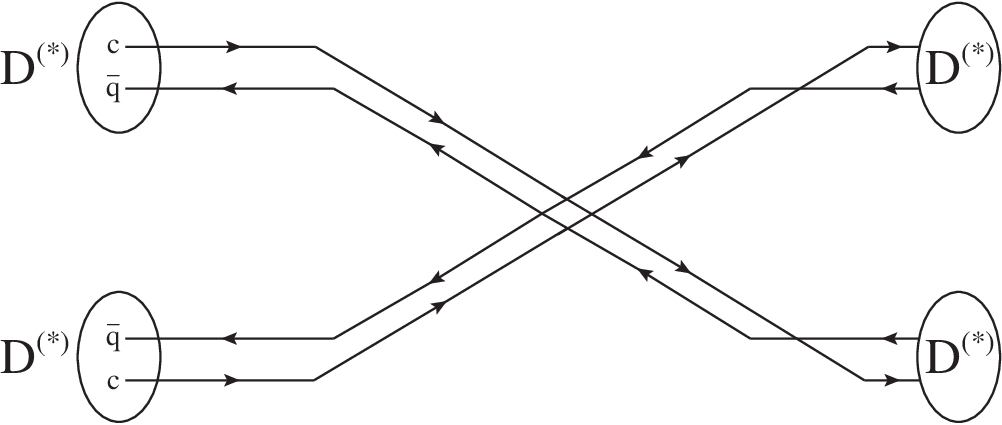}}
\caption{Feynman diagrams between two charmed mesons corresponding to : a) the leading term $\Pi^{D^{(*)}D^{(*)}}_0(x) = \Pi^{D^{(*)}}(x) \times \Pi^{D^{(*)}}(x)$ contributed by two non-correlated charmed mesons, b) the light-quark-exchange interaction $\Pi^{D^{(*)}D^{(*)}}_q$, c) the heavy-quark-exchange interaction $\Pi^{D^{(*)}D^{(*)}}_c$, and d) the interaction $\Pi^{D^{(*)}D^{(*)}}_{qc}$ exchanging both light and heavy quarks. Here $q$ denotes a light up/down quark.}
\label{fig:DD}
\end{center}
\end{figure*}
%

\begin{figure}[hbtp]
\begin{center}
\subfigure[~$DD^*$]{\includegraphics[width=0.2\textwidth]{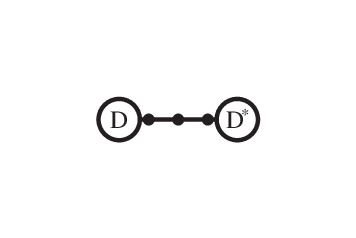}}
\subfigure[~$D\bar B^*$]{\includegraphics[width=0.2\textwidth]{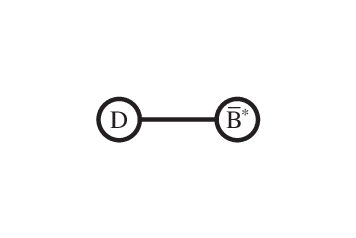}}
\\[3mm]
\subfigure[~$DDD^*$]{\includegraphics[width=0.2\textwidth]{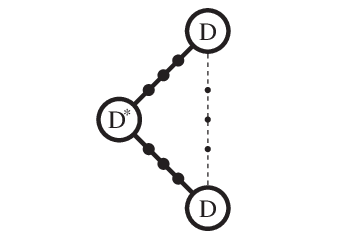}}
\subfigure[~$DD\bar B^*$]{\includegraphics[width=0.2\textwidth]{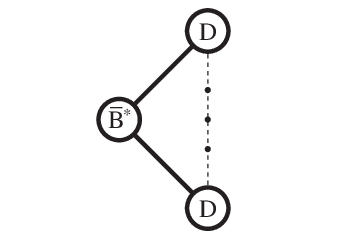}}
\\[3mm]
\subfigure[~$DDD^*D^*$]{\includegraphics[width=0.2\textwidth]{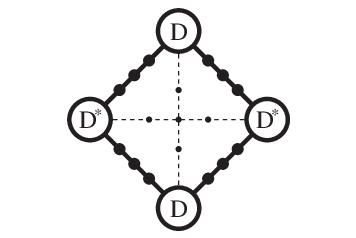}}
\subfigure[~$DD\bar B^*\bar B^*$]{\includegraphics[width=0.2\textwidth]{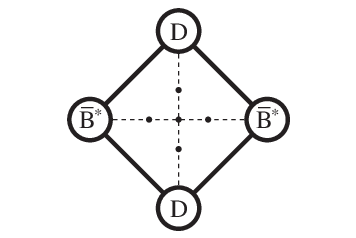}}
\caption{Illustration of the hadronic molecules $DD^*$ and $D\bar B^*$ of $(I)J^P = (0)1^+$, $DDD^*$ and $DD\bar B^*$ of $(I)J^P = (1/2)1^+$, and $DDD^*D^*$ and $DD\bar B^*\bar B^*$ of $(I)J^P = (0)0^+/(0)2^+$ in our model.}
\label{fig:DDs}
\end{center}
\end{figure}

Similar to Sec.~\ref{sec:correlation}, we investigate the $(I)J^P = (0)1^+$ $D D^*$ hadronic molecule through its corresponding current:
\begin{eqnarray}
\sqrt2 J_\alpha^{D D^*} &=& J_\alpha^{D^+ D^{*0}} - J_\alpha^{D^0 D^{*+}}
\label{def:currentDD}
\\ \nonumber &=& J^{D^+} \times J_\alpha^{D^{*0}} - J^{D^0} \times J_\alpha^{D^{*+}}
\\ \nonumber &=& \bar d_a \gamma_5 c_a ~ \bar u_b \gamma_\alpha c_b - \bar u_a \gamma_5 c_a ~ \bar d_b \gamma_\alpha c_b \, .
\end{eqnarray}
Its correlation function
\begin{equation}
\Pi_{\alpha\beta}^{D D^*}(x) = \langle 0 | \mathbb{T}\left[J_\alpha^{D D^*}(x) J_\beta^{D D^* \dagger}(0)\right] | 0 \rangle \, ,
\end{equation}
can be separated into (omitting the subscripts $\alpha\beta$ for simplicity):
\begin{eqnarray}
\Pi^{D D^*}(x) &=& \Pi^{D D^*}_0(x) + \Pi^{D D^*}_{G}(x) + \Pi^{D D^*}_{Q}(x)
\\ \nonumber &=& \Pi^{D D^*}_0(x) + \Pi^{D D^*}_{G}(x)
\\ \nonumber &+& \Pi^{D D^*}_{q}(x) + \Pi^{D D^*}_{c}(x) + \Pi^{D D^*}_{qc}(x) \, ,
\end{eqnarray}
where
\begin{widetext}
\begin{eqnarray}
\Pi^{D D^*}_0(x) &=&  \Pi^{D}(x) \times \Pi^{D^*}(x)
\\ \nonumber &=& - {\bf Tr}\left[{\bf iS}_q^{a^\prime a}(-x) \gamma_5 {\bf iS}_c^{aa^\prime}(x) \gamma_5 \right]~{\bf Tr}\left[{\bf iS}_q^{b^\prime b}(-x) \gamma_\alpha {\bf iS}_c^{bb^\prime}(x) \gamma_\beta \right] \, ,
\\[2mm] \Pi^{D D^*}_{q}(x) &=& - {\bf Tr}\left[{\bf iS}_q^{b^\prime a}(-x) \gamma_5 {\bf iS}_c^{aa^\prime}(x) \gamma_5 {\bf iS}_q^{a^\prime b}(-x) \gamma_\alpha {\bf iS}_c^{bb^\prime}(x) \gamma_\beta \right] \, ,
\\[2mm] \Pi^{D D^*}_{c}(x) &=& + {\bf Tr}\left[{\bf iS}_q^{a^\prime a}(-x) \gamma_5 {\bf iS}_c^{ab^\prime}(x) \gamma_\beta {\bf iS}_q^{b^\prime b}(-x) \gamma_\alpha {\bf iS}_c^{ba^\prime}(x) \gamma_5 \right] \, ,
\\[2mm] \Pi^{D D^*}_{qc}(x) &=& + {\bf Tr}\left[{\bf iS}_q^{b^\prime a}(-x) \gamma_5 {\bf iS}_c^{ab^\prime}(x) \gamma_\beta \right]~{\bf Tr}\left[{\bf iS}_q^{a^\prime b}(-x) \gamma_\alpha {\bf iS}_c^{ba^\prime}(x) \gamma_5 \right] \, .
\end{eqnarray}
\end{widetext}
Their corresponding Feynman diagrams (without condensates) are depicted in Fig.~\ref{fig:DD}. We calculate them using QCD sum rules and find:
\begin{itemize}

\item The term $\Pi^{D D^*}_{qc}(x)$ exchanging both light and heavy quarks simply vanishes, {\it i.e.}, $\Pi^{D D^*}_{qc}(x) = 0$.

\item The term $\Pi^{D D^*}_q(x)$ exchanging light quarks is positive, so its induced interaction is attractive.

\item The term $\Pi^{D D^*}_c(x)$ exchanging heavy charm quarks is negative, so its induced interaction is repulsive.

\item The terms $\Pi^{D D^*}_c(x)$ and $\Pi^{D D^*}_q(x)$ are almost opposite, {\it i.e.}, $\Pi^{D D^*}_{c}(x) \approx -\Pi^{D D^*}_{q}(x)$.

\end{itemize}
Therefore, our QCD sum rule results suggest that the two charm quarks in $D^{(*)}D^{(*)}$ hadronic molecules are capable of forming repulsive cores. Accordingly, we can estimate binding energies of $D^{(*)}D^{(*)}/\bar B^{(*)} \bar B^{(*)}$ hadronic molecules, as illustrated in Fig.~\ref{fig:DDs}(a).

%
%

The $(I)J^P = (0)1^+$ $D D^*$ hadronic molecule contains two shared light antiquarks with the configuration of $(I)J^P = (0)1^+$:
\begin{equation}\nonumber
\begin{array}{cccccc}
\hline
                                     & {\rm color} & {\rm flavor} & {\rm spin} & {\rm orbital} & {\rm total}
\\ \hline
\bar q_1 \leftrightarrow \bar q_2    & {\bf S}     & {\bf A}      & {\bf S}    & {\bf S}       & {\bf A}
\\ \hline
\end{array}
\end{equation}
We estimate its binding energy to be
\begin{equation}
B^{DD^*}_{I=0,J=1} = A - R - 2 \epsilon \sim 1~{\rm MeV} \, ,
\end{equation}
suggesting it possible to interpret the recently observed $T_{cc}^+$~\cite{LHCb:2021auc,LHCb:2021vvq} as the $(I)J^P = (0)1^+$ $D D^*$ covalent hadronic molecule.

The $(I)J^P = (0)1^+$ $D^* D^*/\bar B \bar B^*/\bar B^* \bar B^*$ hadronic molecules have similar binding energies:
\begin{equation}
B^{D^*D^*/\bar B \bar B^*/\bar B^* \bar B^*}_{I=0,J=1} = A - R - 2 \epsilon \sim 1~{\rm MeV} \, .
\end{equation}
However, in the above estimations we have not considered the spin splitting effect, and we have also not considered the long-range light-meson-exchange interaction. These uncertainties prevent us to well determine whether these hadronic molecules exist or not.

Different from $D^{(*)}D^{(*)}$ and $\bar B^{(*)} \bar B^{(*)}$ molecules, the charm and bottom quarks in $D^{(*)}\bar B^{(*)}$ hadronic molecules can not be exchanged, so they are not capable of forming repulsive cores, {\it i.e.}, there does not exist the Feynman diagram corresponding to Fig.~\ref{fig:DD}(c). In this case we include the spin splitting effect described by the parameter $\kappa$, and estimate binding energies of $D^{(*)}\bar B^{(*)}$ hadronic molecules to be
\begin{eqnarray}
\nonumber B^{D \bar B}_{I=0,J=0} &=& A - 2 \epsilon - 1.33 \kappa \sim 1~{\rm MeV} \, ,
\\[1mm] \nonumber B^{D \bar B^*/D^* \bar B}_{I=0,J=1} &=& A - 2 \epsilon + 0.44 \kappa \sim 24~{\rm MeV} \, ,
\\[1mm] B^{D^* \bar B^*}_{I=0,J=0} &=& A - 2 \epsilon - 1.33 \kappa \sim 1~{\rm MeV} \, ,
\\[1mm] \nonumber B^{D^* \bar B^*}_{I=0,J=1} &=& A - 2 \epsilon + 4 \kappa \sim 70~{\rm MeV} \, ,
\\[1mm] \nonumber B^{D^* \bar B^*}_{I=0,J=2} &=& A - 2 \epsilon - 1.33 \kappa \sim 1~{\rm MeV} \, .
\end{eqnarray}
Especially, binding energies of the $(I)J^P = (0)1^+$ $D\bar B^*/D^*\bar B$ covalent hadronic molecules are much larger than those of the $(I)J^P = (0)1^+$ $DD^*/\bar B \bar B^*$ molecules.

In the above estimations we have assumed that the charm and bottom quarks form the symmetric color representation $\mathbf{6}_c$, so that
\begin{equation}
\langle \lambda_c \cdot \lambda_b \rangle = {16 \over 3} \, .
\end{equation}
Besides, we need to perform the spin decomposition from the $| s_{c \bar q} \, , \, s_{b \bar q^\prime} \, ; \, J  \rangle$ basis to the $| s_{c b} \, , \, s_{\bar q \bar q^\prime} \, ; \, J  \rangle$ basis (see Appendix~\ref{app:spin}), and select the components satisfying $s_{\bar q \bar q^\prime} = 1$ to evaluate $\langle s_c \cdot s_b \rangle$. Take the $(I)J^P = (0)1^+$ $D \bar B^*$ molecule as an example, we obtain
\begin{eqnarray}
&& \langle 0_{c \bar q}, 1_{b \bar q^\prime} ; 1 | s_c \cdot s_b | 0_{c \bar q}, 1_{b \bar q^\prime} ; 1 \rangle
\\ \nonumber &\rightarrow& {4\over3} \times \Big( {1\over4} \langle 0_{cb}, 1_{\bar q \bar q^\prime} ; 1 | s_c \cdot s_b | 0_{cb}, 1_{\bar q \bar q^\prime} ; 1 \rangle
\\ \nonumber && ~~~~~~~ + {1\over2} \langle 1_{cb}, 1_{\bar q \bar q^\prime} ; 1 | s_c \cdot s_b | 1_{cb}, 1_{\bar q \bar q^\prime} ; 1 \rangle \Big)
\\ \nonumber &=& - 0.0833 \, .
\end{eqnarray}

We further use $D^{(*)}$ to compose multi-$D^{(*)}$ hadronic molecules, such as the $DDD^*$ and $DDD^*D^*$ molecules, etc. As illustrated in Fig.~\ref{fig:DDs}(c,e), their structures are similar to the $^3$He and $^4$He, respectively. We estimate their binding energies to be:
\begin{eqnarray}
B^{DDD^*}_{I=1/2,J=1} &=& 2 A - 2 R - 3 \epsilon \sim 8~{\rm MeV} \, ,
\\[1mm] \nonumber
B^{DDD^*D^*}_{I=0,J=0,2} &=& 4 A - 4 R - 4 \epsilon \sim 28~{\rm MeV} \, .
\end{eqnarray}
Besides, our model supports the existence of the $D D \bar B^*$ and $D D \bar B^* \bar B^*$ hadronic molecules, etc. As illustrated in Fig.~\ref{fig:DDs}(d,f), they have much larger binding energies:
\begin{eqnarray}
B^{DD\bar B^*}_{I=1/2,J=1} &=& 2 A - 3 \epsilon + 0.89 \kappa \sim 54~{\rm MeV} \, ,
\\[1mm] \nonumber
B^{DD\bar B^*\bar B^*}_{I=0,J=0,2} &=& 4 A - 4 \epsilon + 1.78 \kappa \sim 119~{\rm MeV} \, .
\end{eqnarray}
More examples can be found in Table~\ref{tab:binding}.

\subsection{$\Sigma_c^{(*)}\Sigma_c^{(*)}/\Sigma_c^{(*)} \Sigma_b^{(*)}/\Sigma_b^{(*)} \Sigma_b^{(*)}$ molecules}
\label{sec:SS}

\begin{figure}[hbtp]
\begin{center}
\subfigure[~$\Sigma_c \Sigma_c^*$]{\includegraphics[width=0.2\textwidth]{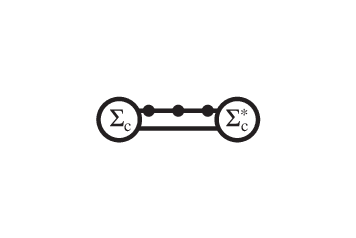}}
\subfigure[~$\Sigma_c \Sigma_b^*$]{\includegraphics[width=0.2\textwidth]{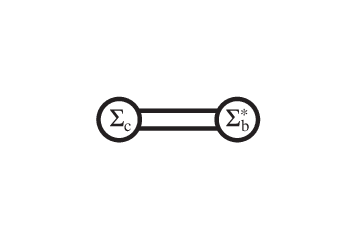}}
\caption{Illustration of the hadronic molecules $\Sigma_c \Sigma_c^*$ and $\Sigma_c \Sigma_b^*$ of $(I)J^P = (0)1^+/(0)2^+$ in our model.}
\label{fig:SS}
\end{center}
\end{figure}

We follow Sec.~\ref{sec:DD} and find that the two charm quarks in $\Sigma_c^{(*)}\Sigma_c^{(*)}$ hadronic molecules also form repulsive cores. Accordingly, we can estimate binding energies of $\Sigma_c^{(*)}\Sigma_c^{(*)}/\Sigma_b^{(*)} \Sigma_b^{(*)}$ hadronic molecules, as illustrated in Fig.~\ref{fig:SS}(a).

%
%

The $(I)J^P = (0)1^+$ $\Sigma_c[q_1q_2c]\Sigma_c[q_3q_4c^\prime]$ hadronic molecule contains four shared light up/down quarks with the configuration of $(I)J^P = (0)2^+$:
\begin{equation}\nonumber
\begin{array}{cccccc}
\hline
                        & {\rm color} & {\rm flavor} & {\rm spin} & {\rm orbital} & {\rm total}
\\ \hline
q_1 \leftrightarrow q_3 & {\bf S}     & {\bf A}      & {\bf S}    & {\bf S}       & {\bf A}
\\
q_2 \leftrightarrow q_4 & {\bf S}     & {\bf A}      & {\bf S}    & {\bf S}       & {\bf A}
\\
q_1 \leftrightarrow q_2 & {\bf A}     & {\bf S}      & {\bf S}    & {\bf S}       & {\bf A}
\\
q_1 \leftrightarrow q_4 & {\bf A}     & {\bf S}      & {\bf S}    & {\bf S}       & {\bf A}
\\
q_2 \leftrightarrow q_3 & {\bf A}     & {\bf S}      & {\bf S}    & {\bf S}       & {\bf A}
\\
q_3 \leftrightarrow q_4 & {\bf A}     & {\bf S}      & {\bf S}    & {\bf S}       & {\bf A}
\\ \hline
\end{array}
\end{equation}
We estimate its binding energy to be
\begin{equation}
B^{\Sigma_c\Sigma_c}_{I=0,J=1} = 2A - R - 2 \epsilon \sim 31~{\rm MeV} \, .
\end{equation}
The $I=0$ $\Sigma_c\Sigma_c^*/\Sigma_c^*\Sigma_c^*$ hadronic molecules have similar binding energies:
\begin{equation}
B^{\Sigma_c\Sigma_c^*/\Sigma_c^*\Sigma_c^*}_{I=0} = 2A - R - 2 \epsilon \sim 31~{\rm MeV} \, .
\end{equation}
So do their corresponding $\Sigma_b^{(*)} \Sigma_b^{(*)}$ hadronic molecules. However, the $(I)J^P = (0)0^+$ $\Sigma_c\Sigma_c/\Sigma_c^*\Sigma_c^*$ and $\Sigma_b\Sigma_b/\Sigma_b^*\Sigma_b^*$ hadronic molecules do not exist due to the Fermi-Dirac statistics.

Different from $\Sigma_c^{(*)} \Sigma_c^{(*)}$ and $\Sigma_b^{(*)} \Sigma_b^{(*)}$ molecules, the charm and bottom quarks in $\Sigma_c^{(*)} \Sigma_b^{(*)}$ hadronic molecules can not be exchanged, so they are not capable of forming repulsive cores. In this case we include the spin splitting effect described by the parameter $\kappa$, and estimate binding energies of $\Sigma_c^{(*)} \Sigma_b^{(*)}$ hadronic molecules to be
\begin{eqnarray}
\nonumber B^{\Sigma_c \Sigma_b}_{I=0,J=1} &=& 2A - 2 \epsilon - 1.33 \kappa \sim 31~{\rm MeV} \, ,
\\[1mm] \nonumber B^{\Sigma_c \Sigma_b^*/\Sigma_c^* \Sigma_b}_{I=0,J=1} &=& 2A - 2 \epsilon - 1.33 \kappa \sim 31~{\rm MeV} \, ,
\\[1mm] B^{\Sigma_c \Sigma_b^*/\Sigma_c^* \Sigma_b}_{I=0,J=2} &=& 2A - 2 \epsilon + 0.8 \kappa \sim 58~{\rm MeV} \, ,
\\[1mm] \nonumber B^{\Sigma_c^* \Sigma_b^*}_{I=0,J=1} &=& 2A - 2 \epsilon - 1.33 \kappa \sim 31~{\rm MeV} \, ,
\\[1mm] \nonumber B^{\Sigma_c^* \Sigma_b^*}_{I=0,J=2} &=& 2A - 2 \epsilon + 4 \kappa \sim 100~{\rm MeV} \, ,
\\[1mm] \nonumber B^{\Sigma_c^* \Sigma_b^*}_{I=0,J=3} &=& 2A - 2 \epsilon - 1.33 \kappa \sim 31~{\rm MeV} \, .
\end{eqnarray}

\subsection{$\bar D^{(*)} \Sigma_c^{(*)}/\bar D^{(*)} \Sigma_b^{(*)}/B^{(*)} \Sigma_c^{(*)}/B^{(*)} \Sigma_b^{(*)}$ molecules}
\label{sec:DS}

The charm and anti-charm quarks in $\bar D^{(*)} \Sigma_c^{(*)}$ hadronic molecules can not be exchanged, so they are not capable of forming repulsive cores. Accordingly, we include the spin splitting effect described by the parameter $\kappa$, and estimate their binding energies to be
\begin{eqnarray}
\nonumber B^{\bar D \Sigma_c}_{I=1/2,J=1/2} &=& A - 2 \epsilon - 0.67 \kappa \sim 9~{\rm MeV} \, ,
\\[1mm] \nonumber B^{\bar D \Sigma_c^*}_{I=1/2,J=3/2} &=& A - 2 \epsilon + 0.33 \kappa \sim 22~{\rm MeV} \, ,
\\[1mm] \nonumber B^{\bar D^* \Sigma_c}_{I=1/2,J=1/2} &=& A - 2 \epsilon - 0.67 \kappa \sim 9~{\rm MeV} \, ,
\\[1mm] B^{\bar D^* \Sigma_c}_{I=1/2,J=3/2} &=& A - 2 \epsilon + 0.33 \kappa \sim 22~{\rm MeV} \, ,
\\[1mm] \nonumber B^{\bar D^* \Sigma_c^*}_{I=1/2,J=1/2} &=& A - 2 \epsilon - 0.67 \kappa \sim 9~{\rm MeV} \, ,
\\[1mm] \nonumber B^{\bar D^* \Sigma_c^*}_{I=1/2,J=3/2} &=& A - 2 \epsilon +1.83 \kappa \sim 42~{\rm MeV} \, ,
\\[1mm] \nonumber B^{\bar D^* \Sigma_c^*}_{I=1/2,J=5/2} &=& A - 2 \epsilon - 0.67 \kappa \sim 9~{\rm MeV} \, .
\end{eqnarray}
Similarly, we can estimate binding energies of $\bar D^{(*)} \Sigma_b^{(*)}/B^{(*)} \Sigma_c^{(*)}/B^{(*)} \Sigma_b^{(*)}$ hadronic molecules, which are the same as their corresponding $\bar D^{(*)} \Sigma_c^{(*)}$ ones.

These hadronic molecules all contain three shared light up/down quarks with the configuration of $(I)J^P = (1/2)3/2^+$:
\begin{equation}\nonumber
\begin{array}{cccccc}
\hline
                        & {\rm color} & {\rm flavor} & {\rm spin} & {\rm orbital} & {\rm total}
\\ \hline
q_1 \leftrightarrow q_2 & {\bf S}     & {\bf A}      & {\bf S}    & {\bf S}       & {\bf A}
\\
q_1 \leftrightarrow q_3 & {\bf A}     & {\bf S}      & {\bf S}    & {\bf S}       & {\bf A}
\\
q_2 \leftrightarrow q_3 & {\bf A}     & {\bf S}      & {\bf S}    & {\bf S}       & {\bf A}
\\ \hline
\end{array}
\end{equation}

%
\begin{figure}[hbtp]
\begin{center}
\subfigure[~$\bar D \Sigma_c$]{\includegraphics[width=0.2\textwidth]{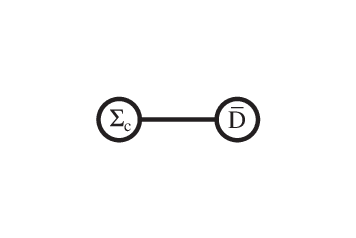}}
\subfigure[~$\bar D \bar D \Sigma_c$]{\includegraphics[width=0.2\textwidth]{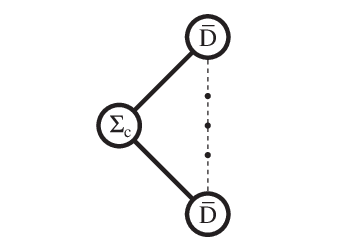}}
\caption{Illustration of the hadronic molecules $\bar D \Sigma_c$ of $(I)J^P = (1/2)1/2^-$ and $\bar D \bar D \Sigma_c$ of $(I)J^P = (0)1/2^+$ in our model.}
\label{fig:DS}
\end{center}
\end{figure}
%

We further use two charmed mesons and one $\Sigma_c^{(*)}$ baryon to compose $\bar D^{(*)} \bar D^{(*)} \Sigma_c^{(*)}$ hadronic molecules. Especially, the $\bar D[\bar c q_1] \bar D[\bar c^\prime q_2] \Sigma_c[q_3 q_4 c]$ illustrated in Fig.~\ref{fig:DS}(b) contains four shared light up/down quarks with the configuration of $(I)J^P = (0)2^+$:
\begin{equation}\nonumber
\begin{array}{cccccc}
\hline
                        & {\rm color} & {\rm flavor} & {\rm spin} & {\rm orbital} & {\rm total}
\\ \hline
q_1 \leftrightarrow q_3 & {\bf S}     & {\bf A}      & {\bf S}    & {\bf S}       & {\bf A}
\\
q_2 \leftrightarrow q_4 & {\bf S}     & {\bf A}      & {\bf S}    & {\bf S}       & {\bf A}
\\
q_1 \leftrightarrow q_2 & {\bf A}     & {\bf S}      & {\bf S}    & {\bf S}       & {\bf A}
\\
q_1 \leftrightarrow q_4 & {\bf A}     & {\bf S}      & {\bf S}    & {\bf S}       & {\bf A}
\\
q_2 \leftrightarrow q_3 & {\bf A}     & {\bf S}      & {\bf S}    & {\bf S}       & {\bf A}
\\
q_3 \leftrightarrow q_4 & {\bf A}     & {\bf S}      & {\bf S}    & {\bf S}       & {\bf A}
\\ \hline
\end{array}
\end{equation}
We estimate its binding energy to be:
\begin{equation}
B^{\bar D \bar D \Sigma_c}_{I=0,J=1/2} = 2 A - 3 \epsilon - 1.33 \kappa \sim 25~{\rm MeV} \, .
\end{equation}
More examples can be found in Table~\ref{tab:binding}.

\subsection{Molecules with strangeness}
\label{sec:strange}

In the previous subsections we only consider the up and down quarks as exchanged light quarks, and in this subsection we further take the light strange quark into account. We introduce another parameter $S$ to describe the attractive interaction induced by the shared strange and up/down quarks with the configuration of either $(I)J^P = (1/2)0^+$ or $(1/2)1^+$:
\begin{equation}\nonumber
\begin{array}{cccccc}
\hline\hline
                           & {\rm color} & {\rm flavor} & {\rm spin} & {\rm orbital} & {\rm total}
\\ \hline\hline
s \leftrightarrow q        & {\bf S}     & {\bf S}      & {\bf A}    & {\bf S}       & {\bf A}
\\ \hline
s \leftrightarrow q        & {\bf S}     & {\bf A}      & {\bf S}    & {\bf S}       & {\bf A}
\\ \hline\hline
\end{array}
\end{equation}
This parameter is estimated to be $S \sim 20$~MeV for each bond, with $N_S$ the number of such bonds. We still use the solid curve to illustrate it, but this solid curve is slightly thinner than that denoting the attractive bond $A$.

\begin{figure}[hbtp]
\begin{center}
\subfigure[~$^3_\Lambda$H]{\includegraphics[width=0.2\textwidth]{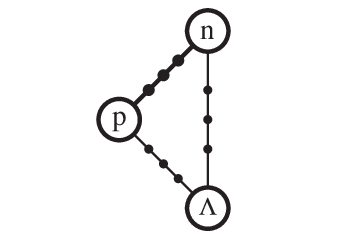}}
\subfigure[~$^4_\Lambda$H]{\includegraphics[width=0.2\textwidth]{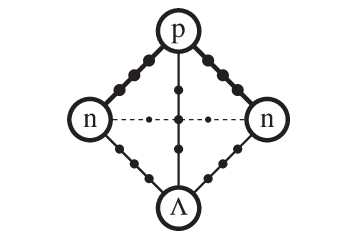}}
\\[3mm]
\subfigure[~$^5_\Lambda$He]{\includegraphics[width=0.2\textwidth]{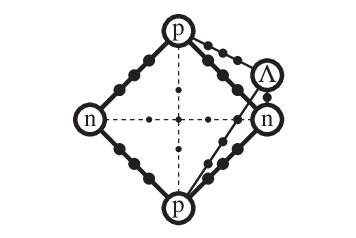}}
\subfigure[~$^{~~6}_{\Lambda\Lambda}$He]{\includegraphics[width=0.2\textwidth]{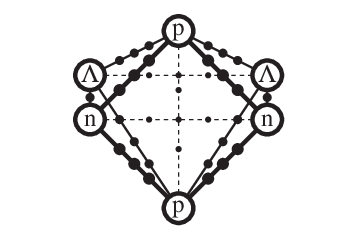}}
\caption{Illustration of the $^3_\Lambda$H, $^4_\Lambda$H, $^5_\Lambda$He, and $^{~~6}_{\Lambda\Lambda}$He in our model. The shape of $ppnn$ in the subfigures (c) and (d) is a tetrahedron other than a square.}
\label{fig:strange}
\end{center}
\end{figure}

Taking the $\Lambda$ hyperon as the combination of an $(I)J^P = (0)0^+$ up-down quark pair together with an strange quark, we can estimate binding energies of some hypernuclei. Considering that there are at most two up, two down, and two strange quarks in the lowest orbit, we obtain:
\begin{eqnarray}
\nonumber B^{^3_\Lambda{\rm H}}_{I=0} &=& A + 2S - 3R - 3 \epsilon \sim 1~{\rm MeV} \, ,
\\[1mm]
B^{^4_\Lambda{\rm H}/^4_\Lambda{\rm He}}_{I=1/2} &=& 2A + 3S - 5R - 4 \epsilon \sim 11~{\rm MeV} \, ,
\\[1mm] \nonumber
B^{^5_\Lambda{\rm He}}_{I=0} &=& 4A + 3S - 7R - 5 \epsilon \sim 31~{\rm MeV} \, ,
\\[1mm] \nonumber
B^{^{~~6}_{\Lambda\Lambda}{\rm He}}_{I=0} &=& 4A + 6S - 10R - 6 \epsilon \sim 34~{\rm MeV} \, .
\end{eqnarray}
We illustrate these hypernuclei in Fig.~\ref{fig:strange}. More possibly-existing covalent hadronic molecules with strangeness can be found in Table~\ref{tab:binding}.
%
%

\subsection{Discussions on the spin splitting effect}
\label{sec:spin}

In the present study we investigate the spin splitting effect through Eqs.~(\ref{eq:spin1}) and (\ref{eq:spin2}) with the same parameter $\kappa \sim 13$~MeV. These two equations are used for $\bar D^{(*)} \Sigma_c^{(*)}$ and $D^{(*)} \bar B^{(*)}/\Sigma_c^{(*)} \Sigma_b^{(*)}$ hadronic molecules, respectively; while we do not include such terms when investigating $D^{(*)}D^{(*)}/\Sigma_c^{(*)} \Sigma_c^{(*)}$ hadronic molecules, since the interaction between two charm quarks has been (partly) taken into account in the repulsive bond energy $R$.

Because we do not well understand the spin splitting effect in hadronic molecules at this moment, we still need to update it with future experiments, and there can be other approaches better describing it. We take  $\bar D^{(*)} \Sigma_c^{(*)}$ hadronic molecules as an example, and discuss several possible improvements.

Firstly, it is possible to use two different $\kappa$'s for Eqs.~(\ref{eq:spin1}) and (\ref{eq:spin2}). It is also possible to use some formulae other than Eqs.~(\ref{eq:spin1}) and (\ref{eq:spin2}), such as the spin-spin interaction,
\begin{equation}
\mathcal{H}^{\prime\prime}_{spin} = - \kappa^\prime ~ \langle s_{\bar D^{(*)}} \cdot s_{\Sigma_c^{(*)}} \rangle \, ,
\end{equation}
or the spin-orbit interaction, etc.

Secondly, in the present study we perform the spin decomposition from the $| s_{\bar c q_1} \, , \,  s_{q_2 q_3 c} \, ; \, J  \rangle_{s_{q_2 q_3}={ 1}}$ basis to the $| s_{\bar c c} \, , \, s_{q_1 q_2 q_3} \, ; \, J  \rangle_{s_{q_2 q_3}={ 1}}$ basis, and select the components satisfying $s_{q_1 q_2 q_3} = 3/2$ to evaluate $\langle s_c \cdot s_{\bar c} \rangle$. This is because the three light up/down quarks of $(I)J^P = (1/2)3/2^+$ are assumed to supply the attraction forming $\bar D^{(*)} \Sigma_c^{(*)}$ hadronic molecules in our model, while the two heavy quarks $c/\bar c$ are assumed to be freely polarised. If this is not the case and $c/\bar c$ are also fully polarised, there might be two $\bar D \Sigma_c^{*}$ hadronic molecules with either $s_{\bar c c} = 0$ or $s_{\bar c c} = 1$ (see Appendix~\ref{app:spin}):
\begin{eqnarray}
\nonumber | s_{\bar c q_1} \, , \,  s_{q_2 q_3 c} \, ; \, J  \rangle_{s_{q_2 q_3}={ 1}} &\to& | s_{\bar c c} \, , \, s_{q_1 q_2 q_3} \, ; \, J  \rangle_{s_{q_2 q_3}={ 1}} \, ,
\\
| { 0}, { 3\over2}; {3\over2} \rangle &\to& | { 0}, { 3\over2}; {3\over2} \rangle \, ,
\\
| { 0}, { 3\over2}; { 3\over2} \rangle &\to& | { 1}, { 3\over2}; {3\over2} \rangle \, .
\end{eqnarray}
It can be seen from the recent LHCb experiment~\cite{LHCb:2021chn} that there might be two peaks near the $\bar D \Sigma_c^{*}$ threshold, and we propose to further study them as well as the $P_c(4440)^+$ and $P_c(4457)^+$ in order to better understand the spin splitting effect of hadronic molecules in future experiments.

To end this section, we further simply our toy model by neglecting the spin splitting effect described by the parameter $\kappa \sim 13$~MeV, and estimate binding energies of some possibly-existing covalent hadronic molecules through the simplified formula,
\begin{equation}
B = N_A A + N_S S - N_R R - N \epsilon \, ,
\end{equation}
which still have four parameters $A \sim 30$~MeV, $S \sim 20$~MeV, $R \sim 17$~MeV, and $\epsilon \sim 6$~MeV. The obtained results are summarized in Table~\ref{tab:binding},

Very quickly, we arrive at the unique feature of our covalent hadronic molecule picture: binding energies of the $(I)J^P = (0)1^+$ $D\bar B^*/D^* \bar B$ hadronic molecules are much larger than those of the $(I)J^P = (0)1^+$ $DD^*/\bar B \bar B^*$ ones, while the $(I)J^P = (1/2)1/2^+$ $\bar D \Sigma_c/\bar D \Sigma_b/B \Sigma_c/B \Sigma_b$ hadronic molecules have similar binding energies. This is due to that the two identical heavy quarks in $DD^*/\bar B \bar B^*$ hadronic molecules form repulsive cores, but the two different heavy quarks in $D\bar B^*/D^* \bar B$ and $\bar D \Sigma_c/\bar D \Sigma_b/B \Sigma_c/B \Sigma_b$ hadronic molecules do not.

\begin{table*}[hptb]
\begin{center}
\renewcommand{\arraystretch}{2}
\caption{Binding energies of some possibly-existing covalent hadronic molecules, estimated in our toy model through the simplified formula $B = N_A A + N_S S - N_R R - N \epsilon$, with $A \sim 30$~MeV, $S \sim 20$~MeV, $R \sim 17$~MeV, and $\epsilon \sim 6$~MeV. We do not take into account the spin splitting effect described by the parameter $\kappa \sim 13$~MeV here.}
\begin{tabular}{c | c || c | c}
\hline\hline
~~~Molecules~~~ & ~~~Binding energies~~~ & ~~~Molecules~~~ & ~~~Binding energies~~~
\\ \hline \hline
$^2$H, $D^*D^{(*)}/\bar B^*\bar B^{(*)}$                                                      &   1~MeV
&
$D^{(*)} \bar B^{(*)}$                                                                        &   18~MeV
\\ \hline
$^3$H/$^3$He, $D^*D^{(*)}D^{(*)}/\bar B^*\bar B^{(*)}\bar B^{(*)}$                            &   8~MeV
&
$D^{(*)}D^{(*)}\bar B^{(*)}/D^{(*)}\bar B^{(*)}\bar B^{(*)}$                                  &   42~MeV
\\ \hline
\multirow{2}{*}{$^4$He, $D^*D^*D^{(*)}D^{(*)}/\bar B^*\bar B^*\bar B^{(*)}\bar B^{(*)}$}      &   \multirow{2}{*}{28~MeV}
&
$D^*D^{(*)}D^{(*)}\bar B^{(*)}/D^{(*)}\bar B^{(*)}\bar B^{(*)}\bar B^*$                       &   62~MeV
\\ \cline{3-4}
& &
$D^{(*)}D^{(*)}\bar B^{(*)}\bar B^{(*)}$                                                      &   96~MeV
\\ \hline\hline
$\Sigma_c^{(*)}\Sigma_c^{(*)}/\Sigma_b^{(*)}\Sigma_b^{(*)}$                                   &   31~MeV
&
$\Sigma_c^{(*)}\Sigma_b^{(*)}$                                                                &   48~MeV
\\ \hline\hline
\multicolumn{2}{c||}{} &
$\bar D^{(*)} \Sigma_c^{(*)}/\bar D^{(*)} \Sigma_b^{(*)}/B^{(*)} \Sigma_c^{(*)}/B^{(*)} \Sigma_b^{(*)}$ & 18~MeV
\\ \cline{3-4}
\multicolumn{2}{c||}{} &
$\bar D^{(*)}\bar D^{(*)}\Sigma_c^{(*)}$                                                      &   42~MeV
\\ \hline\hline
& &
$D^{(*)} \bar B^{(*)}_s$                                                                      &   8~MeV
\\ \hline
$^3_\Lambda$H, $D^*D^{(*)}D_s^{(*)}$                                                          &   1~MeV
&
$D^{(*)}D^{(*)}\bar B_s^{(*)}$                                                                &   35~MeV
\\ \hline
$^4_\Lambda$H/$^4_\Lambda$He, $D^*D^{(*)}D^{(*)}D_s^{(*)}$                                    &   11~MeV
&
$D^*D^{(*)}D^{(*)}\bar B_s^{(*)}$                                                             &   62~MeV
\\ \hline
$^5_\Lambda$He, $D^*D^*D^{(*)}D^{(*)}D_s^{(*)}$                                               &   31~MeV
&
$D^*D^*D^{(*)}D^{(*)}\bar B_s^{(*)}$                                                          &   82~MeV
\\ \hline
$^{~~6}_{\Lambda\Lambda}$He, $D^*D^*D^{(*)}D^{(*)}D_s^{(*)}D_s^{(*)}$                         &   34~MeV
&
$D^*D^*D^{(*)}D^{(*)}\bar B_s^{(*)}\bar B_s^{(*)}$                                            &   136~MeV
\\ \hline\hline
$\Sigma_c^{(*)}\Xi_c^{(\prime*)}$                                                             &   21~MeV
&
$\Sigma_c^{(*)}\Xi_b^{(\prime*)}$                                                             &   38~MeV
\\ \hline
$\Xi_c^{(\prime*)}\Xi_c^{(\prime*)}$                                                          &   11~MeV
&
$\Xi_c^{(\prime*)}\Xi_b^{(\prime*)}$                                                          &   28~MeV
\\ \hline
$\Sigma_c^{(*)}\Xi_c^{(\prime*)}\Xi_c^{(\prime*)}$                                            &   71~MeV
&
$\Sigma_c^{(*)}\Xi_c^{(\prime*)}\Xi_b^{(\prime*)}$                                            &   105~MeV
\\ \hline\hline
\multicolumn{2}{c||}{} &
$\bar D^{(*)} \Xi_c^{(\prime*)}$                                                              &   18~MeV
\\ \cline{3-4}
\multicolumn{2}{c||}{} &
$\bar D^{(*)}\bar D^{(*)}\Xi_c^{(\prime*)}$                                                    &   35~MeV
\\ \cline{3-4}
\multicolumn{2}{c||}{} &
$\bar D^{(*)}\bar D^{(*)}\Xi_c^{(\prime*)}\Xi_b^{(\prime*)}$                                  &   116~MeV
\\ \hline\hline
\end{tabular}
\label{tab:binding}
\end{center}
\end{table*}

\section{Summary and Discussions}
\label{sec:summary}

In this paper we systematically examine Feynman diagrams corresponding to the $\bar D^{(*)} \Sigma_c^{(*)}$, $\bar D^{(*)} \Lambda_c$, $D^{(*)} \bar K^{*}$, and $D^{(*)} \bar D^{(*)}$ hadronic molecular states. Take the $\bar D \Sigma_c$ molecule as an example, first we calculate correlation functions of $\bar D$ and $\Sigma_c$ in the coordinate space to be $\Pi^{\bar D}(x)$ and $\Pi^{\Sigma_c}(x)$, respectively. Then we calculate correlation functions of the $D^-\Sigma_c^{++}$, $\bar D^0 \Sigma_c^{+}$, $I=1/2$ $\bar D \Sigma_c$, and $I=3/2$ $\bar D \Sigma_c$ molecules, separated into:
\begin{eqnarray}
\nonumber \Pi^{D^-\Sigma_c^{++}}(x) &=& \Pi^{\bar D \Sigma_c}_0(x) + \Pi^{\bar D \Sigma_c}_{G}(x) \, ,
\\[2mm] \nonumber \Pi^{\bar D^0 \Sigma_c^{+}}(x) &=& \Pi^{\bar D \Sigma_c}_0(x) + \Pi^{\bar D \Sigma_c}_{G}(x) - \Pi^{\bar D \Sigma_c}_{Q}(x) \, ,
\\[2mm] \nonumber \Pi_{I=1/2}^{\bar D \Sigma_c}(x) &=& \Pi^{\bar D \Sigma_c}_0(x) + \Pi^{\bar D \Sigma_c}_{G}(x) + \Pi^{\bar D \Sigma_c}_{Q}(x) \, ,
\\[2mm] \nonumber \Pi_{I=3/2}^{\bar D \Sigma_c}(x) &=& \Pi^{\bar D \Sigma_c}_0(x) + \Pi^{\bar D \Sigma_c}_{G}(x) - 2\Pi^{\bar D \Sigma_c}_{Q}(x) \, ,
\end{eqnarray}
where $\Pi^{\bar D\Sigma_c}_0(x) = \Pi^{\bar D}(x) \times \Pi^{\Sigma_c}(x)$ is the leading term contributed by non-correlated $\bar D$ and $\Sigma_c$; $\Pi^{\bar D \Sigma_c}_{G}$ describes the double-gluon-exchange interaction between them, but we do not take it into account in the present study, because the other term $\Pi^{\bar D \Sigma_c}_{Q}$ is much larger.

The term $\Pi^{\bar D \Sigma_c}_{Q}$ describes the light-quark-exchange effect between $\bar D$ and $\Sigma_c$, {\it i.e.}, $\bar D$ and $\Sigma_c$ are exchanging and so sharing light up/down quarks. We systematically study it using the method of QCD sum rules, and calculate the mass correction $\Delta M$ induced by this term. Note that the obtained results can be further applied to study production and decay properties of these hadronic molecular states~\cite{Chen:2020pac,Chen:2020opr,Chen:2021erj}. The parameter $\Delta M$ is actually not the binding energy, because we are using local currents in QCD sum rule analyses. We can relate it to some potential $V(r)$ between $\bar D$ and $\Sigma_c$ induced by exchanged/shared light quarks, satisfying:
\begin{eqnarray}
\nonumber V(|r|=0) &=& \Delta M \, ,
\\ \nonumber
V(|r| \rightarrow \infty) &\rightarrow& 0 \, .
\end{eqnarray}

We systematically investigate the light-quark-exchange term $\Pi_{Q}$, and study its contributions to the $\bar D^{(*)} \Sigma_c^{(*)}$, $\bar D^{(*)} \Lambda_c$, $D^{(*)} \bar K^{*}$, and $D^{(*)} \bar D^{(*)}$ hadronic molecular states. We calculate their mass corrections, some of which are listed here:
\begin{eqnarray}
 \nonumber\Delta M^{\bar D \Sigma_c}_{I=1/2,J=1/2} &=& - 95~{\rm MeV} \, ,
\\[1mm] \nonumber \Delta M^{\bar D^* \Sigma_c}_{I=1/2,J=3/2} &=& - 89~{\rm MeV} \, ,
\\[1mm] \nonumber \Delta M^{\bar D \Sigma_c^*}_{I=1/2,J=3/2} &=& - 86~{\rm MeV} \, ,
\\[1mm] \nonumber \Delta M^{\bar D^* \Sigma_c^*}_{I=1/2,J=5/2} &=& -107~{\rm MeV} \, ,
\\[1mm] \nonumber \Delta M^{D \bar K^{*}}_{I=0,J=1} &=& -180~{\rm MeV} \, ,
\\[1mm] \nonumber \Delta M^{D^* \bar K^{*}}_{I=0,J=2} &=& -119~{\rm MeV} \, .
\end{eqnarray}
These results suggest their corresponding light-quark-exchange potentials $V(r)$ to be attractive, so there can be
\begin{itemize}[leftmargin=15pt]

\item the $\bar D \Sigma_c$ covalent molecule of $I=1/2$ and $J=1/2$,

\item the $\bar D^* \Sigma_c$ covalent molecule of $I=1/2$ and $J=3/2$,

\item the $\bar D \Sigma_c^*$ covalent molecule of $I=1/2$ and $J=3/2$,

\item the $\bar D^* \Sigma_c^*$ covalent molecule of $I=1/2$ and $J=5/2$,

\item the $D \bar K^{*}$ covalent molecule of $I=0$ and $J=1$,

\item the $D^* \bar K^{*}$ covalent molecule of $I=0$ and $J=2$.

\end{itemize}
The binding mechanism induced by shared light quarks is somewhat similar to the covalent bond in chemical molecules induced by shared electrons, so we call such hadronic molecules ``covalent hadronic molecules''. Different from the chemical molecules, the internal structure of the hadrons and so the covalent hadronic molecules is much more complicated, since they contain not only the valence quarks but also the sea quarks and gluons. In the present study we consider the covalent hadronic molecules induced only by the shared valence quarks.

Recalling that the two shared electrons must spin in opposite directions (and so totally antisymmetric obeying the Pauli principle) in order to form a chemical covalent bond, our QCD sum rule results indicate a similar hypothesis: {\it the light-quark-exchange interaction is attractive when the shared light quarks are totally antisymmetric so that obey the Pauli principle}.

Its logical chain is quite straightforward. We assume the two light quarks $q_A$ inside $Y$ and $q_B$ inside $Z$ are totally antisymmetric. Hence, $q_A$ and $q_B$ obey the Pauli principle, so that they can be exchanged and shared. By doing this, wave-functions of $Y$ and $Z$ overlap with each other, so that they are attracted and there can be the covalent hadronic molecule $X = | YZ \rangle$. This picture has been depicted in Fig.~\ref{fig:sharing}. We believe it better and more important than our QCD sum rule results, given it to be model independent and more easily applicable.

We apply the above hypothesis to reanalysis the $\bar D^{(*)} \Sigma_c^{(*)}$, $\bar D^{(*)} \Lambda_c$, $D^{(*)} \bar K^{*}$, and $D^{(*)} \bar D^{(*)}$ hadronic molecules, and the obtained results are generally consistent with our QCD sum rule results. However, there can be more hyperfine structures allowed/predictd by the hypothesis, similar to the case of para-hydrogen and ortho-hydrogen with the two protons spinning in opposite directions and in the same direction, respectively. These hyperfine structures can not be differentiated in our QCD sum rule studies, since there we need to sum over polarizations.

We also apply the above hypothesis to predict more possibly-existing covalent hadronic molecules, as summarized in Table~\ref{tab:result}. We build a toy model to formulize this picture and estimate their binding energies. Our model has four parameters, which are fixed by considering the $P_c/P_{cs}$ and the recently observed $T_{cc}^+$ as possible covalent hadronic molecules. Some simplified results neglecting the spin splitting effect are summarized in Table~\ref{tab:binding}. Note that these results are obtained from the light-quark-exchange interaction only, and there can be some other interactions among hadrons, {\it e.g.}, the light-quark-exchange term $\Pi_{Q}(x)$ does not contribute to the $D^{(*)} \bar D^{(*)}$ molecules, suggesting that the $D^{(*)} \bar D^{(*)}$ covalent hadronic molecules do not exist, but there can still be the $D^{(*)} \bar D^{(*)}$ molecules possibly induced by the one-meson-exchange interaction. These interactions can also contribute to the binding energies given in Table~\ref{tab:binding}, which is one source of their theoretical uncertainties.

A unique feature of our covalent hadronic molecule picture is that binding energies of the $(I)J^P = (0)1^+$ $D\bar B^*/D^* \bar B$ hadronic molecules are much larger than those of the $(I)J^P = (0)1^+$ $DD^*/\bar B \bar B^*$ ones, while the $(I)J^P = (1/2)1/2^+$ $\bar D \Sigma_c/\bar D \Sigma_b/B \Sigma_c/B \Sigma_b$ hadronic molecules have similar binding energies. This is due to that the two identical heavy quarks in $DD^*/\bar B \bar B^*$ hadronic molecules form repulsive cores, but the two different heavy quarks in $D\bar B^*/D^* \bar B$ and $\bar D \Sigma_c/\bar D \Sigma_b/B \Sigma_c/B \Sigma_b$ hadronic molecules do not.

To end this paper, we note again that the one-meson-exchange interaction $\Pi_{M}$ at the hadron level and the light-quark-exchange interaction $\Pi_{Q}$ at the quark-gluon level can overlap with each other. Hence, we attempt to understand the nuclear force based on the picture of covalent hadronic molecules very roughly: a) the $(I)J^P = (0)0^+$ up-down quark pairs inside protons/neutrons are in some sense ``saturated'', so they can not be exchanged and form repulsive cores in the nucleus; b) the other up/down quarks inside protons/neutrons can be freely exchanged/shared/moving, inducing some interactions among nucleons; c) in the multi-nucleon nucleus there can be many up/down quarks being shared, so its binding mechanism transfers into the ``metallic'' hadronic bond. Based on these understandings, we estimate binding energies of some nuclei using our toy model, as summarized in Table~\ref{tab:binding}. Finally, we propose another possibly-existing binding mechanism similar to the ``ionic'' bond, but it might only be observable in the quark-gluon plasma.

\section*{Acknowledgments}

We thank Yan-Rui Liu and Li-Ming Zhang for helpful discussions.
This project is supported by the National Natural Science Foundation of China under Grants No.~11722540 and No.~12075019,
the Jiangsu Provincial Double-Innovation Program under Grant No.~JSSCRC2021488,
and
the Fundamental Research Funds for the Central Universities.

\appendix

\begin{widetext}
\section{Spectral densities}
\label{app:ope}
\allowdisplaybreaks[4]

In this appendix we list spectral densities extracted from the currents $J^{D^- \Sigma_c^{++}}$, $J^{\bar D^0 \Sigma_c^+}$, $J^{\bar D \Sigma_c}$ of $I=1/2$, and $J_{I=3/2}^{\bar D \Sigma_c}$. These currents are defined in Eqs.~(\ref{def:currentDm}), (\ref{def:currentD0}), (\ref{def:currentDhalf}), and (\ref{def:currentD3half}), respectively. In the following expressions, $\mathcal{F}(s) = \FF(s)$, $\mathcal{H}(s) = \HH(s)$, and the integration limits are $\alpha_{min}=\frac{1-\sqrt{1-4m_c^2/s}}{2}$, $\alpha_{max}=\frac{1+\sqrt{1-4m_c^2/s}}{2}$, $\beta_{min}=\frac{\alpha m_c^2}{\alpha s-m_c^2}$, and $\beta_{max}=1-\alpha$.

The spectral density $\rho^{\bar D \Sigma_c}(s)$ extracted from the current $J^{\bar D \Sigma_c}$ of $I=1/2$ can be separated into
\begin{equation}
\rho^{\bar D \Sigma_c}(s) = \rho_0^{\bar D \Sigma_c}(s) + \rho_Q^{\bar D \Sigma_c}(s) \, .
\end{equation}
The leading term $\rho_0^{\bar D \Sigma_c}(s)$ is
\begin{eqnarray}
\nonumber \rho_0^{\bar D \Sigma_c}(s) &=& m_c ~ \left( \rho^{pert}_{1}(s) + \rho^{\qq}_{1}(s) + \rho^{\GGa}_{1}(s)+ \rho^{\qGqa}_{1}(s) + \rho^{\qq^2}_{1}(s)  + \rho^{\qq\qGqa}_{1}(s)+ \rho^{\qGqa^2}_{1}(s) + \rho^{\qq^3}_{1}(s) \right)
\\ \nonumber &+& q\!\!\!\slash ~ \left( \rho^{pert}_{2}(s) + \rho^{\qq}_{2}(s) + \rho^{\GGa}_{2}(s)+ \rho^{\qGqa}_{2}(s) + \rho^{\qq^2}_{2}(s)  + \rho^{\qq\qGqa}_{2}(s)+ \rho^{\qGqa^2}_{2}(s) + \rho^{\qq^3}_{2}(s) \right) \, ,
\\ \label{ope:J1L}
\end{eqnarray}
where
\begin{eqnarray}
\nonumber \rho^{pert}_{1}(s) &=& \dab \Bigg\{ \mathcal{F}(s)^5 \times       \frac{(1 - \alpha - \beta)^3}{81920 \pi ^8 \alpha ^5 \beta ^4}                      \Bigg\} \, ,
\non
\rho^{\qq}_{1}(s) &=& {m_c \qq } \dab \Bigg\{ \mathcal{F}(s)^3 \times       \frac{-(1 - \alpha - \beta)^2}{1024 \pi ^6 \alpha ^3 \beta ^3}                  \Bigg\} \, ,
\non
\rho^{\GGa}_{1}(s) &=& {\GGb } \dab\Bigg\{ m_c^2 \mathcal{F}(s)^2    \times  \frac{(1 - \alpha - \beta)^3 \left(\alpha ^3+\beta ^3\right)}{98304 \pi ^8 \alpha ^5 \beta ^4}
\non && ~~~~~~ + \mathcal{F}(s)^3 \times           \frac{(\alpha +\beta -1) \left(3 \alpha ^3+\alpha ^2 (8 \beta -3)+\alpha  (\beta -1) \beta -(\beta -1)^2 \beta \right)}{98304 \pi ^8 \alpha ^5 \beta ^3}  \Bigg\} \, ,
\non
\rho^{\qGqa}_{1}(s) &=& {m_c\qGqb } \dab \Bigg\{ \mathcal{F}(s)^2 \times  \frac{3(1 - \alpha - \beta) (\alpha +2 \beta -1)}{2048 \pi ^6 \alpha ^2 \beta ^3}      \Bigg\} \, ,
\non
\rho^{\qq^2}_{1}(s)&=& {\qq^2 } \dab \Bigg\{ \mathcal{F}(s)^2 \times     \frac{-1}{64 \pi ^4 \alpha ^2 \beta }       \Bigg\} \, ,
\non
\rho^{\qq\qGqa}_{1}(s)&=& {\qq\qGqb } \int^{\alpha_{max}}_{\alpha_{min}}{\rm d}\alpha  \Bigg\{ \int^{\beta_{max}}_{\beta_{min}}{\rm d}\beta \Bigg\{ \mathcal{F}(s) \times  \frac{-1}{128 \pi ^4 \alpha ^2}     \Bigg\}
+  \mathcal{H}(s) \times     \frac{1}{64 \pi ^4 \alpha }     \Bigg\} \, ,
\non
\rho^{\qGqa^2}_{1}(s)&=& {\qGqb^2 } \Bigg\{\int^{\alpha_{max}}_{\alpha_{min}}{\rm d}\alpha \Bigg\{ \frac{(\alpha -1) (2 \alpha -1)}{512 \pi ^4 \alpha } \Bigg\}
+ \int^{1}_{0}{\rm d}\alpha \Bigg\{ m_c^2  \delta\left(s - {m_c^2 \over \alpha(1-\alpha)}\right) \times   \frac{-1}{512 \pi ^4 \alpha }      \Bigg\}\Bigg\} \, ,
\non
\rho^{\qq^3}_{1}(s)&=& {m_c\qq^3 } \int^{\alpha_{max}}_{\alpha_{min}}{\rm d}\alpha \Bigg\{    \frac{1}{24 \pi ^2}     \Bigg\} \, ,
\non
\rho^{pert}_{2}(s) &=& \dab \Bigg\{ \mathcal{F}(s)^5 \times              \frac{(1 - \alpha - \beta)^3}{40960 \pi ^8 \alpha ^4 \beta ^4}        \Bigg\} \, ,
\non
\rho^{\qq}_{2}(s) &=& {m_c \qq } \dab \Bigg\{ \mathcal{F}(s)^3 \times       \frac{-(1 - \alpha - \beta)^2}{512 \pi ^6 \alpha ^2 \beta ^3}              \Bigg\} \, ,
\non
\rho^{\GGa}_{2}(s) &=& {\GGb } \dab\Bigg\{ m_c^2 \mathcal{F}(s)^2    \times  \frac{(1 - \alpha - \beta)^3 \left(\alpha ^3+\beta ^3\right)}{49152 \pi ^8 \alpha ^4 \beta ^4}
\non && ~~~~~~ + \mathcal{F}(s)^3 \times                                    \frac{(\alpha +\beta -1)^2 (2 \alpha +\beta )}{32768 \pi ^8 \alpha ^3 \beta ^3}            \Bigg\} \, ,
\non
\rho^{\qGqa}_{2}(s) &=& {m_c\qGqb } \dab \Bigg\{ \mathcal{F}(s)^2 \times   \frac{3 (1 - \alpha - \beta) (\alpha +2 \beta -1)}{1024 \pi ^6 \alpha  \beta ^3}   \Bigg\} \, ,
\non
\rho^{\qq^2}_{2}(s)&=& {\qq^2 } \dab \Bigg\{ \mathcal{F}(s)^2 \times    \frac{-1}{128 \pi ^4 \alpha  \beta }        \Bigg\} \, ,
\non
\rho^{\qq\qGqa}_{2}(s)&=& {\qq\qGqb } \int^{\alpha_{max}}_{\alpha_{min}}{\rm d}\alpha \Bigg\{ \int^{\beta_{max}}_{\beta_{min}}{\rm d}\beta \Bigg\{ \mathcal{F}(s) \times   \frac{-1}{256 \pi ^4 \alpha }     \Bigg\}
+  \mathcal{H}(s) \times \frac{1}{128 \pi ^4}  \Bigg\} \, ,
\non
\rho^{\qGqa^2}_{2}(s)&=& {\qGqb^2 } \Bigg\{ \int^{\alpha_{max}}_{\alpha_{min}}{\rm d}\alpha \Bigg\{ \frac{(\alpha -1) (2 \alpha -1)}{1024 \pi ^4}  \Bigg\}
+ \int^{1}_{0}{\rm d}\alpha \Bigg\{  m_c^2 \delta\left(s - {m_c^2 \over \alpha(1-\alpha)}\right) \times  \frac{-1}{1024 \pi ^4}  \Bigg\} \Bigg\} \, ,
\non
\rho^{\qq^3}_{2}(s)&=& {m_c\qq^3 } \int^{\alpha_{max}}_{\alpha_{min}}{\rm d}\alpha \Bigg\{   \frac{\alpha}{48 \pi ^2}   \Bigg\} \, .
\end{eqnarray}
The light-quark-exchange term $\rho_Q^{\bar D \Sigma_c}(s)$ is
\begin{eqnarray}
\nonumber \rho_Q^{\bar D \Sigma_c}(s) &=& m_c ~ \left( \rho^{pert}_{3}(s) + \rho^{\qq}_{3}(s) + \rho^{\GGa}_{3}(s)+ \rho^{\qGqa}_{3}(s) + \rho^{\qq^2}_{3}(s)  + \rho^{\qq\qGqa}_{3}(s)+ \rho^{\qGqa^2}_{3}(s) + \rho^{\qq^3}_{3}(s) \right)
\\ \nonumber &+& q\!\!\!\slash ~ \left( \rho^{pert}_{4}(s) + \rho^{\qq}_{4}(s) + \rho^{\GGa}_{4}(s)+ \rho^{\qGqa}_{4}(s) + \rho^{\qq^2}_{4}(s)  + \rho^{\qq\qGqa}_{4}(s)+ \rho^{\qGqa^2}_{4}(s) + \rho^{\qq^3}_{4}(s) \right) \, ,
\\ \label{ope:J1Q}
\end{eqnarray}
where
\begin{eqnarray}
\nonumber \rho^{pert}_{3}(s) &=& \dab \Bigg\{ \mathcal{F}(s)^5 \times       \frac{(1 - \alpha - \beta)^3}{983040 \pi ^8 \alpha ^5 \beta ^4}                 \Bigg\} \, ,
\non
\rho^{\qq}_{3}(s) &=& {m_c \qq } \dab \Bigg\{ \mathcal{F}(s)^3 \times       \frac{-(1 - \alpha - \beta)^2}{3072 \pi ^6 \alpha ^3 \beta ^3}          \Bigg\} \, ,
\non
\rho^{\GGa}_{3}(s) &=& {\GGb } \dab\Bigg\{ m_c^2 \mathcal{F}(s)^2    \times  \frac{(1 - \alpha - \beta)^3 \left(\alpha ^3+\beta ^3\right)}{1179648 \pi ^8 \alpha ^5 \beta ^4}
\non && ~~~~~~ + \mathcal{F}(s)^3 \times          \frac{(\alpha +\beta -1)^2 \left(8 \alpha ^2+6 \alpha  \beta +\alpha -2 (\beta -1) \beta \right)}{2359296 \pi ^8 \alpha ^5 \beta ^3}  \Bigg\} \, ,
\non
\rho^{\qGqa}_{3}(s) &=& {m_c\qGqb } \dab \Bigg\{ \mathcal{F}(s)^2 \times  \frac{(1 - \alpha - \beta) \left(2 \alpha ^2+\alpha  (6 \beta -2)+(\beta -1) \beta \right)}{8192 \pi ^6 \alpha ^3 \beta ^3}    \Bigg\} \, ,
\non
\rho^{\qq^2}_{3}(s)&=& {\qq^2 } \dab \Bigg\{ \mathcal{F}(s)^2 \times     \frac{-5}{1536 \pi ^4 \alpha ^2 \beta }    \Bigg\} \, ,
\non
\rho^{\qq\qGqa}_{3}(s)&=& {\qq\qGqb } \int^{\alpha_{max}}_{\alpha_{min}}{\rm d}\alpha  \Bigg\{ \int^{\beta_{max}}_{\beta_{min}}{\rm d}\beta \Bigg\{ \mathcal{F}(s) \times  \frac{-6 \alpha -5 \beta}{3072 \pi ^4 \alpha ^2 \beta }   \Bigg\}
+  \mathcal{H}(s) \times  \frac{7 }{3072 \pi ^4 \alpha } \Bigg\} \, ,
\non
\rho^{\qGqa^2}_{3}(s)&=& {\qGqb^2 } \Bigg\{\int^{\alpha_{max}}_{\alpha_{min}}{\rm d}\alpha \Bigg\{ \frac{4 \alpha ^2-3 \alpha +5}{12288 \pi ^4 \alpha } \Bigg\}
+ \int^{1}_{0}{\rm d}\alpha \Bigg\{ m_c^2  \delta\left(s - {m_c^2 \over \alpha(1-\alpha)}\right) \times   \frac{-1}{6144 \pi ^4 \alpha }   \Bigg\}\Bigg\} \, ,
\non
\rho^{\qq^3}_{3}(s)&=& {m_c\qq^3 } \int^{\alpha_{max}}_{\alpha_{min}}{\rm d}\alpha \Bigg\{  \frac{1}{288 \pi ^2}  \Bigg\} \, ,
\non
\rho^{pert}_{4}(s) &=& \dab \Bigg\{ \mathcal{F}(s)^5 \times             \frac{(1 -\alpha - \beta)^3}{491520 \pi ^8 \alpha ^4 \beta ^4}     \Bigg\} \, ,
\non
\rho^{\qq}_{4}(s) &=& {m_c \qq } \dab \Bigg\{ \mathcal{F}(s)^3 \times     \frac{-5(1 - \alpha - \beta)^2}{12288 \pi ^6 \alpha ^2 \beta ^3}          \Bigg\} \, ,
\non
\rho^{\GGa}_{4}(s) &=& {\GGb } \dab\Bigg\{ m_c^2 \mathcal{F}(s)^2    \times  \frac{(1 - \alpha - \beta)^3 \left(\alpha ^3+\beta ^3\right)}{589824 \pi ^8 \alpha ^4 \beta ^4}
\non && ~~~~~~ + \mathcal{F}(s)^3 \times                                  \frac{(\alpha +\beta -1) \left(23 \alpha ^2+\alpha  (7 \beta -22)+8 \beta ^2-7 \beta -1\right)}{2359296 \pi ^8 \alpha ^3 \beta ^3}     \Bigg\} \, ,
\non
\rho^{\qGqa}_{4}(s) &=& {m_c\qGqb } \dab \Bigg\{ \mathcal{F}(s)^2 \times  \frac{(1 - \alpha - \beta) \left(14 \alpha ^2+\alpha  (25 \beta -14)+3 (\beta -1) \beta \right)}{32768 \pi ^6 \alpha ^2 \beta ^3}  \Bigg\} \, ,
\non
\rho^{\qq^2}_{4}(s)&=& {\qq^2 } \dab \Bigg\{ \mathcal{F}(s)^2 \times   \frac{-1}{384 \pi ^4 \alpha  \beta }    \Bigg\} \, ,
\non
\rho^{\qq\qGqa}_{4}(s)&=& {\qq\qGqb } \int^{\alpha_{max}}_{\alpha_{min}}{\rm d}\alpha \Bigg\{ \int^{\beta_{max}}_{\beta_{min}}{\rm d}\beta \Bigg\{ \mathcal{F}(s) \times  \frac{-5 \alpha -3 \beta}{3072 \pi ^4 \alpha  \beta }   \Bigg\}
+  \mathcal{H}(s) \times \frac{7}{3072 \pi ^4} \Bigg\} \, ,
\non
\rho^{\qGqa^2}_{4}(s)&=& {\qGqb^2 } \Bigg\{ \int^{\alpha_{max}}_{\alpha_{min}}{\rm d}\alpha \Bigg\{ \frac{6 \alpha ^2-4 \alpha +3}{12288 \pi ^4} \Bigg\}
+ \int^{1}_{0}{\rm d}\alpha \Bigg\{  \frac{-1}{4096 \pi ^4} \Bigg\} \Bigg\} \, ,
\non
\rho^{\qq^3}_{4}(s)&=& {m_c\qq^3 } \int^{\alpha_{max}}_{\alpha_{min}}{\rm d}\alpha \Bigg\{ \frac{\alpha}{576 \pi ^2} \Bigg\} \, .
\end{eqnarray}

The spectral densities $\rho^{D^- \Sigma_c^{++}}(s)$, $\rho^{\bar D^0 \Sigma_c^+}(s)$, and $\rho_{I=3/2}^{\bar D \Sigma_c}(s)$ extracted from the currents $J^{D^- \Sigma_c^{++}}$, $J^{\bar D^0 \Sigma_c^+}$, and $J_{I=3/2}^{\bar D \Sigma_c}$ are related to $\rho^{\bar D \Sigma_c}(s)$ through
\begin{eqnarray}
\rho^{D^- \Sigma_c^{++}}(s) &=& \rho_0^{\bar D \Sigma_c}(s) \, ,
\\ \rho^{\bar D^0 \Sigma_c^+}(s) &=& \rho_0^{\bar D \Sigma_c}(s) - \rho_Q^{\bar D \Sigma_c}(s) \, ,
\\ \rho_{I=3/2}^{\bar D \Sigma_c}(s) &=& \rho_0^{\bar D \Sigma_c}(s) - 2\rho_Q^{\bar D \Sigma_c}(s) \, .
\end{eqnarray}

\section{Spin decompositions}
\label{app:spin}

In this appendix we list some spin decompositions used in the present study. The transitions from the $| s_{c \bar q} \, , \, s_{c^\prime \bar q^\prime} \, ; \, J  \rangle$ basis to the $| s_{c c^\prime} \, , \, s_{\bar q \bar q^\prime} \, ; \, J  \rangle$ basis are:
\begin{eqnarray}
| s_{c \bar q} \, , \, s_{c^\prime \bar q^\prime} \, ; \, J  \rangle &\rightarrow& | s_{c c^\prime} \, , \, s_{\bar q \bar q^\prime} \, ; \, J  \rangle \, ,
\\ \nonumber
| { 0}, { 0}; { 0} \rangle &=& {1\over2} ~ | { 0}, { 0}; { 0} \rangle + {\sqrt3\over2} ~ | { 1}, { 1}; { 0} \rangle  \, ,
\\ \nonumber
| { 0}, { 1}; { 1} \rangle &=& {1\over2} ~ | { 0}, { 1}; { 1} \rangle - {1\over2} ~ | { 1}, { 0}; { 1} \rangle  + {1\over\sqrt2} ~ | { 1}, { 1}; { 1} \rangle  \, ,
\\ \nonumber
| { 1}, { 0}; { 1} \rangle &=& - {1\over2} ~ | { 0}, { 1}; { 1} \rangle + {1\over2} ~ | { 1}, { 0}; { 1} \rangle  + {1\over\sqrt2} ~ | { 1}, { 1}; { 1} \rangle  \, ,
\\ \nonumber
| { 1}, { 1}; { 0} \rangle &=& {\sqrt3\over2} ~ | { 0}, { 0}; { 0} \rangle - {1\over2} ~ | { 1}, { 1}; { 0} \rangle  \, ,
\\ \nonumber
| { 1}, { 1}; { 1} \rangle &=& {1\over\sqrt2} ~ | { 0}, { 1}; { 1} \rangle + {1\over\sqrt2} ~ | { 1}, { 0}; { 1} \rangle  \, ,
\\ \nonumber
| { 1}, { 1}; { 2} \rangle &=& | { 1}, { 1}; { 2} \rangle \, .
\end{eqnarray}
The transitions from the $| s_{\bar c q_1} \, , \,  s_{q_2 q_3 c} \, ; \, J  \rangle_{s_{q_2 q_3}={ 1}}$ basis to the $| s_{\bar c c} \, , \, s_{q_1 q_2 q_3} \, ; \, J  \rangle_{s_{q_2 q_3}={ 1}}$ basis are:
\begin{eqnarray}
| s_{\bar c q_1} \, , \,  s_{q_2 q_3 c} \, ; \, J  \rangle_{s_{q_2 q_3}={ 1}} &\rightarrow& | s_{\bar c c} \, , \, s_{q_1 q_2 q_3} \, ; \, J  \rangle_{s_{q_2 q_3}={ 1}} \, ,
\\ \nonumber
| { 0}, { 1\over2}; { 1\over2} \rangle &=& {1\over2} ~ | { 0}, { 1\over2}; { 1\over2} \rangle - {1\over\sqrt{12}} ~ | { 1}, { 1\over2}; { 1\over2} \rangle - \sqrt{2\over3} ~ | { 1}, { 3\over2}; { 1\over2} \rangle  \, ,
\\ \nonumber
| { 1}, { 1\over2}; { 1\over2} \rangle &=& -{1\over\sqrt{12}} ~ | { 0}, { 1\over2}; { 1\over2} \rangle + {5\over6} ~ | { 1}, { 1\over2}; { 1\over2} \rangle - {\sqrt2\over3} ~ | { 1}, { 3\over2}; { 1\over2} \rangle  \, ,
\\ \nonumber
| { 1}, { 1\over2}; { 3\over2} \rangle &=& {1\over\sqrt3} ~ | { 0}, { 3\over2}; { 3\over2} \rangle + {1\over3} ~ | { 1}, { 1\over2}; { 3\over2} \rangle + {\sqrt5\over3} ~ | { 1}, { 3\over2}; { 3\over2} \rangle  \, ,
\\ \nonumber
| { 0}, { 3\over2}; { 3\over2} \rangle &=& {1\over2} ~ | { 0}, { 3\over2}; { 3\over2} \rangle + {1\over\sqrt3} ~ | { 1}, { 1\over2}; { 3\over2} \rangle - \sqrt{5\over12} ~ | { 1}, { 3\over2}; { 3\over2} \rangle  \, ,
\\ \nonumber
| { 1}, { 3\over2}; { 1\over2} \rangle &=& -\sqrt{2\over3} ~ | { 0}, { 1\over2}; { 1\over2} \rangle - {\sqrt2\over3} ~ | { 1}, { 1\over2}; { 1\over2} \rangle - {1\over3} ~ | { 1}, { 3\over2}; { 1\over2} \rangle  \, ,
\\ \nonumber
| { 1}, { 3\over2}; { 3\over2} \rangle &=& -\sqrt{5\over12} ~ | { 0}, { 3\over2}; { 3\over2} \rangle + {\sqrt5\over3} ~ | { 1}, { 1\over2}; { 3\over2} \rangle + {1\over6} ~ | { 1}, { 3\over2}; { 3\over2} \rangle  \, ,
\\ \nonumber
| { 1}, { 3\over2}; { 5\over2} \rangle &=& | { 1}, { 3\over2}; { 5\over2} \rangle \, .
\end{eqnarray}
The transitions from the $| s_{c_1 q_3 q_4} \, , \,  s_{c_2 q_5 q_6} \, ; \, J  \rangle_{s_{q_3 q_4}={ 1},s_{q_5 q_6}={ 1}}$ basis to the $| s_{c_1 c_2} \, , \,  s_{q_3 q_4 q_5 q_6} \, ; \, J  \rangle_{s_{q_3 q_4}={ 1},s_{q_5 q_6}={ 1}}$ basis are:
\begin{eqnarray}
| s_{c_1 q_3 q_4} \, , \,  s_{c_2 q_5 q_6} \, ; \, J  \rangle_{s_{q_3 q_4}={ 1},s_{q_5 q_6}={ 1}} &\rightarrow& | s_{c_1 c_2} \, , \,  s_{q_3 q_4 q_5 q_6} \, ; \, J  \rangle_{s_{q_3 q_4}={ 1},s_{q_5 q_6}={ 1}} \, ,
\\ \nonumber
| { 1\over2}, { 1\over2}; { 0} \rangle &=& {1\over\sqrt3} ~ | { 0}, { 0}; { 0} \rangle + \sqrt{2\over3} ~ | { 1}, { 1}; { 0} \rangle \, ,
\\ \nonumber
| { 1\over2}, { 1\over2}; { 1} \rangle &=& {\sqrt2\over3} ~ | { 0}, { 1}; { 1} \rangle - {1\over\sqrt{27}} ~ | { 1}, { 0}; { 1} \rangle + \sqrt{20\over27} ~ | { 1}, { 2}; { 1} \rangle \, ,
\\ \nonumber
| { 1\over2}, { 3\over2}; { 1} \rangle &=& {1\over3} ~ | { 0}, { 1}; { 1} \rangle - \sqrt{8\over{27}} ~ | { 1}, { 0}; { 1} \rangle + {1\over\sqrt2} ~ | { 1}, { 1}; { 1} \rangle - \sqrt{5\over54} ~ | { 1}, { 2}; { 1} \rangle \, ,
\\ \nonumber
| { 1\over2}, { 3\over2}; { 2} \rangle &=& -{1\over\sqrt3} ~ | { 0}, { 2}; { 2} \rangle + {1\over\sqrt6} ~ | { 1}, { 1}; { 2} \rangle + {1\over\sqrt2} ~ | { 1}, { 2}; { 2} \rangle \, ,
\\ \nonumber
| { 3\over2}, { 1\over2}; { 1} \rangle &=& {1\over3} ~ | { 0}, { 1}; { 1} \rangle - \sqrt{8\over{27}} ~ | { 1}, { 0}; { 1} \rangle - {1\over\sqrt2} ~ | { 1}, { 1}; { 1} \rangle - \sqrt{5\over54} ~ | { 1}, { 2}; { 1} \rangle \, ,
\\ \nonumber
| { 3\over2}, { 1\over2}; { 2} \rangle &=& {1\over\sqrt3} ~ | { 0}, { 2}; { 2} \rangle - {1\over\sqrt6} ~ | { 1}, { 1}; { 2} \rangle + {1\over\sqrt2} ~ | { 1}, { 2}; { 2} \rangle \, ,
\\ \nonumber
| { 3\over2}, { 3\over2}; { 0} \rangle &=& \sqrt{2\over3} ~ | { 0}, { 0}; { 0} \rangle - {1\over\sqrt3} ~ | { 1}, { 1}; { 0} \rangle \, ,
\\ \nonumber
| { 3\over2}, { 3\over2}; { 1} \rangle &=& {\sqrt5\over3} ~ | { 0}, { 1}; { 1} \rangle + \sqrt{10\over{27}} ~ | { 1}, { 0}; { 1} \rangle - \sqrt{2\over27} ~ | { 1}, { 2}; { 1} \rangle \, ,
\\ \nonumber
| { 3\over2}, { 3\over2}; { 2} \rangle &=& {1\over\sqrt3} ~ | { 0}, { 2}; { 2} \rangle + \sqrt{2\over3} ~ | { 1}, { 1}; { 2} \rangle \, ,
\\ \nonumber
| { 3\over2}, { 3\over2}; { 3} \rangle &=& | { 1}, { 2}; { 3} \rangle \, .
\end{eqnarray}

\end{widetext}

\end{document}